\documentclass[longauth]{aa}  
\usepackage{graphicx}
\usepackage{txfonts}
\usepackage[colorlinks=true,allcolors=blue]{hyperref}
\usepackage{color}
\usepackage{multirow}
\usepackage{ulem}
\usepackage{booktabs}

\usepackage{xcolor}

\begin{document} 
\title{CHEX-MATE: Are we getting cluster thermodynamics right?}% A simulation-based assessment of X-ray profile recovery}
%\subtitle{AAA}
\author{
R. Seppi\inst{\ref{unige}}\thanks{E-mail: riccardo.seppi@unige.ch} \and
D. Eckert\inst{\ref{unige}} \and
E. Rasia\inst{\ref{inaftrieste}, \ref{umich}} \and
S. T. Kay\inst{\ref{umanchester}} \and
K. Dolag\inst{\ref{origins}, \ref{LMU}, \ref{mpa}} \and
V. Biffi\inst{\ref{inaftrieste}, \ref{ifpu}} \and \\
Y. E. Bahar\inst{\ref{inafBO}} \and 
H. Bourdin \inst{\ref{UniR2},\ref{INFN-R2}} \and
F. De Luca \inst{\ref{UniR2},\ref{INFN-R2}} \and
M. De Petris\inst{\ref{Sapienza}} \and
S. Ettori\inst{\ref{inafBO}, \ref{INFN}} \and
M. Gaspari\inst{\ref{unimodena}} \and
F. Gastaldello\inst{\ref{inafMI}} \and
V. Ghirardini\inst{\ref{inafBO}} \and
L. Lovisari\inst{\ref{inafMI},\ref{cfa}} \and
P. Mazzotta \inst{\ref{UniR2},\ref{INFN-R2}} \and
G. W. Pratt\inst{\ref{saclay}} \and
E. Pointecouteau\inst{\ref{cnes}} \and
M. Rossetti\inst{\ref{inafMI}} \and
J. Sayers\inst{\ref{caltech}} \and
M. Sereno\inst{\ref{inafBO},\ref{INFN}} \and
G. Yepes\inst{\ref{uniMadrid}, \ref{ciaff}}
}
\institute{
Department of Astronomy, University of Geneva, Ch. d’Ecogia 16, CH-1290 Versoix, Switzerland \label{unige} \and
INAF – Osservatorio Astronomico di Trieste, via Tiepolo 11, I-34131 Trieste, Italy \label{inaftrieste} \and
Department of Physics, University of Michigan, Ann Arbor, MI 48109, USA \label{umich} \and
Jodrell Bank Centre for Astrophysics, Department of Physics and Astronomy, The University of Manchester, Oxford Road, Manchester M13 9PL, UK \label{umanchester}\and
Excellence Cluster ORIGINS, Boltzmannstr. 2, D-85748 Garching bei M\"{u}nchen, Germany \label{origins} \and
Universit\"{a}ts-Sternwarte, Fakult\"{a}t f\"{u}r Physik, Ludwig-Maximilians-Universit\"{a}t M\"{u}nchen, Scheinerstr.1, 81679 M\"{u}nchen, Germany \label{LMU} \and
Max-Planck-Institut f\"{u}r Astrophysik, Karl-Schwarzschildstr. 1, 85741 Garching bei M\"{u}nchen, Germany \label{mpa} \and
IFPU, Institute for Fundamental Physics of the Universe, Via Beirut 2, I-34014 Trieste, Italy \label{ifpu} \and
INAF, Osservatorio di Astrofisica e Scienza dello Spazio, via Piero Gobetti 93/3, I-40129 Bologna, Italy \label{inafBO}\and
INFN, Sezione di Bologna, viale Berti Pichat 6/2, I-40127 Bologna, Italy \label{INFN} \and
Université Paris-Saclay, Université Paris Cité, CEA, CNRS, AIM de Paris-Saclay, 91191 Gif-sur-Yvette, France \label{saclay} \and
IRAP, CNRS, Université de Toulouse, CNES, Toulouse, France \label{cnes} \and
INAF, Istituto di Astrofisica Spaziale e Fisica Cosmica di Milano, via A. Corti 12, 20133 Milano, Italy \label{inafMI} \and
Dipartimento di Fisica, Sapienza Università di Roma, Piazzale Aldo Moro 5, I-000185 Rome, Italy \label{Sapienza} \and
Department of Physics, Informatics and Mathematics, University of Modena and Reggio Emilia, 41125 Modena, Italy \label{unimodena} \and
Center for Astrophysics | Harvard $\&$ Smithsonian, 60 Garden Street, Cambridge, MA 02138, USA \label{cfa} \and
Dipartimento di Fisica, Università di Roma ‘Tor Vergata’, Via della Ricerca Scientifica 1, I-00133 Roma, Italy.\label{UniR2} \and
INFN, Sezione di Roma ‘Tor Vergata’, Via della Ricerca Scientifica, 1, 00133, Roma, Italy.\label{INFN-R2} \and
California Institute of Technology, 1200 E. California Blvd., Pasadena, CA 91125, USA \label{caltech} \and
Departamento de Física Teórica M-8, Universidad Autónoma de Madrid, Cantoblanco, E-28049 Madrid, Spain \label{uniMadrid} \and
Centro de Investigación Avanzada en Física Fundamental (CIAFF), Universidad Autónoma de Madrid, E-28049 Madrid, Spain \label{ciaff}
}

% These dates will be filled out by the publisher
\date{Accepted XXX. Received YYY; in original form ZZZ}

\titlerunning{CHEX-MATE cluster thermodynamics insights through simulations}
\authorrunning{Seppi, Eckert, Rasia et al.}

\abstract{
% Context.  
Galaxy clusters offer powerful insights into the large-scale structure of the Universe and the physics of baryons in hot state. Their scientific exploitation, however, hinges on our ability to accurately measure key thermodynamic properties. }
{
%
% Aims.
 In this work, we aim to assess the reliability of current analysis techniques in reconstructing these properties, with particular focus on samples similar to those observed in the Cluster HEritage project with XMM-Newton (CHEX-MATE).}
{
% 
% Methods 
We develop a suite of dedicated end-to-end simulations of CHEX-MATE-like clusters selected from large scale hydrodynamical simulations, and processed through a newly developed realistic XMM-Newton simulator. We apply a full X-ray data analysis pipeline to the mock datasets, including imaging, spectral fitting, and profile reconstruction.}{% 
% Results
The gas density profiles can be robustly recovered across a wide radial range, when using azimuthal mean surface brightness profiles. Our reconstruction techniques are able to reproduce the intrinsic density profile with the correct scatter, with deviations of at most 10$\%$ between 0.1 and 1$\times$R$_{\rm 500c}$. The gas mass is reconstructed with better than 1$\%$ accuracy. Accurate measurement of temperature profiles is more challenging and possibly subject to biases, particularly in the presence of azimuthal variations and multi-temperature gas along the line of sight, which dominate over projection effects.}% and are likely responsible for underestimating thermal pressure, which shows inconsistencies in the 5-30$\%$ range depending on the simulation and the reconstruction method.
%}
{
%Conclusions
Our results highlight the need for caution in interpreting cluster temperature measurements and underscore the value of tailored mock observations for understanding observational systematics. These findings also suggest that biases in X-ray temperature measurements may alter the interpretation of the thermodynamical state of the intra-cluster medium, an outlook particularly relevant in light of recent low velocity measurements from the XRISM mission.
}

\maketitle

\section{Introduction}
\label{sec:intro}

%\rs{Clusters}
Clusters of galaxies are the end point of the structure formation process throughout the history of the Universe and are located in the nodes of the cosmic web \citep[][]{MoWhite2002MNRAS.336..112M, Springel2005}. They encode precious cosmological information about dark matter, driving the shaping of the large scale structure of the Universe, and dark energy, driving its accelerated expansion at late times \citep[][]{Allen2004MNRAS.353..457A, Kravtsov2012ARA&ABorgani, Clerc2022arXiv220311906C_review, Ghirardini2024_erass1cosmo, Lesci2025arXiv250714285L}. 
Massive galaxy clusters benefit from high signal to noise ratio observations at various wavelengths. In the optical band they are seen as a collection of their galaxy members \citep[][]{Rykoff2014redmapper, Abbott2020DESY1_clusters}. However, only about 1$\%$ of their total mass resides in the galaxy population. Instead, about 90$\%$ is in form of dark matter, whose gravitational effect shows up as peaks in weak lensing convergence maps \citep[][]{Miyazaki2018PASJ_WLclu}. Finally, the majority of baryons is located in the hot gas that constitutes the intra-cluster medium (ICM), heated up by the process of gravitational collapse to high temperatures around 10$^8$ K. This allows detecting clusters in the millimeter band via the Sunyaev-Zeldovich (SZ) effect \citep[][]{Staniszewski2009ApJspt_clusters, Planck2014}, and in X-rays thanks to direct emission via thermal bremsstrahlung \citep[][]{Bohringer2004A&Areflex, Pratt2019SSRv..215...25P}. SZ surveys such as the Atacama Cosmology Telescope \citep[ACT,][]{Hilton2021ApJACT}, the South Pole Telescope \citep[SPT,][]{Bleem2020ApJS_sptEXT}, and \textit{Planck} \citep[][]{Planck2014A&A...571A..20P} are sensitive to most massive clusters up to high redshift. X-ray surveys from ROSAT \citep[][]{Bohringer2004A&Areflex}, XMM-XXL \citep[][]{Pierre2016XLL}, and eROSITA \citep[][]{Predehl2021A&Aerosita, Bulbul2024} are better suited to detect the low mass cluster population, but their sensitivity quickly drops at high redshift. The combination of multi-wavelength data is essential to obtain a clear view of galaxy clusters in the Universe \citep[e.g.,][]{Beauchesne2024MNRAS_lensingXraykine}.

%\rs{CHEXMATE}
The Cluster HEritage project with XMM-Newton: Mass Assembly
and Thermodynamics at the Endpoint of structure formation
\citep[CHEX-MATE\footnote{\url{xmm-heritage.oas.inaf.it}},][]{CHEX-MATE2021A&A_intro} is a Heritage program to follow up a sample of 118 galaxy clusters selected from \textit{Planck} with X-ray observations using XMM-Newton. The program covers three mega-seconds, with a median exposure time of 40 ks per object. The goal is to study the final products of structure formation, focusing on the most recent objects formed in time (Tier-1: 0.05<z<0.2, 2$\times$10$^{14}$<M$_{\rm 500c}$/M$_\odot$<9$\times$10$^{14}$), and the most massive clusters in the Universe (Tier-2: z<0.6, M$_{\rm 500c}$>7.25$\times$10$^{14}$ M$_\odot$). CHEX-MATE aims to combine the deep XMM observations with archival and follow up lensing and SZ data to tackle some open questions in cluster science, such as mass calibration, different selection processes, or the evolution of cluster properties throughout cosmic time.
The CHEX-MATE collaboration has already produced several results relevant to this work. 
\citet{Campitiello2022A&A_chexmate_morph} derived a dynamical-state indicator from X-ray morphological parameters, finding that CHEX-MATE clusters are generally more disturbed than X-ray–selected samples. 
\citet{Bartalucci2023A&A_chexmateSB} showed that surface-brightness profiles exhibit large diversity in the cores but converge with minimal scatter at $0.4$–$0.8\,R_{\rm 500c}$. 
Temperature profiles for a representative subsample were presented by \citet{Rossetti2024A&A_TxprofCXM}, demonstrating the potential for statistical studies once the full sample is analysed. 
A pilot study of entropy profiles by \citet{Riva2024A&Achexmate_icm} revealed correlations between core entropy and dynamical state and deviations from self-similar scaling. 
CHEX-MATE data quality also enables advanced mass-modelling approaches: \citet{Kim2024A&A_clump3d} combining XMM-Newton and \textit{Planck} data to obtain a 3D elliptical models \citep[][]{Kim2024A&A_clump3d, Chappuis2025A&A_1689}, or from galaxy dynamics \citep[][]{Sereno2025A&A_chexmate_masses}.
Finally, \citet{Miren2025_pressureprof} introduced a joint fitting of universal pressure profiles and cluster masses to break their degeneracy, a method well suited to CHEX-MATE.

%\rs{Simulations}
Rigorous validation of the X-ray analysis pipeline further strengthens the accuracy and precision of the recovered observables and the cluster properties inferred from them. The need for tools capable of producing realistic mock X-ray observations from hydrodynamical simulations emerged as a crucial step toward bridging the gap between theory and observation. Early works \citep[][]{Evrard1996ApJ_xrayMASS, Mathiesen2001ApJ...546..100M} already emphasized that meaningful comparisons with X-ray data require simulated quantities that incorporate projection and instrumental effects, rather than purely theoretical profiles. This concept was operationally realized with the development of dedicated X-ray map simulators, such as X-MAS \citep[][]{gardini.etal.2004}, which for the first time reproduced the full observing process—including telescope response, background noise, and photon statistics—yielding realistic event files for instruments such as Chandra and XMM-Newton. Subsequent studies \citep[e.g.][]{Rasia2006MNRAS.369.2013R, Nagai2007ApJ_xrayCLUsims, Rasia2008ApJ_xmas, Biffi2012MNRASphox} demonstrated that such simulated observations are essential to quantify observational biases, test data analysis pipelines, and ensure that theoretical predictions and observed cluster properties are compared on consistent grounds. These developments have firmly established X-ray map simulators as a cornerstone in the modern analysis of the ICM. The literature is rich with such examples, especially in the X-ray band \citep[see also][]{Lau2009ApJ_hydrobias, Battaglia2013ApJ_fgassims, Nelson2014ApJ_hydrob, Rasia2015ApJCCNCCsims, Biffi2016HE}, but also in combination with other probes like lensing \citep[][]{Meneghetti2010A&Axraylenisng_clumass, Meneghetti2011A&A_cluobssim, Rasia2012NJPh_xraylensing, Giocolieuclid2024A&A_WL, giocoli2025_300} or the SZ effect \citep[][]{Gupta2017_pressureprof_magneticum, Gianfagna2021MNRAS_hydrob, Wicker2023A&A_hydrobOBS}. Most of these studies focus on recovering intrinsic properties such as halo mass, providing estimates of the hydrostatic mass bias ranging between 10-30$\%$ \citep[][]{Gianfagna2023MNRAS_hydrob_300, Jennings2023MNRAS_simbahydrob, Miren2024A&A_bhe}, with the bias increasing with cluster mass \citep[][]{Braspenning2025MNRAS_bhe}. Other works also suggest that part of the mass bias is encoded in X-ray temperature measurements \citep[][]{Henson2017MNRAS.465.3361H, Pearce2020MNRAS.491.1622P, Barnes2021MNRAS_bhe}. They assume various levels of complexity concerning the realisticness of the X-ray mock, starting from the simplest approach of analysing the hot gas particles in hydrodynamical simulations with dark matter and baryons. 

%\rs{what we do}
Following the examples of \citet{Rasia2008ApJ_xmas, Biffi2013MNRAS}, our goal is to take one step further by selecting twin CHEX-MATE samples from three different hydrodynamical simulations: \texttt{The Three Hundred (The300,} hereafter) project \citep[][]{Cui2018MNRAS_The300}, \texttt{Magneticum} \citep[][]{Dolag2015IAUGA..2250156D}, and \texttt{MACSIS} \citep[][]{Barnes2017MNRAS_macsis}. We generated realistic end to end XMM-like mock observations and analysed them with standard tools and pipelines widely used in the X-ray community. The high resolution of the simulations, together with realistic X-ray processing, enables a robust comparison between recovered and input thermodynamic profiles. This is essential for assessing whether our models can accurately reproduce X-ray measurements and recover intrinsic cluster properties without bias. We find that the gas density and gas mass reconstruction is robust within a few percent. The reconstruction of X-ray temperature is in agreement with expectations, but its direct link to mass is not straightforward: the results depend on the hydrodynamical simulation and temperature biases propagate to pressure and entropy at the $10-20 \%$ level. 

%\rs{info}
Radial profiles are normalized to the true cluster radius, providing a consistent scale for input-output comparison, while centring and reconstruction are carried out independently through the X-ray analysis. A detailed investigation of the impact on mass and radius estimates is deferred to future work. The best-fit results are always median values accompanied by their 16th-84th percentile points.
This article is organized as follows. In Sect. \ref{sec:hydro_sims} we present the simulations used in this work. In Sect. \ref{sec:mock_gen} we explain the generation of the mock XMM-Newton data. In Sect. \ref{sec:xray_analysis} we describe the X-ray analysis of the mock data. We present our results about X-ray observables in Sect. \ref{sec:results}. We further discuss them and summarize our work in Sect. \ref{sec:conclusions}. 

\section{Hydrodynamical simulations}
\label{sec:hydro_sims}
We briefly describe  the hydrodynamical simulations used in this work, but refer to their presentation papers for a more in depth description. A summary is reported in Table \ref{tab:simulation_parameters}.

\begin{table}
    \caption{Cosmological and numerical parameters describing the \texttt{The300}, \texttt{Magneticum}, and \texttt{MACSIS} simulations. For \texttt{Magneticum} the numbers in parentheses refer to Box2b, the others to Box2. }
    \vskip.1cm
    \centering
    \begin{tabular}{|c|c|c|c|}
    \hline
    \rule{0pt}{2.2ex} &  \textbf{\texttt{The300}} & \textbf{\texttt{Magneticum}} & \textbf{\texttt{MACSIS}} \\
    \hline
    \rule{0pt}{2.2ex} Box [Gpc $h^{-1}$] & 1.0 & 0.352 (0.640) & 2.17 \\
    $\Omega_{\rm M}$ & 0.307 & 0.272 & 0.307 \\
    $\Omega_{\rm B}$ & 0.048 & 0.0456 & 0.04825 \\
    $\Omega_{\rm \Lambda}$ & 0.693 & 0.728 & 0.693\\
    $\sigma_{\rm 8}$ & 0.823 & 0.809 & 0.8288 \\
    $H_{\rm 0}$ & 67.8 & 70.4 & 67.77 \\
    $n_{\rm s}$ & 0.96 & 0.963 & 0.9611 \\
    N particles & 3840$^3$ & 2$\times$1584$^3$ (2$\times$2880$^3$) & 2520$^3$ \\
    M$_{\rm DM}$ [M$_\odot\ h^{-1}$] & 1.27$\times$10$^9$ & 6.9$\times$10$^8$ & 4.4$\times$10$^9$ \\
    M$_{\rm gas}$ [M$_\odot\ h^{-1}$] & 2.36$\times$10$^8$ & 1.4$\times$10$^8$ & 8.0$\times$10$^8$ \\
    \hline
    \end{tabular}
    \vskip.1cm
    \footnotesize{\textbf{Notes.} Box size: comoving length of the box size covered by the simulation, $\Omega_{\rm M}$: total matter density parameter, $\Omega_{\rm B}$: baryonic matter density parameter, $\Omega_{\rm \Lambda}$: dark energy density parameter, $\sigma_{\rm 8}$: normalization of the linear matter power spectrum, $H_{\rm 0}$: Hubble constant, n$_{\rm s}$: initial slope of the linear matter power spectrum, N particles: total number of dark matter particles in the simulation, M$_{\rm DM}$: mass of the dark matter particles, M$_{\rm gas}$: initial mass of the gas matter particles.}
    \label{tab:simulation_parameters}
\end{table}

\subsection{\texttt{The300}}
\label{subsec:the300}
\texttt{The300} simulations consists of 324 regions centred on large galaxy clusters simulated with full hydrodynamical processes  \citep{Cui2018MNRAS_The300}. The clusters have been initially identified by the \texttt{Rockstar} halo finder \citep{Behroozi2013} in the 1 $h^{-1}$ Gpc dark-matter-only MDPL2 box \citep{Klypin2016}\footnote{\url{https://skiesanduniverses.iaa.es/Simulations/MultiDark}, \url{https://www.cosmosim.org}}. 
%MultiDark simulations \citep{Klypin2016}\footnote{\url{https://skiesanduniverses.iaa.es/Simulations/MultiDark}}. In particular, the authors used the dark-matter-only MDPL2 box, a cube of 1 $h^{-1}$ Gpc, containing 3840$^3$ particles, with a particle mass equal to 1.5$\times$10$^9$ M$_\odot$/h. 
%The re-simulation process allows simulating a large portion of the Universe containing hundreds of massive clusters, resolving them with more than 10$^5$ particles. 
The dark matter haloes are selected to have virial mass larger than 8$\times$10$^{14} h^{-1}$ M$_\odot$\footnote{The virial mass is the total mass enclosed within the virial radius, i.e. a region encompassing an average overdensity that is equal to the critical density of the Universe at the cluster redshift multiplied by the virial overdensity \citep[see][]{BryanNorman1998}. At z=0 $\Delta_{\rm vir}$ is about 99 times larger than the critical density and 330 times larger than the background matter density in vanilla $\Lambda$CDM.}. 

In the resimulation process, the dark matter particles within each of the selected region of radius $\sim$15 $h^{-1}$Mpc, are split into dark matter and gas, according to the cosmological baryon fraction and with an initial gas mass resolution of 2.36$\times$10$^{8} h^{-1}$ M$_\odot$. The re-simulation was performed with \texttt{GADGET-X} \citep[][]{Beck2016MNRAS_gadgetX} using smooth-particle hydrodynamics (SPH) and halos and sub-halos were identified by the AMIGA Halo Finder \citep[AHF,][]{Knollmann2009_AHF}, which accounts for the baryonic components in the halo finding process.
%Galaxy clusters are identified with the AMIGA Halo Finder \citep[AHF,][]{Knollmann2009_AHF}, which accounts for gas and stars in the halo finding process.
%
%Within each cluster region of radius $\sim$15 $h^{-1}$Mpc, 
%
%the dark matter particles were split into dark matter and gas, according to the cosmological baryon fraction. The gas mass resolution is 2.36$\times$10$^{8}$ M$_\odot$/h. The re-simulation was performed with \texttt{GADGET-X} \citep[][]{Beck2016MNRAS_gadgetX} using smooth-particle hydrodynamics (SPH). 
The baryonic physics includes models for gas cooling \citep[][]{Wiersma2009MNRAS.399..574W}, stellar evolution \citep[][]{Tornatore2007MNRAS.382.1050T}, stellar feedback \citep[][]{Springel2003MNRAS_SF}, black hole growth and AGN feedback \citep[][]{Steinborn2015MNRAS.448.1504S}. Cosmological parameters are from \citet{Planck2016A&A..cosmopars}. \texttt{The300} project reproduces the baryon fraction and scaling relations of local galaxy clusters down to 10$^{13}$ M$_\odot$. Recent work from \cite{Rasia2025arXiv_300fbar} highlighted the importance of properly modelling the hot gas fraction in galaxy clusters. In particular, \texttt{The300} compares well with CHEX-MATE in terms of emission measure profiles \citep[][]{Bartalucci2023A&A_chexmateSB}, temperature profiles and their inhomogeneities \citep[][]{Rossetti2024A&A_TxprofCXM, Lovisari2024A&A_CXMATETX}, and gas pressure and entropy for the most massive systems \citep[][]{Riva2024A&Achexmate_icm}.

\subsection{\texttt{Magneticum}}
\label{subsec:magneticum}
The \texttt{Magneticum} suite\footnote{\url{http://www.magneticum.org}} is a collection of full cosmological hydrodynamical simulations \citep[][]{Dolag2025arXiv_magneticum}. They are performed with the \texttt{P-GADGET3} code \citep[][]{Springel2005}. They include baryonic process such as AGN feedback \citep[][]{Fabjan2010MNRAS_AGNfeedback}, star formation, supernovae explosions, galactic winds \citep[][]{Springel2003MNRAS_SF}, gas cooling \citep[][]{Wiersma2009MNRAS.399..574W}, and enrichment \citep[][]{Tornatore2007MNRAS.382.1050T}. The cosmological parameters are taken from the WMAP results \citep{Komatsu2011ApJS..192...18K}. In particular, we select galaxy clusters from the \texttt{Box2} and \texttt{Box2b}, respectively with size of 352 and 640 Mpc $h^{-1}$, and dark-matter and gas particle mass of 6.9$\times$10$^{8}$ M$_\odot$ and 1.4$\times$10$^{8}$ M$_\odot$, respectively. 
%dark matter particle mass of 6.9$\times$10$^{8}$ M$_\odot$, with gas mass resolution equal to 1.4$\times$10$^{8}$ M$_\odot$. 
Galaxies and clusters are identified using a Friends-of-Friends (FoF) algorithm combined with \texttt{Subfind} \citep[][]{Dolag2009MNRAS_subhalomf}. 

\texttt{Magneticum} successfully reproduces the AGN luminosity function \citep[][]{Hirschmann2014MNRAS.442.2304H}, morphological properties of clusters \citep[][]{Teklu2015ApJ...812...29T, Remus2017MNRAS_Magneticum, Gupta2017_pressureprof_magneticum}, thermodynamical profiles and features of galaxy groups \citep[][]{Bahar2024A&A_grpsFeedback, Popesso2024arXiv_grpsprop}. It was exploited to study the hydrostatic mass bias and recovery of X-ray observables in the context of eROSITA \citep[][]{Scheck2023A&A_hydrobiasMAG, ZuHone2023A&A_LXMAGerosita}.

\subsection{\texttt{MACSIS}}
\label{subsec:macsis}

The MAssive ClusterS and Intercluster Structures
\citep[\texttt{MACSIS},][]{Barnes2017MNRAS_macsis} dataset is a collection of 390 re-simulated regions carried out with full hydrodynamical models. The concept is similar to the one presented by \texttt{The300} project in Sect. \ref{subsec:the300}, with an initial selection on a large $N$-body simulation followed by a resimulation that includes the baryonic components. The parent simulation is a dark-matter-only cube of 3.2 Gpc, run with \texttt{GADGET3} \citep[][]{Springel2005}. The cosmological parameters are taken from \citet{Planck2014}. The dark matter particle mass is 5.43$\times$10$^{10} h^{-1}$ M$_\odot$. The \texttt{MACSIS} sample was selected from all haloes more massive than 10$^{15}$ M$_\odot$ at z=0, identified by a friend-of-friend algorithm. These haloes were grouped into mass bins of 0.2 dex. If a bin contained less than 100 haloes, all of them were selected, otherwise the bin was refined to 
0.02 dex from which 10 objects were randomly selected.
%0.02 dex, selecting 10 objects randomly from each subdivison. %The result is an ensamble of 390 haloes that is mass complete above 10$^{15.6}$ M$_\odot$. 

These haloes were re-simulated with full hydrodynamical models following the prescriptions from BAHAMAS \citep[][]{McCarthy2017MNRAS_bahamas}. Similarly to \texttt{The300} and \texttt{Magneticum}, the baryonic model includes radiative cooling from different elements \citep[][]{Wiersma2009MNRAS.399..574W}, star formation and feedback \citep[][]{Schaye2008MNRAS_SFmodels}, as well as black hole seeding, growth, and feedback \citep[][]{Booth2009MNRAS_agnmodel}. The hot gas profiles from \texttt{MACSIS} show good agreement with observational data \citep[][]{Barnes2017MNRAS_macsis, Riva2024A&Achexmate_icm}.

\subsection{Simulated cluster sample}
\label{subsec:clusample}

\begin{figure}
    \centering
    \includegraphics[width=\columnwidth]{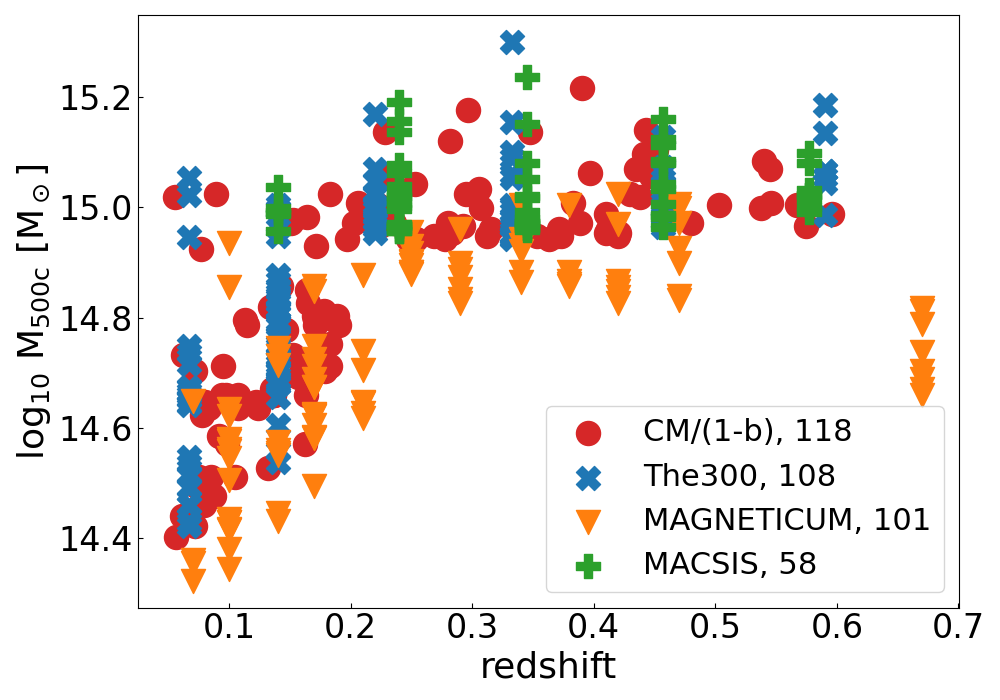}
    \caption{Mass and redshift distribution of the CHEX-MATE clusters compared to the twin selected systems from \texttt{The300}, \texttt{Magneticum}, and \texttt{MACSIS}. The CHEX-MATE masses include the hydrostatic mass bias of 0.2.}
    \label{fig:Mz_distribution}
\end{figure}

We select clusters by picking a twin for each real CHEX-MATE cluster with the closest possible mass in the various redshift snapshots. For \texttt{The300} and \texttt{MACSIS}, a 20\% hydrostatic mass bias is applied when matching observed clusters to simulated analogues. Although this introduces heterogeneous assumptions on $b_{\rm HE}$, it has the advantage that our analogue sample spans the full range from no bias to 20\%. As a result, the reconstruction tests naturally assess the robustness of our methods across the plausible range of hydrostatic mass biases, rather than relying on a single fixed value. As explained in Sect. \ref{sec:intro}, CHEX-MATE is split into Tier 1 and Tier 2 clusters. 
%Because of the lower dark matter particle resolution in \texttt{Macsis}, we ignore the Tier 1 sample of low mass clusters, and only focus on Tier 2. 
%
Because of its construction containing exclusively massive systems, \texttt{MACSIS} is associated only to Tier 2 objects.
Instead, for \texttt{The300} and \texttt{Magneticum} we include matched samples for both Tier 1 and Tier 2.
%The result is shown in 
Fig. \ref{fig:Mz_distribution} %. It 
shows the distribution of CHEX-MATE clusters (in red) in the mass-redshift space, together with the twin selected systems from \texttt{The300}  (in blue), \texttt{Magneticum} (in orange), and \texttt{MACSIS} (in green).

Because the snapshots of the simulations are saved at different redshifts, the matching with the redshifts of real CHEX-MATE objects may be more or less precise, especially at high redshift where the snapshot of the \texttt{Magneticum} Box2b are sparse, and the systems around redshift of 0.6 are matched to the snapshot at 0.67. This is not a limitation for our goals, as we do not aim at studying the properties of individual clusters, but we focus on a comparison at the population level to input properties using a global sample that is representative of CHEX-MATE. %Similarly, the differences in assumed hydrostatic bias do not affect our conclusions, but they rather allow us to test the robustness and flexibility of our reconstruction methods across a range of mass calibration assumptions, while still being within the mass and redshift range of CHEX-MATE.

\section{Mock generation}
\label{sec:mock_gen}

We use the individual gas particles in the simulations to construct an emissivity model projected along the line of sight in a $30^{\prime} \times 30^{\prime}$ field of view (FoV). We fold the model through the response of XMM-Newton and generate mock events assuming a clean 25 ks exposure time. We then analyse the mock X-ray data extracting surface brightness and temperature profiles and applying a deprojection technique (details in Sect. \ref{sec:xray_analysis}).

\subsection{Input data}
\label{subsec:input_data}

%\rs{input from Elena?}
To generate the X-ray emission from each gas particles we use the first unit of the code X-ray Map Simulator described in \cite{gardini.etal.2004} with some modification as follows. The gas particles are selected within a temperature range of $[0.3-40]$ keV and with a gas density below the star formation threshold to guarantee that multi-phase particles are not included. We stress that this does not mean that emission below 0.3 keV is ignored, because gas elements at different temperatures also shine in the softest X-ray band. The field of view covers 30x30 arcmin$^2$ and it is sampled with a grid of 512$\times$512 pixels. We create 495 narrow energy channels, linearly spaced between 0.1 and 10 keV and we sum and project along one random axis the emissivity of the individual particles. Specifically, the projected spectra assume thermal emission from a collisionally-ionized diffuse gas with a fixed metallicity of 0.3 Z$_\odot$ \citep[][]{Asplund2009ARA&A..47..481A} and is corrected by an absorption term assuming a hydrogen column density of 5$\times$10$^{20}$ cm$^{-2}$ (i.e. in \texttt{Xspec}: \texttt{phabs(apec)}). 
The final projected flux is stored in a data cube of size 512$\times$512$\times$495. 
Effectively, this information is similar to that of an ideal integral field unit (IFU), where for each pixel we have access to the input global spectrum.

\subsection{XMM simulation}
\label{subsec:xmm_simulator}

%\rs{input from Dominique?}
We use the data cube obtained in the previous section as an input to the \texttt{xmm\_simulator}\footnote{\url{github.com/domeckert/xmm_simulator}} software to generate realistic XMM-Newton simulations.
%Using the input boxes described in the previous section, w
We build a 2D model image for each energy channel. 
We compute the effective area on the pixel grid accounting for the detector quantum efficiency, filter transmission, CCD gaps, and the telescope vignetting by combining each component in the XMM current calibration files (CCF)\footnote{\url{https://www.cosmos.esa.int/web/xmm-newton/current-calibration-files}}. We also include the non-X-ray background by loading filter wheel closed data from the CCF for each camera. Its intensity is constant on the detector surface, i.e. there are no soft protons. We model the sky background including contributions from the foreground, with the local hot bubble (an unabsorbed \texttt{Apec} model with a temperature of 0.11 keV), the galactic halo (an absorbed \texttt{Phabs(Apec)} model with a temperature of 0.22 keV), and the cosmic X-ray background with contribution from the faint, undetected AGN (an absorbed power law \texttt{Phabs(Power)} with a spectral index $\Gamma$=1.46). Finally we include the emission of AGN by randomly generating their position within the FoV. This means that no AGN clustering is present, but this is negligible given the size of the XMM FoV, and 1D cluster profile studies are not affected. The fluxes are drawn from the logN-logS distribution of \citet{Lehmer2012ApJ_logNlogSChandra}. The AGN spectral model is an absorbed power law with column density randomly drawn in the range $[10^{20}-10^{23.5}]$ cm$^{-2}$ 
%in the 20-23.5 range (in cm$^{-2}$ and $\log_{\rm 10}$ units)
and a slope drawn from a Gaussian distribution with mean of 1.9 and variance 0.2 \citep[][]{Ueda2014ApJ_AGN}. 

For each energy channel, the total model image with the source, sky background, and AGN, is recast into XMM pixels, convolved with the PSF, and the Response Matrix File (RMF). Individual events are generated as a Poisson realization of the total model, including a separate particle background photon list that is merged with the event file. We generate mock EPIC events with an exposure time of 25 ks, which is representative for the CHEX-MATE observations. 

\section{X-ray analysis}
\label{sec:xray_analysis}

We reduce the mock data with dedicated software to analyse XMM-Newton data XMM\_SAS version 21.0\footnote{\url{cosmos.esa.int/web/xmm-newton/sas-news}}, routines from \texttt{pyproffit}\footnote{\url{pyproffit.readthedocs.io}} \citep{Eckert2020_pyproffit} and \texttt{hydromass}\footnote{\url{hydromass.readthedocs.io}} \citep{Eckert2022_hydromass}. We note that this pipeline is not identical to that used for CHEX-MATE, although they share most of the underlying methodology. We ignored 4 (3, 3) clusters in \texttt{The300} (\texttt{Magneticum}, \texttt{MACSIS}) with complex shape due to recent major mergers, where assumptions such as spherical symmetry fail. %\rs{differences between this and chexmate pipeline?}

\subsection{Extraction of Images and Spectra}
\label{subsec:data_process}

First we create images in the 0.7-1.2 keV band by combining the individual detectors MOS1, MOS2, and PN into a single EPIC XMM image. We also generate a single exposure map by summing together individual exposure maps, while multiplying the one relative to the PN detector by 3.42, that is the ratio of the PN to MOS effective area in this energy band. We use the \textit{ewavelet} task to run a wavelet source detection algorithm on the mock images. This allows us to identify individual point sources to be masked during the X-ray analysis. We carefully visually inspect each region file obtained by the source detection process. Obviously the central detection relative to the simulated cluster needs to be removed. This is straightforward. However, sometimes the algorithm splits the clusters into multiple fake sources, or does not include the tails of the point source emission within the aperture corresponding to AGN detections. We manually modify each region mask to include the most amount of clean emission from the cluster, while minimizing the impact of subhaloes, nearby systems, and AGN leakage to the best of our abilities. This step does not involve any prior knowledge from the simulation and the mock is treated exactly as a real observation. From the simulation perspective, we can identify whether masked detections correspond to injected AGN or to non-AGN features, such as substructure or background fluctuations. Across the three simulations we find no significant variation in the substructure fraction: approximately 75$\%$ of masked sources are bona fide AGN with 5$\times$10$^{-14}$ erg/s/cm$^2$.

\begin{figure}
    \centering
    \includegraphics[width=\columnwidth]{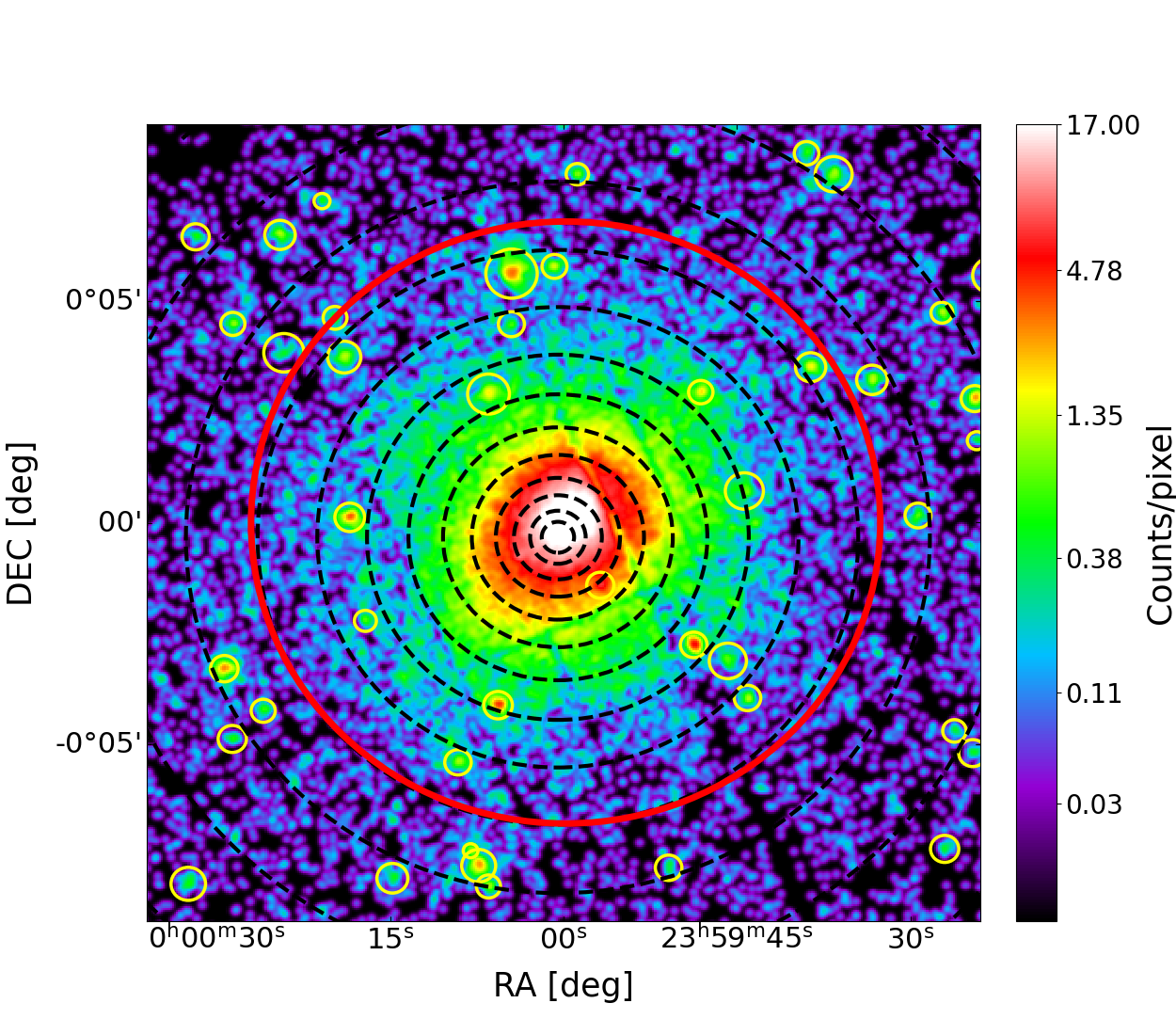}
    \caption{Example of a simulated XMM-Newton EPIC image for one cluster. The black lines denote the regions used for spectral extraction to measure its temperature profile. The red circle denotes R$_{\rm 500c}$.}
    \label{fig:CLUexample}
\end{figure}

Secondly we prepare the regions to extract spectra to measure the radial temperature profile. By default, the radial profiles consist of one inner bin from 0 to 0.04$\times$R$_{\rm 500c}$, in addition to 12 more bins spanning from 0.04 to 1.1$\times$R$_{\rm 500c}$. We extract the background spectrum in a circular region located between 1.5 and 2.0$\times$R$_{\rm 500c}$. If the upper boundary overshoots the XMM FoV we manually set it to 15 arcmin. We carefully inspect each region automatically generated by our pipeline and modify it if needed. In some cases the background region is too large and falls out of the FOV. In other cases it ends up including a filament or part of a secondary nearby structure. For such cases, we extract the background spectrum in one of the FOV corners using a circular aperture of about 3 arcminutes according to the needs of each specific case. 
An example of the end result of the whole procedure is shown in Fig. \ref{fig:CLUexample}. It displays the EPIC count map for one of the \texttt{The300} clusters. The yellow circles denote the sources identified by the wavelet detection, the dashed black lines are the radial bins used for spectral extraction, and the red circle corresponds to R$_{\rm 500c}$.

Finally, we fit each spectrum using the X-COP pipeline \citep{Ghirardini2019A&A...621A..41G}. The global model accounts for a high energy particle background (a broken power law with several Gaussian lines), the sky background including the CXB (an absorbed power law with photon index of 1.46), the Galactic halo (an absorbed apec model with temperature between 0.15 and 0.6 keV), and the local hot bubble (an unabsorbed apec model with temperature of 0.11 keV). We turned off the soft protons (an additional broken power law), since we do not simulate them. The source model is an absorbed \texttt{apec} model with temperature, normalization, and abundance free to vary. 
In the X-COP pipeline, background modelling is improved by simultaneously fitting the background extracted from the XMM field with ROSAT All-Sky Survey (RASS) background data, which are automatically retrieved in the region surrounding the source. To replicate this approach in our simulations, we generate synthetic X-ray background spectra composed of the same physical components used in the pipeline, i.e. the cosmic X-ray background (CXB), Galactic halo emission, and the local hot bubble. Each component is simulated with the same temperatures, abundances, and normalizations adopted in the \texttt{xmm\_simulator}, and convolved with the ROSAT PSPC response files.
The simulated background spectrum is normalized to the extraction area, which by default corresponds to a circular annulus between 60 and 90 arcminutes from the source centre. To remain consistent with the simulation setup, we omit the cross-calibration correction factor between the ROSAT PSPC and the XMM-Newton EPIC camera that is otherwise applied in the X-COP pipeline. This procedure ensures a consistent background treatment and yields more robust temperature measurements, particularly in the outer radial bins where the signal is background-dominated. We use \texttt{X-SPEC} \citep[v12.13.1][]{Arnaud1996ASPC..101...17A} with C-stat \citep[][]{Cash1979ApJ...228..939C}. We refer the reader to \citet{Ghirardini2019A&A...621A..41G} for additional details on the X-COP pipeline.

\subsection{Measurement of X-ray properties}
\label{subsec:xray_meas_analysis}
We extract surface brightness profiles from the EPIC mosaic and from its voronoi binned image. We refer to them as mean and median profiles respectively.
The median profile is derived from the Voronoi-tessellated map, where each adaptive bin is assigned the median value of its constituent pixels before computing the radial profile.
The mean profile is extracted directly from the original image by averaging the native pixels in each radial annulus, without any adaptive binning.
Thus, the terms "median" and "mean" describe the construction of the underlying maps rather than different statistics applied to the same pixel set. 
We use the tessellation scheme from \citet{Diehl2006MNRAS._voronoi}, grouping the count maps into cells containing 25 events each to create the final voronoi image. We use the latter to generate azimuthal median profiles. This alleviates the issue of dense regions biasing the recovered gas density high due to its squared relation to emissivity. The surface brightness profile is modelled as a collection of Kings functions, which allows the computation of the 2D projected profile analytically. In particular, the cluster X-ray emissivity follows:
$\epsilon_{\rm X} = \sum_{\rm n=1}^N \alpha_n \Phi_n(R) \propto n_e^2\Lambda(T,Z)$,
where $\Phi_n(R) = \Big(1+ \frac{R^2}{R_{c,n}^2} \Big)^{-3 \beta}$ and $\Lambda(T,Z)$ is the cooling function, which depends on temperature and metallicity \citep{Sutherland1993_coolfunc}.
The model is convolved with the PSF, and superimposed to a constant background profile that is estimated by computing a count rate value from the spectral model of the background obtained in Sect. \ref{subsec:data_process}. We refer the reader to \citet{Eckert2020_pyproffit} for a detailed description of the surface brightness measurement and modelling.

We then model the gas and mass profiles under the assumptions of hydrostatic equilibrium and spherical symmetry, which allows linking the gas pressure to the total mass. Gas density, temperature, and thermal pressure are related by the ideal gas equation of state. The formalism reads:

\begin{align}
    \frac{dP_{\rm gas}}{dr} &= -\rho_{\rm gas} \frac{GM_{\rm TOT}(<r)}{r^2}, \nonumber \\
    P_{\rm gas} &= \frac{k_{\rm B}}{\mu m_{\rm p}}\rho_{\rm gas}T, \nonumber \\
    M_{\rm TOT}(<r) &= - \frac{rk_{\rm B}T(r)}{G\mu m_{\rm p}} \Big( \frac{d \text{log}T}{d \text{log}r} + \frac{d \text{log}n_{\rm gas}}{d \text{log}r} \Big).
    \label{eq:hydroeq_idealgas}
\end{align}

One can then model the data using Eq. \ref{eq:hydroeq_idealgas} from different points of view. \\
(i) Navarro-Frenk-White model \citep[NFW,][]{Navarro_Frenk_White_1996} or model of the mass profile: given the definition of a mass model, it is possible to derive pressure by integrating the hydrostatic equilibrium equation. Temperature is inferred using the ideal gas equation of state, while gas density is related to the surface brightness via the cooling function. \\
(ii) Non-parameteric model (NP) or model of the temperature profile: a linear combination of log-normal functions is used to describe the 3D temperature profile. It is then projected along the line of sight and fitted to the measurement. \\
(iii) Forward model (FM) of the pressure profile: a generalized NFW is used to describe the pressure profile, which allows computing analytically its gradient and the mass profile. \\
In any of these three cases, the fit is performed on gas density and temperature. The model is fitted to total (gas and dark matter) mass. We refer the reader to \citet{Eckert2022_hydromass} for full details on the modelling. Finally, one can derive the entropy from the 3D temperature and density profiles as $K = k_{\rm B} T n_{\rm gas}^{-2/3}$.  
%wavelet detection, spectral extraction \citep{Rossetti2024A&A_TxprofCXM} 

\section{Results}
\label{sec:results}

In this section we compare the profiles obtained from the analysis in Sect. \ref{sec:xray_analysis} to the input quantities measured directly using the properties of the gas particles in the hydrodynamical simulations.

\subsection{Gas density profiles}

\begin{figure}
    \centering
    \includegraphics[width=\columnwidth]{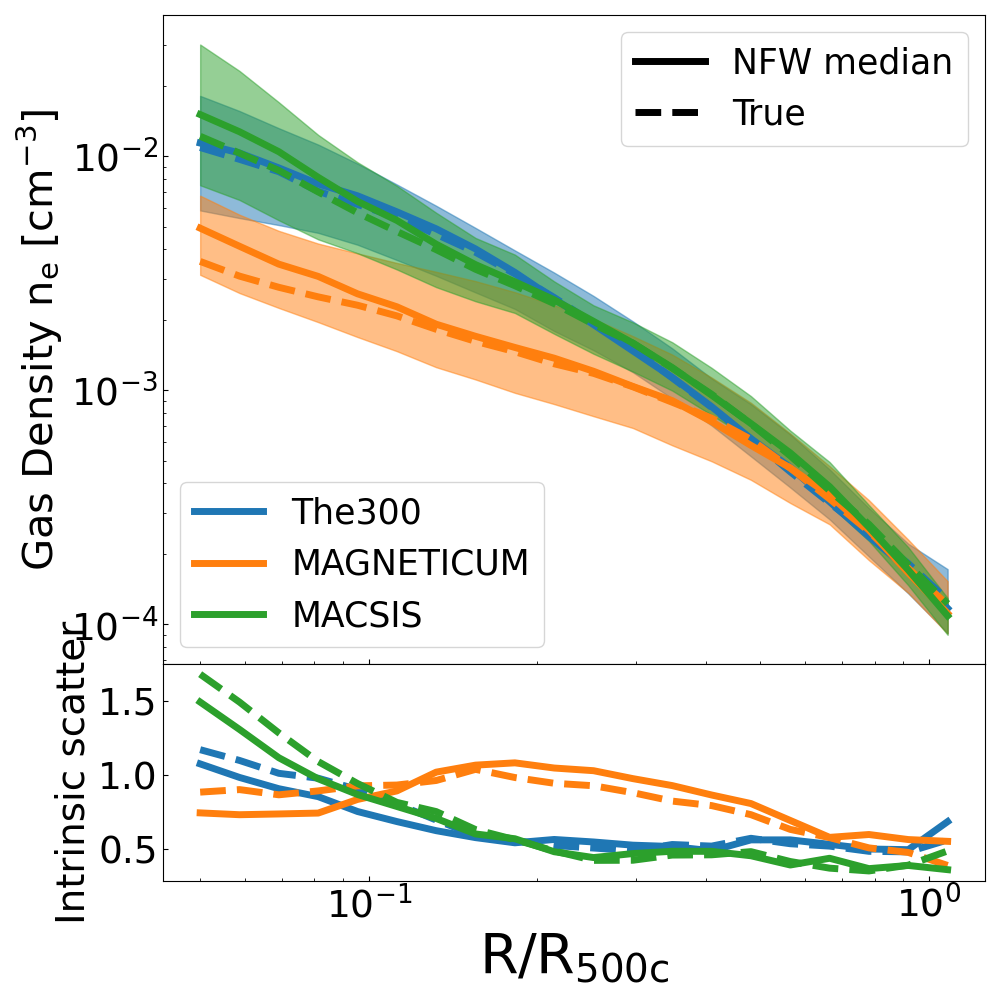}
    \caption{Gas density profiles in our simulations. Top panel: comparison between the measured (solid lines and shaded areas) and true (dashed lines) gas density profiles in different simulations (in various colours), using the NFW reconstruction. Bottom panel: intrinsic scatter as a function of radius.}
    \label{fig:density_profiles_all}
\end{figure}

Gas density is the most fundamental thermodynamic quantity accessible in X-rays, as the surface brightness directly probes $n_{\rm e}^2$ and is therefore less model dependent than temperature. It is key to infer cluster properties such as gas mass, pressure, and entropy, and is essential for comparisons with SZ and lensing.
We study the density reconstruction on a population level. We compare input-output profiles on a per-cluster basis, i.e. we always show the distribution of individual-cluster ratios, rather than the ratio of the median reconstructed and true profiles. This holds for all results across the article.

Figure \ref{fig:density_profiles_all} shows the recovered and true density profiles for the three simulation suites. Overall, the agreement is good: discrepancies are limited to the innermost $0.1\,R_{500c}$ in \texttt{Magneticum}, and remain within the scatter of the measurements. \texttt{The300} and \texttt{MACSIS} exhibit similar core shapes and normalisations, while \texttt{Magneticum} shows flatter and lower-density cores, due to the lower typical masses and the distinct baryonic-physics implementation \citep[see][]{Rasia2025arXiv_300fbar}.
A simple beta model\footnote{$n_e(r) = n_0 \Big(1 + \frac{r^2}{r_c^2}\Big)^{-3/2 \beta}$} fit is close to the expectation $\beta$=2/3, but the flatter profile in \texttt{Magneticum} causes a larger core radius in the best fit. We perform the fitting with the \texttt{curve\_fit} package in \texttt{scipy} \citep[][]{Virtanen2020SciPy-NMeth}. We report the values in Table \ref{tab:betamodel_pars}. The best fit parameters are always compatible within uncertainties between the input and recovered profiles. A single $\beta$-model cannot capture the full complexity of all clusters, so the fitted parameters should be regarded as descriptive summaries rather than ground truths, used to illustrate the flexibility of the reconstruction. Small differences in the core are likely due to miscentering (see Appendix \ref{appendix:miscentering}): the observed profiles are computed from using the peak of the X-ray emission in the XMM mocks, while the input profiles are computed from the position of the most bound particle in each cluster.

The bottom panel of Fig. \ref{fig:density_profiles_all} shows the intrinsic scatter of the profiles presented in the top panel. We define the intrinsic scatter as the difference between the 84th and 16th percentiles of the profile distribution, divided by the median. Among the simulations, \texttt{MACSIS} exhibits the largest scatter in the core region (within 0.1$\times$R$_{\rm 500c}$), reaching values up to 1.5, whereas \texttt{The300} remains below 1 across the same range. \texttt{Magneticum} displays the most diverse behaviour at intermediate radii, with a scatter peaking at 1.2 around 0.2$\times$R$_{\rm 500c}$, while \texttt{The300} and \texttt{MACSIS} are closer to 0.6 at this radius. In the outskirts beyond 0.8$\times$R$_{\rm 500c}$, all simulations converge to a similar relative scatter of about 0.5. Importantly, the dashed lines (true profiles) closely track the solid ones (measured profiles): the reconstructed gas density profiles not only match the true median profiles, but also reproduce the true population scatter as a function of radius. 

We find excellent agreement between different reconstruction methods, with the median profiles from the voronoi images showing the greatest precision (see Fig. \ref{fig:ne_profiles} and discussion in \ref{appendix:gas_mass}).
In Appendix \ref{appendix:gas_mass} we also show that the integration of gas density profiles to measure gas masses also recovers the input gas mass accurately with the voronoi approach to better than 1$\%$ precision, while the standard processing tends to overestimate gas mass by about 5$\%$. 

\subsection{Temperature profiles}
\label{subsec:Temperature_profiles}

\begin{figure}
    \centering
    \includegraphics[width=\columnwidth]{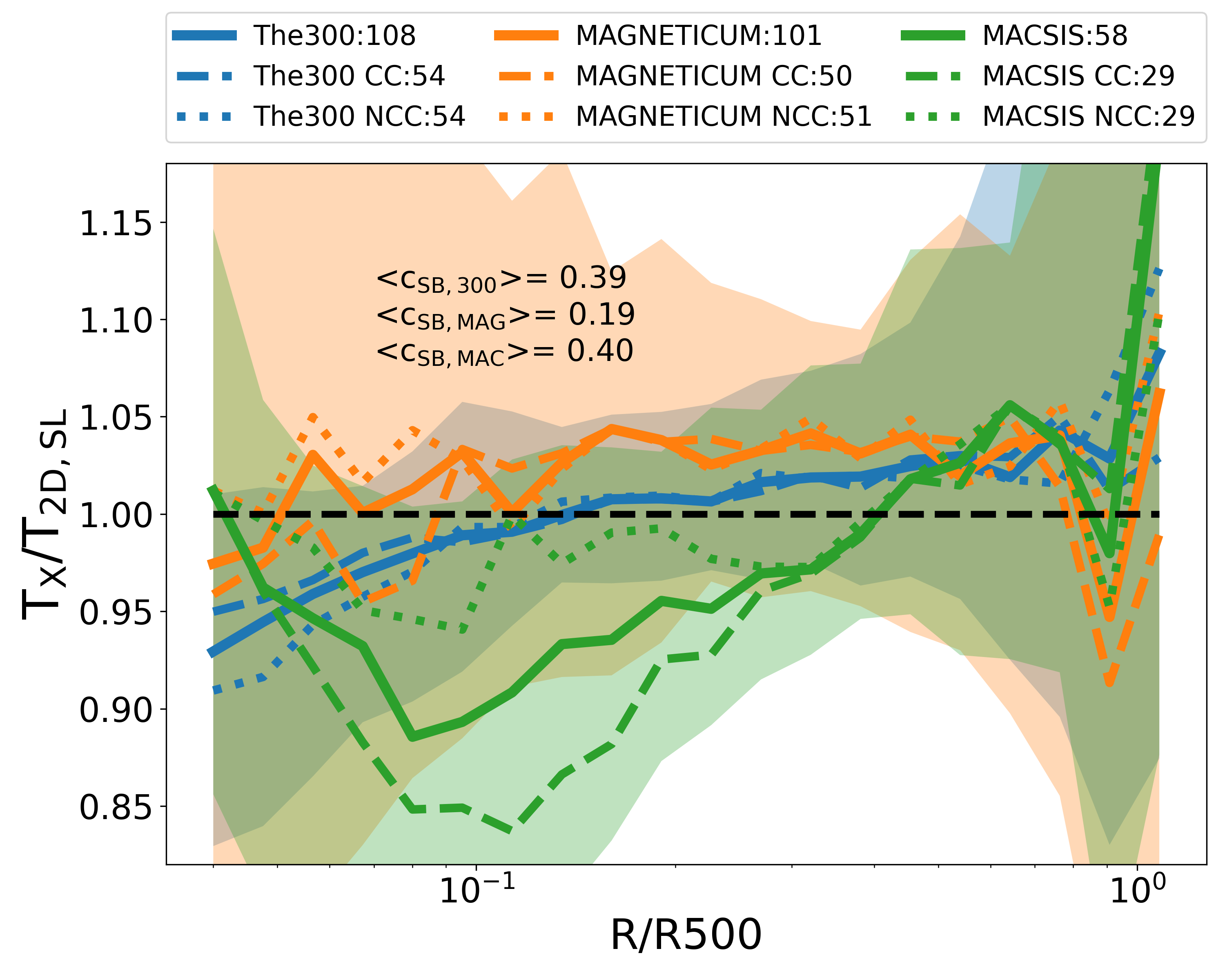}   
    \caption{Ratio between the result of the X-ray spectral fitting and the input spectroscopic-like temperature profile. The profiles are also split between cool core (CC) and non cool core (NCC) according to the median surface brightness concentration reported as text in the panel.}
    \label{fig:Tspecfit_Tsl}
\end{figure}

\begin{figure*}
    \centering
    \includegraphics[width=0.66\columnwidth]{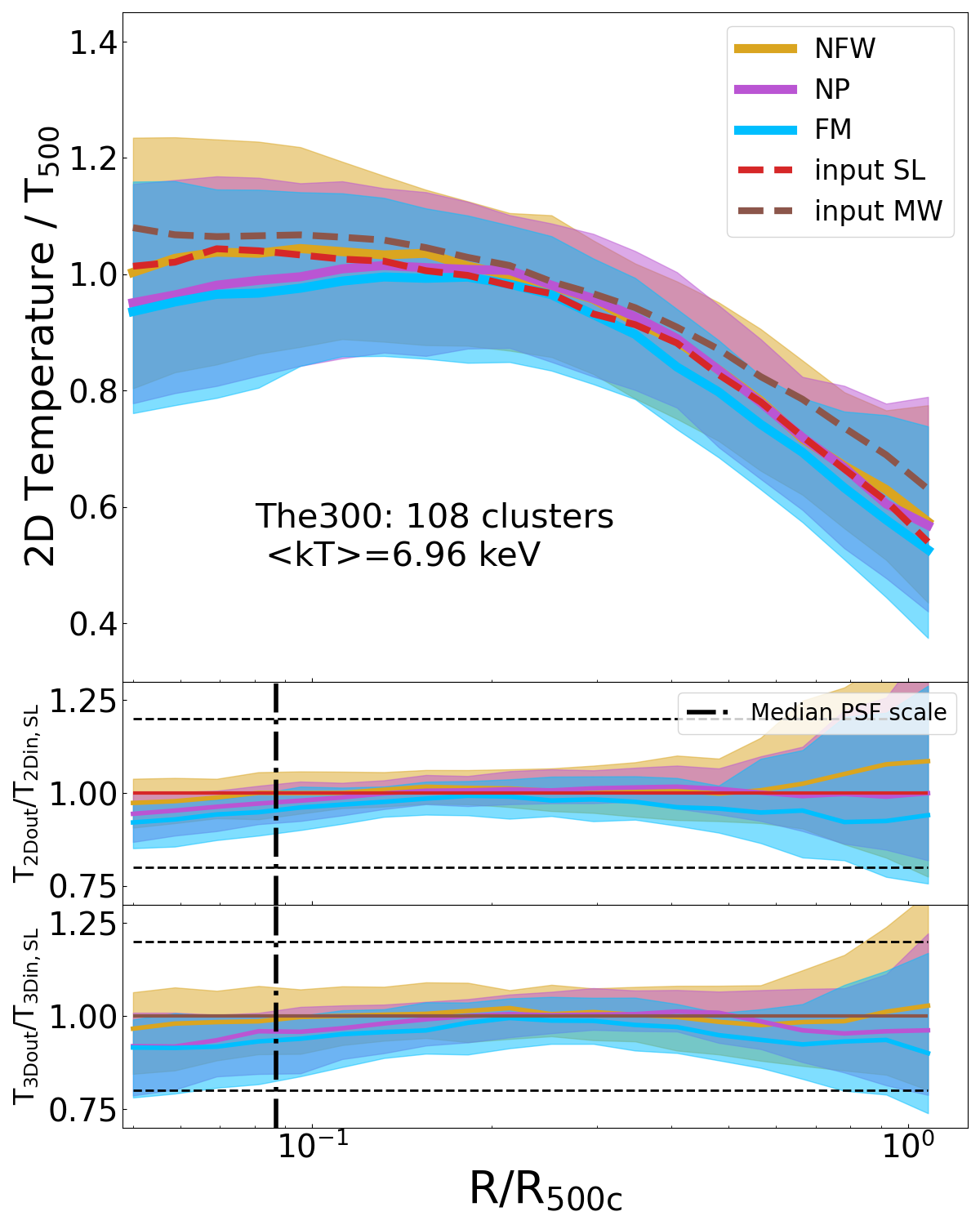}
    \includegraphics[width=0.66\columnwidth]{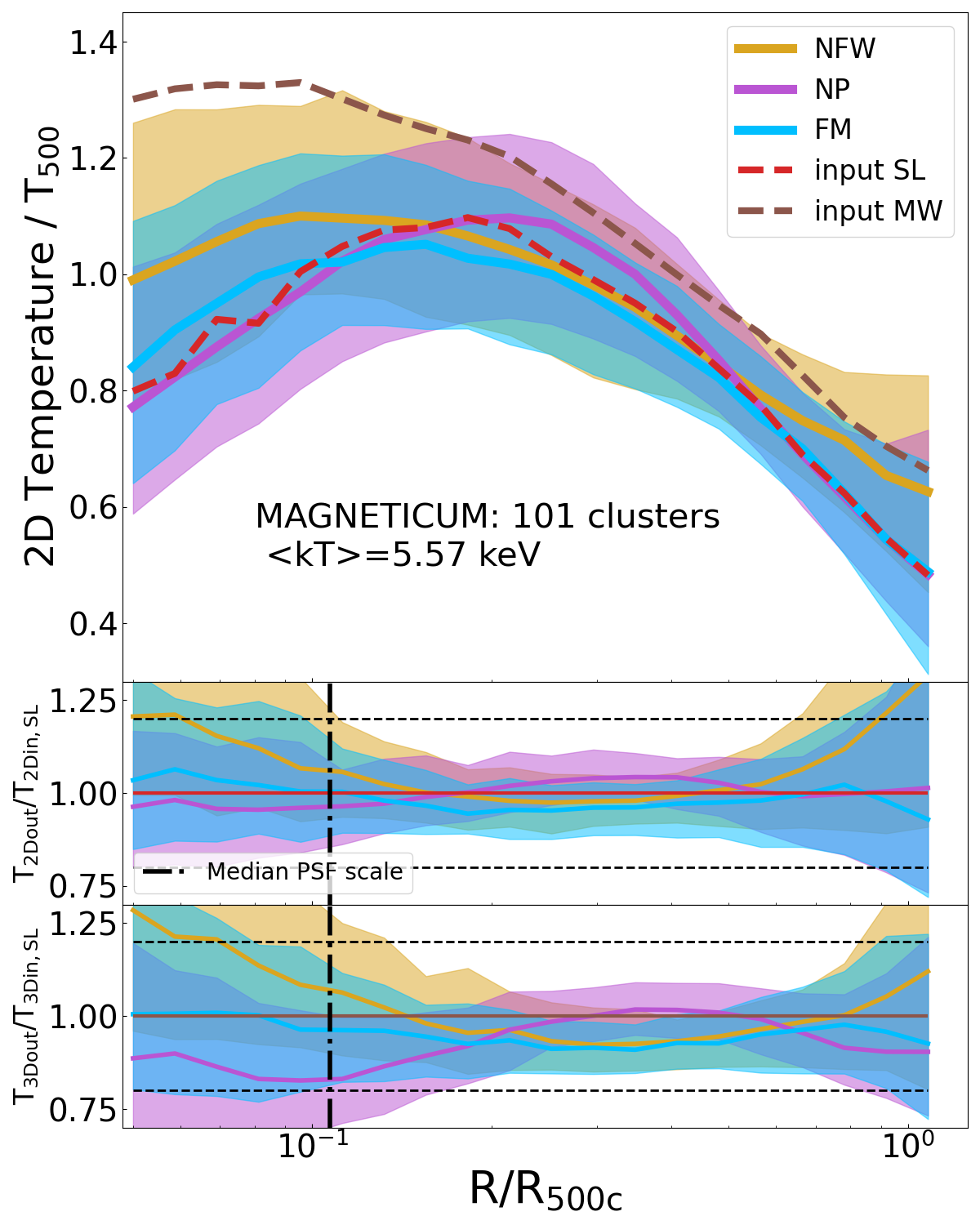}    
    \includegraphics[width=0.66\columnwidth]{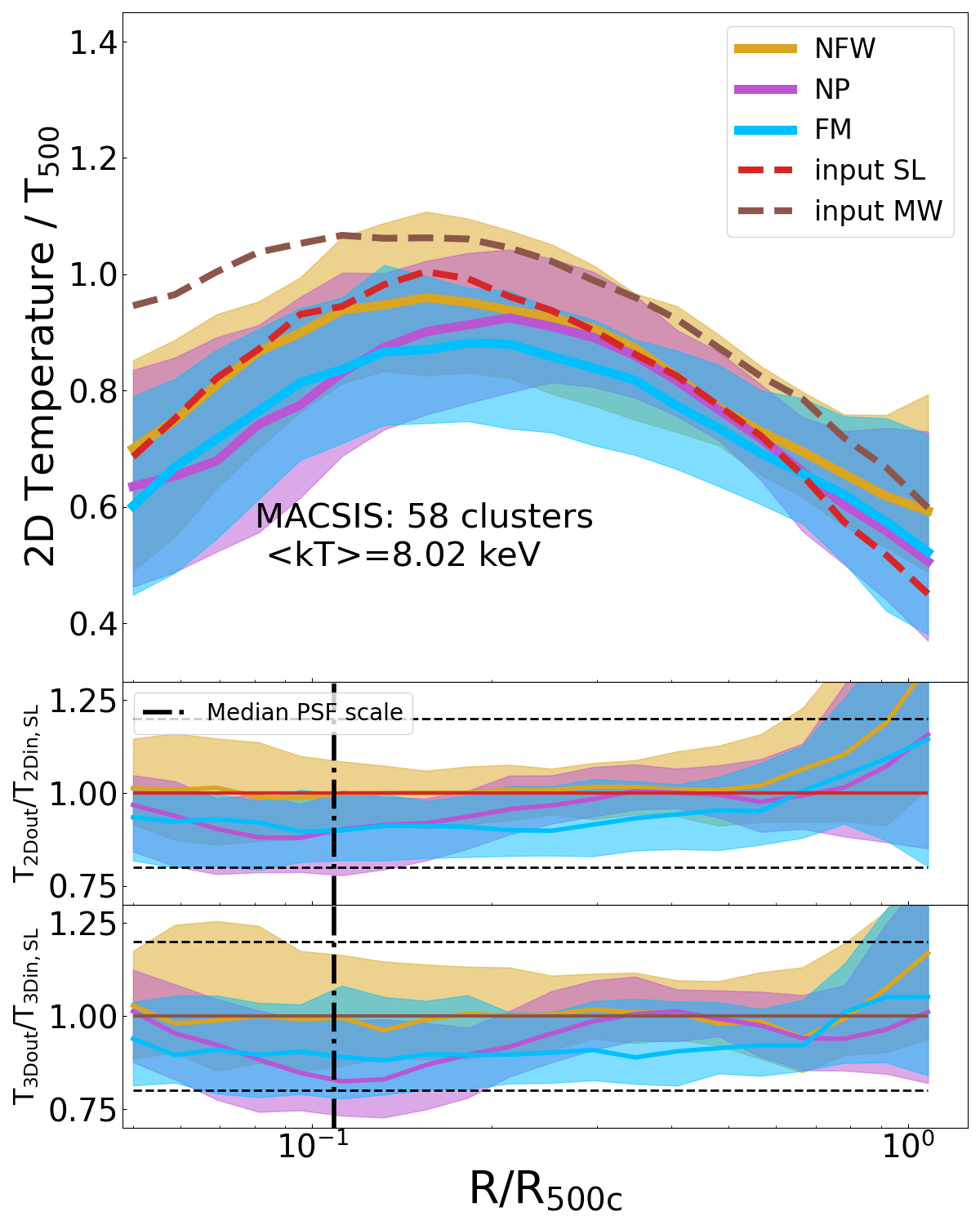}
  
    \caption{Gas temperature profiles. Each panel corresponds to one simulation: \texttt{The300} on the left, \texttt{Magneticum} in the centre, and \texttt{MACSIS} on the right. The colours denote the three reconstruction methods. The smaller panels show the ratio between the reconstructed 2D and deprojected 3D profiles.}
    \label{fig:Tx_profiles}
\end{figure*}

Temperature is a fundamental X-ray observable of galaxy clusters. As a quantity obtained directly from spectral fitting, it traces the depth of the gravitational potential and underpins hydrostatic mass estimates. Its radial profile constrains the thermodynamic structure of the ICM and, together with pressure, calibrates the scaling relations used in cluster cosmology. Biases in the recovered temperature arising from multi-temperature structure, projection, or instrumental effects therefore propagate directly into mass estimates and the interpretation of ICM physics. 

\subsubsection{Temperature weighting schemes}
\label{subsubsec:weighting schemes}

%In this section we assess the accuracy of the temperature profiles recovered from the synthetic XMM-Newton observations (Sect. \ref{sec:xray_analysis}) by comparing them to the corresponding reference values from the simulations. 
Unlike gas density, for which the true profile is easily defined using the particles, the notion of a true temperature profile is inherently ambiguous: although each gas element has a well-defined internal energy and thus a temperature, any radial profile or integrated quantity requires adopting a weighting scheme. The inferred temperature is sensitive to this choice \citep{Mazzotta2004MNRAS_TxSL, Rasia2014ApJ_Tstruc}, especially in the presence of multiphase gas. Observationally, spectral fitting preferentially weights cooler, denser phases of the ICM with stronger X-ray emission. To reflect these effects in our comparison, we consider two standard definitions. The mass-weighted (MW) temperature follows the assumption that more massive gas elements contribute proportionally more to the thermal energy budget. It is a physically motivated thermodynamic average linked to the gravitational potential and should be the benchmark for comparisons to observations based on combinations to the SZ effect. However, it does not fully capture the complex dependence of X-ray emissivity on both gas density and temperature, particularly in the presence of temperature inhomogeneities or multiphase structures \citep{Gaspari2020multiscale}, also because the typical response of X-ray instruments is higher at soft energies around 2 keV and lower for hotter gas above 3 keV. Instead, the spectroscopic-like (SL) scheme, introduced by \citet{Mazzotta2004MNRAS_TxSL} is constructed to emphasize denser, cooler gas phases that dominate the observed X-ray emission in XMM-Newton and Chandra data. It yields robust estimates for high-temperature plasma hotter than about 3 keV, making it particularly suitable for galaxy clusters in the mass range considered here. They are defined as follows:
\begin{align}
    T_{\rm MW} &= \frac{\sum_{i} m_i T_i}{\sum m_i}, \nonumber \\
    T_{\rm SL} &= \frac{\sum_i n_i^2/T_i^{\rm 3/4} T_i}{\sum n_i^2/T_i^{\rm 3/4}},
    \label{eq:SL_MW}
\end{align}
where $m$, $n$ are respectively the mass and the density of each individual gas particle, and the index $i$ identifies each particle within its volume, i.e. a spherical shell in case of a 3D profile, and a cylindrical shell for 2D profiles.
In this work, we test these weighting schemes with a full forward modelling of the XMM observations for the first time.

First we compare the radial profile of the measured temperature obtained by spectral fitting in annuli (T$_{\rm X}$) to the expected T$_{\rm SL}$. The result is shown in Fig. \ref{fig:Tspecfit_Tsl} with solid lines. This is close to one for \texttt{The300} and \texttt{Magneticum}, but it is underestimated in the cores of \texttt{MACSIS}, within about 0.2$\times$R$_{\rm 500c}$ by at most about 10$\%$ (although the one-to-one line is still within the scatter of \texttt{MACSIS} profiles). We exclude that this is due to PSF effects, because the median PSF scale of 30 arcsec corresponds to about 0.11$\times$R$_{\rm 500c}$ in \texttt{MACSIS}, compared to about 0.08 (0.105) in \texttt{The300} (\texttt{Magneticum}). Instead, we find this effect to correlate with cool cores (see dashed and dotted lines in Fig. \ref{fig:Tspecfit_Tsl}). Following \citet{Campitiello2022A&A_chexmate_morph}, we quantify surface‐brightness concentration\footnote{$c_{\rm SB}=\mathrm{SB}(<0.15 R_{\rm 500c})/\mathrm{SB}(<R_{\rm 500c})$} (their Eq. 1 and Sect. 4.1) and classify clusters as CC or NCC using the sample median. We obtain median $c_{\rm SB}$ values of 0.39, 0.19, and 0.54 for \texttt{The300}, \texttt{Magneticum}, and \texttt{MACSIS}, respectively. The300 agrees with \citet{Campitiello2022A&A_chexmate_morph}, while the lower value in \texttt{Magneticum} reflects its flatter cores. We then compare the T${\rm X}$/T${\rm SL}$ ratio for CC and NCC subsamples. In \texttt{Magneticum}, concentrations are generally low (quartiles Q1-Q3=0.15–0.22), and the two populations show no significant difference. In \texttt{The300}, the distribution is broader (Q1-Q3=0.23–0.51), but temperature profiles are flat (Fig. \ref{fig:Tx_profiles}), resulting in only a $\sim$5$\%$ core difference. In \texttt{MACSIS}, both the concentration distribution (Q1-Q3=0.27–0.59) and the temperature gradients are larger, producing a clear T${\rm X}$/T${\rm SL}$ offset: NCC resemble \texttt{The300}, but CC are underestimated by up to $\sim$15$\%$. Indeed, \citet{Mazzotta2004MNRAS_TxSL} defined the weight of 0.75 in T$_{\rm SL}$ to accommodate a combination of various plasma phases, although some variations are expected: their Fig. 9 shows that the larger the difference between the temperature of two plasma, the larger the deviations in T$_{\rm SL}$ compared to T$_{\rm X}$. This is particularly relevant for \texttt{MACSIS}: the temperature profiles are steep, so the 2D projection of the core contains plasma phases around 4–5 keV as well as 8–9 keV, where deviations around the 10$\%$ level are expected. Finally, \citet{Rossetti2024A&A_TxprofCXM} showed that the CHEX-MATE temperature profiles are in agreement with \texttt{The300} and do not exhibit the steep profiles that are typical in \texttt{MACSIS}. This suggests that T$_{\rm SL}$ is in an excellent proxy for T$_{\rm X}$ in the real CHEX-MATE sample. For more complex systems and strong cool cores, multi-temperature fits are required.
%Because \texttt{MACSIS} predominantly comprises Tier-2, higher-redshift systems with steep cool cores (median z=0.347), a 30 arcsec PSF corresponds to about 151 kpc or 0.11R$_{\rm 500c}$. The resulting beam smearing redistributes photons from the very cool central regions into larger annuli, biasing the recovered temperature profile low. This effect is much weaker in the other simulations, which include Tier-1 objects with higher spatial resolution and less centrally peaked profiles. 
In any case, the agreement is excellent also for \texttt{MACSIS} outside of 0.2$\times$R$_{\rm 500c}$, so that total mass measurements more sensitive to local temperature and temperature gradient around R$_{\rm 500c}$ are not affected by the discrepancy in the core. Nonetheless, we verified that the temperature modelling from the observer's perspective is indeed able to reconstruct the measured temperature profile (see Fig. \ref{fig:Tspecfit_Tsl_macsis}).

In the main panels of Fig. \ref{fig:Tx_profiles} we report the reconstructed two-dimensional temperature profiles from the NFW, NP, and FM models, compared to the true T$_{\rm SL}$ and T$_{\rm MW}$ profiles for each simulation. The profiles were rescaled to T$_{\rm 500c}$ adopting the X-COP cluster sample calibration, following \citet{Ghirardini2019A&A...621A..41G} (see their Eq. 10).
The smaller panels show output-input ratios: the top rows compare 2D reconstructed temperature to T$_{\rm SL}$, and the bottom rows show the corresponding ratios for the deprojected 3D profiles. In each panel we also report the median T$_{\rm 500c}$ for each sample. We find this to be low for \texttt{Magneticum} at 5.57 keV compared to 6.96 keV for \texttt{The300} and 8.02 keV for \texttt{MACSIS}. 
%This is expected, because of the lack of hydrostatic mass bias in sample selection 
This is expected because the sample from \texttt{Magneticum} has intrinsically less massive systems 
(Sect. \ref{subsec:clusample}). The black lines denote a median PSF scale of 30 arcsec given the radius and redshift of each system.%, so that the \texttt{Magneticum} sample is composed by less massive and therefore colder systems by construction. 

Similarly to gas density, the three simulations show different behaviours also in the temperature profiles. \texttt{The300} exhibits flat cores with temperatures very close to T$_{\rm 500c}$ within 0.3$\times$R$_{\rm 500c}$, while \texttt{MACSIS} and \texttt{Magneticum} produce clusters with lower temperatures in the core. In particular, \texttt{MACSIS} has strong cool cores with temperature being about 30$\%$ lower than T$_{\rm 500c}$ at 0.05$\times$R$_{\rm 500c}$. We find evidence that the stronger cooling in \texttt{MACSIS} suppresses the global ICM temperature, in fact, using the T$_{\rm 500c}$ normalization from \citet{Ghirardini2019A&A...621A..41G}, the average T(r)/T$_{\rm 500c}$ profile in \texttt{MACSIS} remains below unity over the full radial range examined. In comparison, \texttt{The300} and \texttt{Magneticum} reach (and locally exceed) unity, though with different profile shapes. These trends imply that, at fixed halo mass, \texttt{MACSIS} yields cooler clusters overall—consistent with enhanced cooling efficiency lowering the thermal energy of the ICM. 

The top panels of Fig. \ref{fig:Tx_profiles} also show that there is better agreement between the reconstructed profiles with the SL input models compared to the MW ones. This holds both for the 2D and deprojected 3D profiles. The largest differences are in the core within 0.15$\times$R$_{\rm 500c}$, where the MW scheme overestimates the reconstructed profiles by more than 20$\%$ (40$\%$) in \texttt{MACSIS} (\texttt{Magneticum}). The disagreement is not as pronounced in \texttt{The300}, which points to a smoother ICM temperature distribution, as we further explore in Sect. \ref{subsec:multiT}. Instead, the SL scheme shows a good agreement with the one-to-one ratio to the reconstructed profiles. The best result is provided by the NP model, which is the one that reconstructs temperature directly. Its flexibility provides almost a perfect one-to-one recovery of $T_{\rm SL}$ in \texttt{The300} and \texttt{Magneticum}. In \texttt{MACSIS} instead the agreement is excellent outside of 0.2$\times$R$_{\rm 500c}$, but not in the core, where we find that the temperature is underestimated by 5-10$\%$. This is due to the discrepancy in the temperature measurement investigated in Fig. \ref{fig:Tspecfit_Tsl}. \texttt{The300} shows the best consistency between different reconstruction methods, whereas \texttt{Magneticum} exhibits the largest scatter between NFW, NP, or FM reconstructions. Such differences may be caused by general departures from spherical symmetry or specific assumptions within each model, such as the NFW parametrisation, or likely different levels of multi-temperature structures in these simulations (see Sect. \ref{subsec:multiT}). 

Finally, the trends observed in the ratios between the 2D and deprojected 3D temperature profiles are consistent across the different simulations and reconstruction methods, for both the T$_{\rm SL}$ and T$_{\rm MW}$. This consistency suggests that the deprojection procedure does not introduce any significant bias in the reconstructed 3D profiles. %With this work we confirm that SL weights do not only provide an accurate and precise estimate of the global cluster temperature inferred from X-ray observations, but also of its profile.

We provide full details and figures about pressure and entropy in Appendix \ref{appendix:K_P_profs}.
Pressure is systematically underestimated in all simulations, although at different levels. At 
0.3$\times$R500 the bias is about 5$\%$ in \texttt{The300} and increases to about 20$\%$ in \texttt{Magneticum} and \texttt{MACSIS}. Comparable trends are seen near R500, albeit with some method-dependent scatter. Entropy is recovered to within 
5$\%$ across most radii in \texttt{The300}, whereas \texttt{Magneticum} and \texttt{MACSIS} show a larger deficit (15-20$\%$) in the inner regions. This reduces to 5-10$\%$ near R$_{\rm 500c}$. We also study the 2D cluster temperature distribution by generating and analysing voronoi binned temperature maps. Similarly to the density case, we find that the voronoi technique is less affected by clumpiness on a population level. In particular, after removing outliers, temperatures are raised by about 15-20$\%$, underscoring once again the effect of multi-temperature structure on radial temperature profiles. The same analysis on a 2 Ms simulation shows that the individual profiles are compatible with the radial spectral fitting within uncertainties (see Appendix \ref{appendix:2Ms}).

\subsubsection{Multi-temperature gas}
\label{subsec:multiT}

\begin{figure}
    \centering
    \includegraphics[width=\columnwidth]{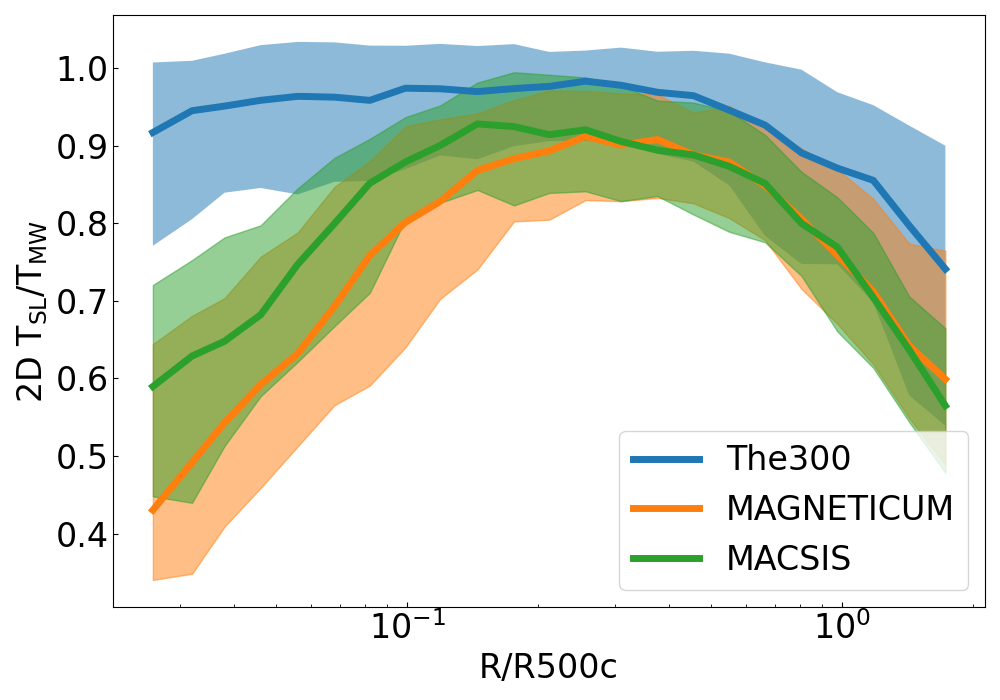}
    \caption{Ratio between the spectroscopic-like and mass-weighted temperature profiles in different simulations.}
    \label{fig:Tsl_TMW}
\end{figure}

\begin{figure}
    \centering
    \includegraphics[width=\columnwidth]{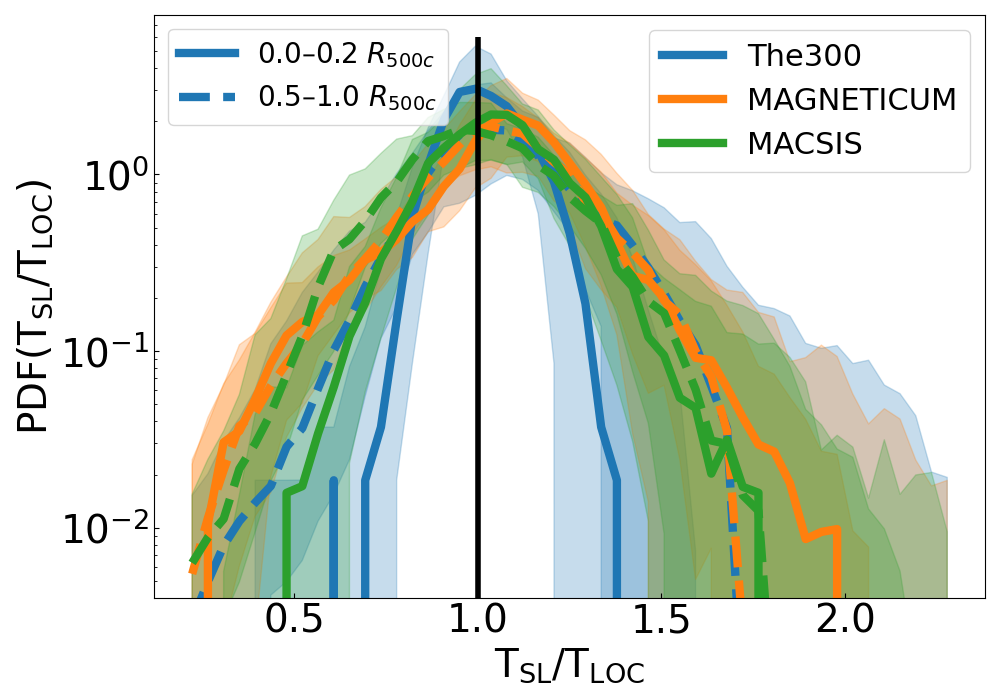}
    \caption{Comparison between the distribution of the local  spectroscopic-like temperature fluctuations in 2D maps within different apertures for different simulations. T$_{\rm LOC}$ is the 1D radial profile at the position of each pixel in the map.}
    \label{fig:Tsl_medTsl}
\end{figure}

\begin{figure}
    \centering
    \includegraphics[width=\columnwidth]{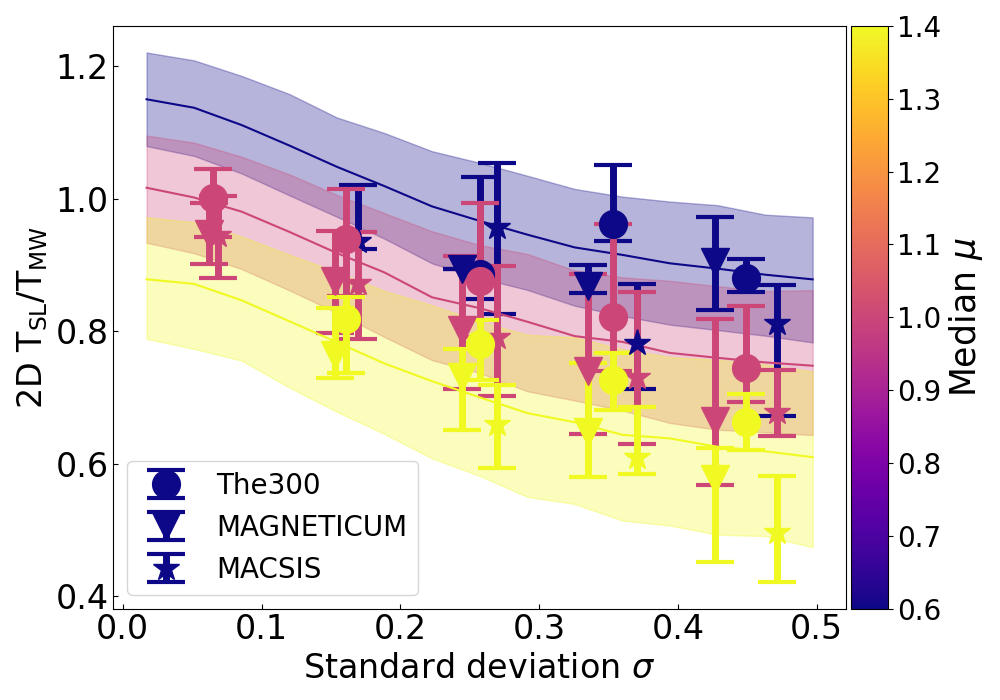}   
    \caption{Ratio between the spectroscopic-like and mass weighted temperature profiles as a function of the standard deviation of the spectroscopic-like temperature fluctuations. The points are colour coded by the median of the fluctuation distributions.}
    \label{fig:TSLTMW_sigma}
\end{figure}

The discrepancies observed between the reconstructed temperature profiles with T$_{\rm SL}$ and T$_{\rm MW}$, as well as the intrinsic differences between these schemes themselves, can arise from several factors: the presence of multi-phase gas due to mixing or substructures, which induces azimuthal temperature variations within a given radial bin, and projection effects affecting the 3D reconstruction of the temperature profile. In Appendix \ref{appendix:toy_model} we used a toy model with one single radial profile distribution and find that projection effects should yield a ratio between SL and MW temperatures below one in the core.
However, in Fig. \ref{fig:Tsl_TMW}, we observe a different trend: T$_{\rm MW}$ remains consistently higher than T$_{\rm SL}$ across all radii. While the discrepancy in the core is expected, the growing divergence in the outskirts suggests that systems simulated in a cosmological environment exhibit significant temperature inhomogeneities, departing from a simple individual temperature profile. This indicates that multi-temperature gas, driven by substructures, mixing, or accretion-related shocks, has a non-negligible impact on the projected temperature profiles of galaxy clusters. This is also in agreement with the picture described by \citet{Lovisari2024A&A_CXMATETX}, who found a small effect on the temperature profile of CHEX-MATE clusters due to highly deviating regions in the core, but larger effects up to 20$\%$ moving towards the outskirts. This also relevant for estimates of the total cluster mass, that depends on the temperature normalisation and gradient at a given radius (Eq. \ref{eq:hydroeq_idealgas}) \citep[see also discussions in][]{Kawahara2007ApJ...659..257K, Rasia2006MNRAS.369.2013R, Rasia2012NJPh_xraylensing, Rasia2014ApJ_Tstruc, Pearce2020MNRAS.491.1622P, Barnes2021MNRAS_bhe}. .

To further study azimuthal temperature variations in the ICM, we use 2D maps of T$_{\rm SL}$ to analyse its distribution normalized by the local 1D radial profile in a given radial annulus interpolated at the position of each pixel of the 2D map (T$_{\rm LOC}$). This approach isolates relative temperature fluctuations within clusters, allowing for a direct comparison of the spatial variability of the thermal structure, independent of the overall temperature normalization. The use of this normalized quantity is motivated by several key trends observed in the simulations. In particular, \texttt{Magneticum} exhibits a more pronounced difference between $T_{\rm SL}$ and $T_{\rm MW}$ compared to \texttt{The300} (see Fig. \ref{fig:Tsl_TMW}). Similarly, \texttt{Magneticum} shows a stronger deviation between temperature profiles reconstructed from median voronoi temperature maps and those obtained via spectral fitting in radial bins (see Fig. \ref{fig:voronoiTxprofs}). These discrepancies suggest that \texttt{Magneticum} may exhibit stronger thermal inhomogeneities and multiphase structure in the ICM, which can enhance azimuthal variations and bias projected temperatures. This is consistent with differences in the underlying hydrodynamic schemes: SPH formulations with limited mixing are known to preserve cool substructures and generate larger temperature contrasts \citep[e.g.][]{Rasia2014ApJ_Tstruc}. More recent SPH implementations, such as the improved scheme in GADGET-X used for \texttt{The300}, or the OWLS-based subgrid physics adopted in \texttt{MACSIS} \citep[][]{Schaye2010MNRAS_owls} promote more efficient mixing and therefore tend to produce smoother temperature fields.
By computing the PDF of the normalised $T_{\rm SL}$ across clusters as a function of radius, we obtain a population-level diagnostic that is sensitive to such internal fluctuations. A broader distribution in this statistic reflects stronger local deviations from the median, and thus captures the degree of thermal asymmetry, clumpiness, or substructure. This method removes the need for assuming a fixed hydrostatic mass bias or one-to-one matching between clusters, it provides a flexible and robust way to compare the structural complexity of the ICM across different simulation sets and physical models. The result is shown in Fig. \ref{fig:Tsl_medTsl}, focusing on the core and outskirts in the three simulations. We find that \texttt{The300} has the tightest distribution in the core, where fluctuations are at the 40$\%$ level at most. The distribution is broader in \texttt{MACSIS} and especially in \texttt{Magneticum}, reaching also a factor of two. In the outskirts there is better agreement between the simulations, especially at the positive fluctuations end. We also notice that the PDF in \texttt{Magneticum} tend to peak at values slightly larger than one, possibly pointing to non-gaussian temperature fluctuation distribution.

\setlength{\tabcolsep}{3.0pt} % default is 6pt
\begin{table}[]
    \centering
    \caption{Best fit parameters of the relation between the ratio of SL and MW temperatures and the width of the temperature distribution (see Eq. \ref{eq:Tratio_model}).}
    \begin{tabular}{|c|c|c|c|c|}
        \hline
        \hline
        \textbf{Simulation}   & \textbf{A} & \textbf{b} & \centering\textbf{$\log P$} & \textbf{$\sigma_{\rm intr}$} \\
        \hline
        \rule{0pt}{2ex}  
        \texttt{The300}  & 0.20$\pm$0.01 & -0.35$\pm$0.01 & 1.09$\pm$0.06 & 0.055$\pm$0.001 \\
        \texttt{Magneticum} & 0.26$\pm$0.02 & -0.38$\pm$0.01 & 1.08$\pm$0.08 & 0.082$\pm$0.001 \\
        \texttt{MACSIS} & 0.31$\pm$0.03 & -0.37$\pm$0.01 & 0.85$\pm$0.09 & 0.083$\pm$0.001 \\
        \texttt{COMBINED} & 0.31$\pm$0.01 & -0.33$\pm$0.01 & 0.76$\pm$0.05 & 0.083$\pm$0.001 \\
        \hline
    \end{tabular}
    \label{tab:TSLTMW_sigma_pars}
\end{table}
\setlength{\tabcolsep}{6.0pt} % default is 6pt

We take a further step by directly measuring the first, second, and third orders of the temperature fluctuations, that are the median, the standard deviation, and the skewness, in 30 bins equally distributed between 0 and 1.05$\times$R$_{\rm 500c}$. In each radial bin we also keep track of the ratio between $T_{\rm SL}$ and $T_{\rm MW}$.
We find an excellent correlation between the standard deviation of these distributions with the ratio between the 2D $T_{\rm SL}$ and $T_{\rm MW}$ in different bins, 
%it is clear in Fig. \ref{fig:TSLTMW_sigma}. 
as shown in Fig. \ref{fig:TSLTMW_sigma}. 
Combining the three simulations we get a Pearson correlation coefficient of -0.76. This is due to the presence of gas chunks with different temperatures compared to the mean, increasing the difference between $T_{\rm SL}$ and $T_{\rm MW}$. In addition, we find a secondary trend in the temperature ratio as a function of the median $\mu$ at fixed standard deviation $\sigma$. The Pearson correlation coefficient is -0.49. The ratio is closer to one for small $\mu$ and decreases as $\mu$ increases. Although we do not find a clear correlation between the skewness of the distribution and the ratio between $T_{\rm SL}$ and $T_{\rm MW}$ (Pearson correlation coefficient of -0.04), this secondary $\mu$-trend is partially encoded in the skewness as well: distributions with a long tail at low temperatures tend to peak at higher values, so that a PDF with mean larger than one describes a gas distribution skewed to colder temperature, as we can see from the core of \texttt{Magneticum} clusters in Fig. \ref{fig:Tsl_medTsl}. The SL scheme increases the weight of the cold gas, reducing the ratio to the $T_{\rm MW}$. The opposite happens for distributions that are more symmetric or even skewed towards hotter gas, with the ratio being closer to one. In general, for tight and symmetric distributions, with $\mu$ close to one and low $\sigma$, $T_{\rm SL}$ is an unbiased tracer of $T_{\rm MW}$. As the distribution broadens, gas chunks that deviate from the median are weighted differently by the SL and MW schemes, which progressively lowers the temperature ratio. We build a model that includes the scatter of the temperature ratio around the expectation value computed from the width and the median. It reads:
\begin{equation}
    \frac{T_{SL}}{T_{MW}} (A, b, p) = \mathcal{N}(1.35 + b\mu - A\frac{\sigma^p}{\sigma^p + 0.22^p}, \sigma_{\rm intr}),
    \label{eq:Tratio_model}
\end{equation}
where $\mathcal{N}$ is a Gaussian distribution, and $\sigma_{\rm intr}$ the relative scatter around the expectation value.
The best fit parameters are reported in Table \ref{tab:TSLTMW_sigma_pars} and their posterior distributions are shown in Fig. \ref{fig:cornerTSLTMW_sigma_mu}. Given a measurement of the median and the standard deviation of the temperature distribution at a given radial distance, this model provides a direct mapping from the spectroscopic-like temperature to the mass weighted one, which is the temperature more directly related to the gravitational potential itself. The correction provided by this model allows X-ray analyses to account for multi-temperature structure, reducing the expected hydrostatic mass bias that is not due to departures from hydrostatic equilibrium but to spectroscopic temperature bias in multi-phase gas, as already pointed out by \citet{Pearce2020MNRAS.491.1622P, Barnes2021MNRAS_bhe}. %For example, around 0.5$\times$R$_{\rm 500c}$ typical values of these parameters are $\sigma=0.19, \mu=1.01$, that corresponds a correction of 0.88, while at R$_{\rm 500c}$ with typical values of $\sigma=0.26, \mu=1.05$, we predict a correction of 0.82. Under hydrostatic equilibrium (Eq. \ref{eq:hydroeq_idealgas}), the enclosed mass scales approximately linearly with the local temperature at fixed gradients, so this temperature deficit translates directly into an 18$\%$ underestimate of the hydrostatic mass. This component of the mass bias arises purely from unresolved multi-temperature structure, rather than from departures from hydrostatic support. We stress that is not a robust and accurate prediction, because the scatter of $\sigma$ ($\mu$) is around 40$\%$ (12$\%$) around R$_{\rm 500c}$. 
We leave the quantitative evaluation of this effect to future \citep[see also][]{Ansarifard2020A&A_300bhse}.

In conclusion, identifying and quantifying multi-temperature structures is crucial for accurately interpreting observed temperature profiles and understanding ICM thermodynamics. One promising avenue is the use of high-resolution X-ray spectroscopy from missions like the ongoing XRISM or Athena \citep[][]{Cruise2025NatAs_newathena} in the next decades. In addition to fitting a global temperature from the thermal bremsstrahlung continuum, the spectral resolution of the microcalorimeters onboard these missions allows for detailed diagnostics using line ratios, particularly of the FeXXV and FeXXVI lines around 6.4–6.7 keV. These lines are sensitive to different ionization states and thus to the plasma temperature: in a hot, single-phase plasma the FeXXVI line is prominent and comparable to FeXXV, while in cooler plasma the FeXXV line dominates \citep[][]{Xrism2025ApJ_coma}. Importantly, the effective area of XRISM is sharply peaked at soft energies, where the folded continuum spectra of hot and cool clusters can appear deceptively similar due to instrumental response. In contrast, the iron line ratio method probes a narrow energy window where the ARF is approximately constant, reducing the impact of effective area variation and making it a more direct diagnostic of the thermal structure of the plasma. A discrepancy between temperatures inferred from the continuum and from line ratios would potentially indicate the presence of multi-T gas components. Such works would still need a careful treatment of systematics, as new spectral data with high signal to noise ratio and resolution may uncover limitation of the current models \citep[see][]{Chatzigiannakis2025arXiv_TZ_ICM}.
An alternative and complementary approach involves combining X-ray and SZ observations \citep[][De Luca et al. subm]{Pointecouteau2002A&AXraySZ, Kitayama2004PASJ_SZXrayclu, Ruppin2018A&A_NIKA2, Eckert2019xcop, Deluca2021MNRAS_300MAPS, Chappuis2025A&A_1689, Gavidia2025_clump3d}. The SZ signal is proportional to the integrated electron pressure, whereas the X-ray emission is proportional to the emission measure, i.e. the gas density squared. As a result, X-ray temperatures are weighted more heavily toward denser, cooler regions, while SZ-derived temperatures reflect a more volume-averaged (or mass-weighted) thermal state. Comparing the SZ-inferred temperature, obtained by dividing the SZ pressure by the X-ray electron density or directly from the relativistic SZ signal, to the X-ray spectroscopic temperature can thus reveal discrepancies arising from multi-T structure, clumping, or non-thermal pressure support \citep[][]{Kay2024MNRAS_flamingo}. In both approaches, a consistent difference between the temperature diagnostics offers a valuable observational pathway to identifying thermal complexity in the ICM.

\section{Summary and conclusions}
\label{sec:conclusions}

Within the context of the CHEX-MATE project \citep{CHEX-MATE2021A&A_intro}, our primary goal is to assess whether standard X-ray analysis techniques can accurately recover the true thermodynamic properties of massive galaxy clusters. To this end, we selected CHEX-MATE-like cluster samples from three state-of-the-art hydrodynamical simulations: \texttt{The300} \citep[][]{Cui2018MNRAS_The300}, \texttt{Magneticum} \citep{Dolag2017_Magneticum_metals}, and \texttt{MACSIS} \citep[][]{Barnes2017MNRAS_macsis}. Using the gas particles in each simulated cluster, we generated idealised X-ray emission maps and folded them through the instrumental response of XMM-Newton using our newly developed tool, \texttt{xmm\_simulator}, to create high-fidelity, end-to-end synthetic EPIC observations (see Sect. \ref{sec:mock_gen}).

We then applied standard X-ray analysis tools, including routines from \texttt{pyproffit} \citep{Eckert2020_pyproffit} and \texttt{hydromass} \citep{Eckert2022_hydromass}, to extract surface brightness and temperature profiles, perform deprojection, and derive intrinsic 3D thermodynamic profiles (see Sect. \ref{sec:xray_analysis}). This workflow allows for a direct comparison between the output of our X-ray analysis and the corresponding "true" profiles from the hydrodynamical simulations.

We find that the gas density profiles are robustly reconstructed across all methods and simulations, with the correct intrinsic scatter. In particular, using azimuthal median surface brightness profiles derived from voronoi-tessellated images improves accuracy: other than some discrepancies in the core because of mis-centering effects between the peak of the X-ray emitting gas and the dark matter, the gas density profiles are reconstructed within at most 2$\%$ in all simulations. The gas masses are reconstructed to better than 1$\%$ uncertainty. The voronoi technique suppresses the influence of localised surface brightness enhancements due to cool gas clumps, which otherwise bias the inferred density high by a few percent. %These findings hold across simulations with diverse physical prescriptions for baryon physics, and the analysis successfully reproduces the intrinsic scatter of the profiles.

While the gas density is well-defined in simulations, defining a representative true temperature is more complex. We compare two standard weighting schemes: the spectroscopic-like (SL) and mass-weighted (MW) temperatures. The SL weighting provides an accurate approximation of temperatures derived from X-ray spectral fitting in annular bins. However, we observe more scatter in the temperature reconstruction compared to the density case, and a systematic difference between $T_{\rm SL}$ and $T_{\rm MW}$ profiles, a difference that varies across simulations. This suggests that these weighting schemes carry information about the underlying physical state of the ICM.
If these effects were only due to projection effects, we would expect the $T_{\rm MW}$ to be higher in the core and lower in the outskirts compared to $T_{\rm SL}$. However, in our simulations, we consistently find $T_{\rm MW}$ to be higher than $T_{\rm SL}$ at all radii. This deviation follows expectations and points to the presence of unresolved multi-temperature gas structures, which have a significant impact on the reconstructed profiles. Indeed, even though the input and output density and temperature profiles appear consistent, the derived thermal pressure and entropy profiles differ. This discrepancy may result from an unaccounted multi-temperature distribution in the X-ray spectral modelling, combined with possible non-thermal pressure contributions.

%We also generated voronoi-tessellated temperature maps for each simulated cluster and extracted azimuthal median profiles \citep[see e.g.,][]{Lovisari2024A&A_CXMATETX}. We find that in the cluster cores the voronoi-based temperature profiles are systematically higher than those from annular spectral fitting, which is biased low by dense, cool substructures. Conversely, in the 0.2–0.6 R$_{\rm 500c}$ range, the voronoi temperatures are lower, possibly because spectral fitting is more sensitive to high-emissivity, hotter regions, while the voronoi median averages over broader azimuthal variation. Near R$_{\rm 500c}$, larger voronoi bins (required for sufficient S/N) introduce noise and increase the scatter in the distributions (see Appendix \ref{appendix:2Ms}).
%To verify the robustness of this approach, we simulated a deep 2 Ms XMM-Newton observation of a single cluster without AGN contamination. We repeated the full analysis, including the voronoi map construction, and confirmed that the trends observed in the 25 ks simulations persist. The result is presented in Appendix \ref{appendix:2Ms}.
These findings raise important implications for the interpretation of hydrostatic mass estimates in galaxy clusters \citep[see discussions in][]{Rasia2006MNRAS.369.2013R, Rasia2012NJPh_xraylensing, Biffi2016HE, Pearce2020MNRAS.491.1622P}. A long-standing issue in X-ray cluster cosmology is the hydrostatic mass bias, where masses derived under the assumption of hydrostatic equilibrium are systematically low. This discrepancy has often been attributed to non-thermal pressure support from bulk motions or turbulence in the ICM. However, recent high-resolution spectroscopic measurements from XRISM have revealed remarkably low gas velocity dispersions in relaxed systems, suggesting that non-thermal pressure may not be sufficient to account for the full bias \citep{XRISM2025Natur_centaurus, Xrism2025ApJ_coma, Xrism2025ApJ_a2029, XRISM2025arXiv_A2029II, Fujita2025arXiv_xrismOphiucus}, although constrained to central regions so far. Our results suggest an alternative and complementary explanation in line with findings from \citet{Henson2017MNRAS.465.3361H, Pearce2020MNRAS.491.1622P, Barnes2021MNRAS_bhe}: the presence of unresolved multi-temperature structures in the ICM can bias spectroscopic temperature measurements, leading to a suppression of the inferred thermal pressure. If the temperature is underestimated due to projection effects or local cooling structures, the derived hydrostatic mass will also be biased low, even in systems that are otherwise in equilibrium. These results highlight the importance of temperature reconstruction methods that mitigate these biases, and stress the importance of multi-wavelength, spatially resolved, and dynamical analyses to disentangle the true thermodynamic state and equilibrium conditions of the ICM. Looking ahead, our findings emphasise the importance of isolating the distinct sources of bias affecting X-ray-derived cluster mass estimates. In future work, we will extend this analysis to directly compare hydrostatic mass estimates with true masses, enabling a detailed assessment of how much bias arises from assumptions of hydrostatic equilibrium, from the use of single-temperature models in a multi-phase medium, and from projection effects such as triaxiality and line-of-sight substructure \citep{Kim2024A&A_clump3d, Saxena2025arXiv_triaxial, Chappuis2025A&A_1689}. This will be key to understanding whether the observed hydrostatic mass bias truly reflects a breakdown of equilibrium in the ICM, or whether it stems, at least in part, from systematic limitations in our current X-ray analysis methodologies.

\begin{acknowledgements}
RS and DE are supported by Swiss National Science Foundation project grant \#200021\_212576. ER is partially supported by the NASA -EGIP Grant 80NSSC25K8009.
EP acknowledges support of CNES, the French space agency and from the French Agence Nationale de la Recherche (ANR), under grant ANR-22-CE31-0010.
VB acknowledges partial support from the INAF Grant 2023 “Origins of the ICM metallicity in galaxy clusters”.
This research was supported by the International Space Science Institute (ISSI) in Bern, through ISSI International Team project \#565 ({\it Multi-Wavelength Studies of the Culmination of Structure Formation in the Universe}).
We acknowledge the financial contribution from the contracts Prin-MUR 2022 supported by Next Generation EU (M4.C2.1.1, n.20227RNLY3 {\it The concordance cosmological model: stress-tests with galaxy clusters}), and from the European Union’s Horizon 2020 Programme under the AHEAD2020 project (grant agreement n. 871158).
MDP acknowledges financial support from PRIN-MUR grant 20228B938N {\it Mass and selection biases of galaxy clusters: a multi-probe approach} funded by the European Union Next generation EU, Mission 4 Component 2 CUP B53D23004790006.
MS acknowledges financial contributions from contract ASI-INAF n.2017-14-H.0, contract INAF mainstream project 1.05.01.86.10, INAF Theory Grant 2023: Gravitational lensing detection of matter distribution at galaxy cluster boundaries and beyond (1.05.23.06.17).
MR acknowledges the financial contribution from INAF grant 1.05.24.02.10
LL acknowledges the financial contribution from the INAF grant 1.05.12.04.01.
MG acknowledges support from the ERC Consolidator Grant \textit{BlackHoleWeather} (101086804).
JS was supported by NASA Astrophysics Data Analysis Program (ADAP) Grant 80NSSC21K1571.
HB, FDL, PM acknowledge the support by INFN through the InDark initiative.

\end{acknowledgements}

\bibliographystyle{aa}
\bibliography{biblio}

@ARTICLE{Cash1979ApJ...228..939C,
       author = {{Cash}, W.},
        title = "{Parameter estimation in astronomy through application of the likelihood ratio.}",
      journal = {\apj},
         year = 1979,
        month = mar,
       volume = {228},
        pages = {939-947},
          doi = {10.1086/156922},
       adsurl = {https://ui.adsabs.harvard.edu/abs/1979ApJ...228..939C}
}

@ARTICLE{Sutherland1993_coolfunc,
       author = {{Sutherland}, Ralph S. and {Dopita}, M.~A.},
        title = "{Cooling Functions for Low-Density Astrophysical Plasmas}",
      journal = {\apjs},
         year = 1993,
        month = sep,
       volume = {88},
        pages = {253},
          doi = {10.1086/191823},
       adsurl = {https://ui.adsabs.harvard.edu/abs/1993ApJS...88..253S}
}

@ARTICLE{Evrard1996ApJ_xrayMASS,
       author = {{Evrard}, August E. and {Metzler}, Christopher A. and {Navarro}, Julio F.},
        title = "{Mass Estimates of X-Ray Clusters}",
      journal = {\apj},
         year = 1996,
        month = oct,
       volume = {469},
        pages = {494},
          doi = {10.1086/177798},
archivePrefix = {arXiv},
       eprint = {astro-ph/9510058},
 primaryClass = {astro-ph},
       adsurl = {https://ui.adsabs.harvard.edu/abs/1996ApJ...469..494E}
}

@INPROCEEDINGS{Arnaud1996ASPC..101...17A,
       author = {{Arnaud}, K.~A.},
        title = "{XSPEC: The First Ten Years}",
    booktitle = {Astronomical Data Analysis Software and Systems V},
         year = 1996,
       editor = {{Jacoby}, George H. and {Barnes}, Jeannette},
       series = {Astronomical Society of the Pacific Conference Series},
       volume = {101},
        month = jan,
        pages = {17},
       adsurl = {https://ui.adsabs.harvard.edu/abs/1996ASPC..101...17A}
}

@ARTICLE{Navarro_Frenk_White_1996,
       author = {{Navarro}, Julio F. and {Frenk}, Carlos S. and {White}, Simon D.~M.},
        title = "{The Structure of Cold Dark Matter Halos}",
      journal = {\apj},
         year = 1996,
        month = may,
       volume = {462},
        pages = {563},
          doi = {10.1086/177173},
archivePrefix = {arXiv},
       eprint = {astro-ph/9508025},
 primaryClass = {astro-ph},
       adsurl = {https://ui.adsabs.harvard.edu/abs/1996ApJ...462..563N}
}

@ARTICLE{BryanNorman1998,
       author = {{Bryan}, Greg L. and {Norman}, Michael L.},
        title = "{Statistical Properties of X-Ray Clusters: Analytic and Numerical Comparisons}",
      journal = {\apj},
         year = 1998,
        month = mar,
       volume = {495},
       number = {1},
        pages = {80-99},
          doi = {10.1086/305262},
archivePrefix = {arXiv},
       eprint = {astro-ph/9710107},
 primaryClass = {astro-ph},
       adsurl = {https://ui.adsabs.harvard.edu/abs/1998ApJ...495...80B}
}

@ARTICLE{Mathiesen2001ApJ...546..100M,
       author = {{Mathiesen}, B.~F. and {Evrard}, A.~E.},
        title = "{Four Measures of the Intracluster Medium Temperature and Their Relation to a Cluster's Dynamical State}",
      journal = {\apj},
         year = 2001,
        month = jan,
       volume = {546},
       number = {1},
        pages = {100-116},
          doi = {10.1086/318249},
archivePrefix = {arXiv},
       eprint = {astro-ph/0004309},
 primaryClass = {astro-ph},
       adsurl = {https://ui.adsabs.harvard.edu/abs/2001ApJ...546..100M}
}

@ARTICLE{Pointecouteau2002A&AXraySZ,
       author = {{Pointecouteau}, E. and {Hattori}, M. and {Neumann}, D. and {Komatsu}, E. and {Matsuo}, H. and {Kuno}, N. and {B{\"o}hringer}, H.},
        title = "{SZ and X-ray combined analysis of a distant galaxy cluster, RX J2228+2037.}",
      journal = {\aap},
         year = 2002,
        month = may,
       volume = {387},
        pages = {56-62},
          doi = {10.1051/0004-6361:20020394},
archivePrefix = {arXiv},
       eprint = {astro-ph/0203268},
 primaryClass = {astro-ph},
       adsurl = {https://ui.adsabs.harvard.edu/abs/2002A&A...387...56P}
}

@ARTICLE{MoWhite2002MNRAS.336..112M,
       author = {{Mo}, H.~J. and {White}, S.~D.~M.},
        title = "{The abundance and clustering of dark haloes in the standard {\ensuremath{\Lambda}}CDM cosmogony}",
      journal = {\mnras},
         year = 2002,
        month = oct,
       volume = {336},
       number = {1},
        pages = {112-118},
          doi = {10.1046/j.1365-8711.2002.05723.x},
archivePrefix = {arXiv},
       eprint = {astro-ph/0202393},
 primaryClass = {astro-ph},
       adsurl = {https://ui.adsabs.harvard.edu/abs/2002MNRAS.336..112M}
}

@ARTICLE{Mazzotta2004MNRAS_TxSL,
       author = {{Mazzotta}, P. and {Rasia}, E. and {Moscardini}, L. and {Tormen}, G.},
        title = "{Comparing the temperatures of galaxy clusters from hydrodynamical N-body simulations to Chandra and XMM-Newton observations}",
      journal = {\mnras},
         year = 2004,
        month = oct,
       volume = {354},
       number = {1},
        pages = {10-24},
          doi = {10.1111/j.1365-2966.2004.08167.x},
archivePrefix = {arXiv},
       eprint = {astro-ph/0404425},
 primaryClass = {astro-ph},
       adsurl = {https://ui.adsabs.harvard.edu/abs/2004MNRAS.354...10M}
}

@ARTICLE{Springel2005,
   author = {{Springel}, V.},
    title = "{The cosmological simulation code GADGET-2}",
  journal = {\mnras},
   eprint = {astro-ph/0505010},
     year = 2005,
    month = dec,
   volume = 364,
    pages = {1105-1134},
      doi = {10.1111/j.1365-2966.2005.09655.x},
   adsurl = {http://adsabs.harvard.edu/abs/2005MNRAS.364.1105S}
}

@ARTICLE{Springel2003MNRAS_SF,
       author = {{Springel}, Volker and {Hernquist}, Lars},
        title = "{Cosmological smoothed particle hydrodynamics simulations: a hybrid multiphase model for star formation}",
      journal = {\mnras},
         year = 2003,
        month = feb,
       volume = {339},
       number = {2},
        pages = {289-311},
          doi = {10.1046/j.1365-8711.2003.06206.x},
archivePrefix = {arXiv},
       eprint = {astro-ph/0206393},
 primaryClass = {astro-ph},
       adsurl = {https://ui.adsabs.harvard.edu/abs/2003MNRAS.339..289S}
}

@ARTICLE{Allen2004MNRAS.353..457A,
       author = {{Allen}, S.~W. and {Schmidt}, R.~W. and {Ebeling}, H. and {Fabian}, A.~C. and {van Speybroeck}, L.},
        title = "{Constraints on dark energy from Chandra observations of the largest relaxed galaxy clusters}",
      journal = {\mnras},
         year = 2004,
        month = sep,
       volume = {353},
       number = {2},
        pages = {457-467},
          doi = {10.1111/j.1365-2966.2004.08080.x},
archivePrefix = {arXiv},
       eprint = {astro-ph/0405340},
 primaryClass = {astro-ph},
       adsurl = {https://ui.adsabs.harvard.edu/abs/2004MNRAS.353..457A}
}

@ARTICLE{Kitayama2004PASJ_SZXrayclu,
       author = {{Kitayama}, Tetsu and {Komatsu}, Eiichiro and {Ota}, Naomi and {Kuwabara}, Takeshi and {Suto}, Yasushi and {Yoshikawa}, Kohji and {Hattori}, Makoto and {Matsuo}, Hiroshi},
        title = "{Exploring Cluster Physics with High-Resolution Sunyaev--Zel'dovich Effect Images and X-Ray Data: The Case of the Most X-Ray-Luminous Galaxy Cluster RX J1347-1145}",
      journal = {\pasj},
         year = 2004,
        month = feb,
       volume = {56},
        pages = {17-28},
          doi = {10.1093/pasj/56.1.17},
archivePrefix = {arXiv},
       eprint = {astro-ph/0311574},
 primaryClass = {astro-ph},
       adsurl = {https://ui.adsabs.harvard.edu/abs/2004PASJ...56...17K}
}

@ARTICLE{Bohringer2004A&Areflex,
       author = {{B{\"o}hringer}, H. and {Schuecker}, P. and {Guzzo}, L. and {Collins}, C.~A. and {Voges}, W. and {Cruddace}, R.~G. and {Ortiz-Gil}, A. and {Chincarini}, G. and {De Grandi}, S. and {Edge}, A.~C. and {MacGillivray}, H.~T. and {Neumann}, D.~M. and {Schindler}, S. and {Shaver}, P.},
        title = "{The ROSAT-ESO Flux Limited X-ray (REFLEX) Galaxy cluster survey. V. The cluster catalogue}",
      journal = {\aap},
         year = 2004,
        month = oct,
       volume = {425},
        pages = {367-383},
          doi = {10.1051/0004-6361:20034484},
archivePrefix = {arXiv},
       eprint = {astro-ph/0405546},
 primaryClass = {astro-ph},
       adsurl = {https://ui.adsabs.harvard.edu/abs/2004A&A...425..367B}
}

@ARTICLE{Diehl2006MNRAS._voronoi,
       author = {{Diehl}, Steven and {Statler}, Thomas S.},
        title = "{Adaptive binning of X-ray data with weighted Voronoi tessellations}",
      journal = {\mnras},
         year = 2006,
        month = may,
       volume = {368},
       number = {2},
        pages = {497-510},
          doi = {10.1111/j.1365-2966.2006.10125.x},
archivePrefix = {arXiv},
       eprint = {astro-ph/0512074},
 primaryClass = {astro-ph},
       adsurl = {https://ui.adsabs.harvard.edu/abs/2006MNRAS.368..497D}
}

@ARTICLE{Rasia2006MNRAS.369.2013R,
       author = {{Rasia}, E. and {Ettori}, S. and {Moscardini}, L. and {Mazzotta}, P. and {Borgani}, S. and {Dolag}, K. and {Tormen}, G. and {Cheng}, L.~M. and {Diaferio}, A.},
        title = "{Systematics in the X-ray cluster mass estimators}",
      journal = {\mnras},
         year = 2006,
        month = jul,
       volume = {369},
       number = {4},
        pages = {2013-2024},
          doi = {10.1111/j.1365-2966.2006.10466.x},
archivePrefix = {arXiv},
       eprint = {astro-ph/0602434},
 primaryClass = {astro-ph},
       adsurl = {https://ui.adsabs.harvard.edu/abs/2006MNRAS.369.2013R}
}

@ARTICLE{Kawahara2007ApJ...659..257K,
       author = {{Kawahara}, Hajime and {Suto}, Yasushi and {Kitayama}, Tetsu and {Sasaki}, Shin and {Shimizu}, Mamoru and {Rasia}, Elena and {Dolag}, Klaus},
        title = "{Radial Profile and Lognormal Fluctuations of the Intracluster Medium as the Origin of Systematic Bias in Spectroscopic Temperature}",
      journal = {\apj},
         year = 2007,
        month = apr,
       volume = {659},
       number = {1},
        pages = {257-266},
          doi = {10.1086/512231},
archivePrefix = {arXiv},
       eprint = {astro-ph/0611018},
 primaryClass = {astro-ph},
       adsurl = {https://ui.adsabs.harvard.edu/abs/2007ApJ...659..257K}
}

@ARTICLE{Tornatore2007MNRAS.382.1050T,
       author = {{Tornatore}, L. and {Borgani}, S. and {Dolag}, K. and {Matteucci}, F.},
        title = "{Chemical enrichment of galaxy clusters from hydrodynamical simulations}",
      journal = {\mnras},
         year = 2007,
        month = dec,
       volume = {382},
       number = {3},
        pages = {1050-1072},
          doi = {10.1111/j.1365-2966.2007.12070.x},
archivePrefix = {arXiv},
       eprint = {0705.1921},
 primaryClass = {astro-ph},
       adsurl = {https://ui.adsabs.harvard.edu/abs/2007MNRAS.382.1050T}
}

@ARTICLE{Nagai2007ApJ_xrayCLUsims,
       author = {{Nagai}, Daisuke and {Vikhlinin}, Alexey and {Kravtsov}, Andrey V.},
        title = "{Testing X-Ray Measurements of Galaxy Clusters with Cosmological Simulations}",
      journal = {\apj},
         year = 2007,
        month = jan,
       volume = {655},
       number = {1},
        pages = {98-108},
          doi = {10.1086/509868},
archivePrefix = {arXiv},
       eprint = {astro-ph/0609247},
 primaryClass = {astro-ph},
       adsurl = {https://ui.adsabs.harvard.edu/abs/2007ApJ...655...98N}
}

@ARTICLE{Croston2008A&A_gasProfs_rexcess,
       author = {{Croston}, J.~H. and {Pratt}, G.~W. and {B{\"o}hringer}, H. and {Arnaud}, M. and {Pointecouteau}, E. and {Ponman}, T.~J. and {Sanderson}, A.~J.~R. and {Temple}, R.~F. and {Bower}, R.~G. and {Donahue}, M.},
        title = "{Galaxy-cluster gas-density distributions of the representative XMM-Newton cluster structure survey (REXCESS)}",
      journal = {\aap},
         year = 2008,
        month = aug,
       volume = {487},
       number = {2},
        pages = {431-443},
          doi = {10.1051/0004-6361:20079154},
archivePrefix = {arXiv},
       eprint = {0801.3430},
 primaryClass = {astro-ph},
       adsurl = {https://ui.adsabs.harvard.edu/abs/2008A&A...487..431C}
}

@ARTICLE{Rasia2008ApJ_xmas,
       author = {{Rasia}, E. and {Mazzotta}, P. and {Bourdin}, H. and {Borgani}, S. and {Tornatore}, L. and {Ettori}, S. and {Dolag}, K. and {Moscardini}, L.},
        title = "{X-MAS2: Study Systematics on the ICM Metallicity Measurements}",
      journal = {\apj},
         year = 2008,
        month = feb,
       volume = {674},
       number = {2},
        pages = {728-741},
          doi = {10.1086/524345},
archivePrefix = {arXiv},
       eprint = {0707.2614},
 primaryClass = {astro-ph},
       adsurl = {https://ui.adsabs.harvard.edu/abs/2008ApJ...674..728R}
}

@ARTICLE{Schaye2008MNRAS_SFmodels,
       author = {{Schaye}, Joop and {Dalla Vecchia}, Claudio},
        title = "{On the relation between the Schmidt and Kennicutt-Schmidt star formation laws and its implications for numerical simulations}",
      journal = {\mnras},
         year = 2008,
        month = jan,
       volume = {383},
       number = {3},
        pages = {1210-1222},
          doi = {10.1111/j.1365-2966.2007.12639.x},
archivePrefix = {arXiv},
       eprint = {0709.0292},
 primaryClass = {astro-ph},
       adsurl = {https://ui.adsabs.harvard.edu/abs/2008MNRAS.383.1210S}
}

@ARTICLE{Asplund2009ARA&A..47..481A,
       author = {{Asplund}, Martin and {Grevesse}, Nicolas and {Sauval}, A. Jacques and {Scott}, Pat},
        title = "{The Chemical Composition of the Sun}",
      journal = {\araa},
         year = 2009,
        month = sep,
       volume = {47},
       number = {1},
        pages = {481-522},
          doi = {10.1146/annurev.astro.46.060407.145222},
archivePrefix = {arXiv},
       eprint = {0909.0948},
 primaryClass = {astro-ph.SR},
       adsurl = {https://ui.adsabs.harvard.edu/abs/2009ARA&A..47..481A}
}

@ARTICLE{Lau2009ApJ_hydrobias,
       author = {{Lau}, Erwin T. and {Kravtsov}, Andrey V. and {Nagai}, Daisuke},
        title = "{Residual Gas Motions in the Intracluster Medium and Bias in Hydrostatic Measurements of Mass Profiles of Clusters}",
      journal = {\apj},
         year = 2009,
        month = nov,
       volume = {705},
       number = {2},
        pages = {1129-1138},
          doi = {10.1088/0004-637X/705/2/1129},
archivePrefix = {arXiv},
       eprint = {0903.4895},
 primaryClass = {astro-ph.CO},
       adsurl = {https://ui.adsabs.harvard.edu/abs/2009ApJ...705.1129L}
}

@ARTICLE{Booth2009MNRAS_agnmodel,
       author = {{Booth}, C.~M. and {Schaye}, Joop},
        title = "{Cosmological simulations of the growth of supermassive black holes and feedback from active galactic nuclei: method and tests}",
      journal = {\mnras},
         year = 2009,
        month = sep,
       volume = {398},
       number = {1},
        pages = {53-74},
          doi = {10.1111/j.1365-2966.2009.15043.x},
archivePrefix = {arXiv},
       eprint = {0904.2572},
 primaryClass = {astro-ph.CO},
       adsurl = {https://ui.adsabs.harvard.edu/abs/2009MNRAS.398...53B}
}

@ARTICLE{Knollmann2009_AHF,
       author = {{Knollmann}, Steffen R. and {Knebe}, Alexander},
        title = "{AHF: Amiga's Halo Finder}",
      journal = {\apjs},
         year = 2009,
        month = jun,
       volume = {182},
       number = {2},
        pages = {608-624},
          doi = {10.1088/0067-0049/182/2/608},
archivePrefix = {arXiv},
       eprint = {0904.3662},
 primaryClass = {astro-ph.CO},
       adsurl = {https://ui.adsabs.harvard.edu/abs/2009ApJS..182..608K}
}

@ARTICLE{Dolag2009MNRAS_subhalomf,
       author = {{Dolag}, K. and {Borgani}, S. and {Murante}, G. and {Springel}, V.},
        title = "{Substructures in hydrodynamical cluster simulations}",
      journal = {\mnras},
         year = 2009,
        month = oct,
       volume = {399},
       number = {2},
        pages = {497-514},
          doi = {10.1111/j.1365-2966.2009.15034.x},
archivePrefix = {arXiv},
       eprint = {0808.3401},
 primaryClass = {astro-ph},
       adsurl = {https://ui.adsabs.harvard.edu/abs/2009MNRAS.399..497D}
}

@ARTICLE{Wiersma2009MNRAS.399..574W,
       author = {{Wiersma}, Robert P.~C. and {Schaye}, Joop and {Theuns}, Tom and {Dalla Vecchia}, Claudio and {Tornatore}, Luca},
        title = "{Chemical enrichment in cosmological, smoothed particle hydrodynamics simulations}",
      journal = {\mnras},
         year = 2009,
        month = oct,
       volume = {399},
       number = {2},
        pages = {574-600},
          doi = {10.1111/j.1365-2966.2009.15331.x},
archivePrefix = {arXiv},
       eprint = {0902.1535},
 primaryClass = {astro-ph.CO},
       adsurl = {https://ui.adsabs.harvard.edu/abs/2009MNRAS.399..574W}
}

@ARTICLE{Staniszewski2009ApJspt_clusters,
       author = {{Staniszewski}, Z. and {Ade}, P.~A.~R. and {Aird}, K.~A. and {Benson}, B.~A. and {Bleem}, L.~E. and {Carlstrom}, J.~E. and {Chang}, C.~L. and {Cho}, H. -M. and {Crawford}, T.~M. and {Crites}, A.~T. and {de Haan}, T. and {Dobbs}, M.~A. and {Halverson}, N.~W. and {Holder}, G.~P. and {Holzapfel}, W.~L. and {Hrubes}, J.~D. and {Joy}, M. and {Keisler}, R. and {Lanting}, T.~M. and {Lee}, A.~T. and {Leitch}, E.~M. and {Loehr}, A. and {Lueker}, M. and {McMahon}, J.~J. and {Mehl}, J. and {Meyer}, S.~S. and {Mohr}, J.~J. and {Montroy}, T.~E. and {Ngeow}, C. -C. and {Padin}, S. and {Plagge}, T. and {Pryke}, C. and {Reichardt}, C.~L. and {Ruhl}, J.~E. and {Schaffer}, K.~K. and {Shaw}, L. and {Shirokoff}, E. and {Spieler}, H.~G. and {Stalder}, B. and {Stark}, A.~A. and {Vanderlinde}, K. and {Vieira}, J.~D. and {Zahn}, O. and {Zenteno}, A.},
        title = "{Galaxy Clusters Discovered with a Sunyaev-Zel'dovich Effect Survey}",
      journal = {\apj},
         year = 2009,
        month = aug,
       volume = {701},
       number = {1},
        pages = {32-41},
          doi = {10.1088/0004-637X/701/1/32},
archivePrefix = {arXiv},
       eprint = {0810.1578},
 primaryClass = {astro-ph},
       adsurl = {https://ui.adsabs.harvard.edu/abs/2009ApJ...701...32S}
}

@ARTICLE{Meneghetti2010A&Axraylenisng_clumass,
       author = {{Meneghetti}, M. and {Rasia}, E. and {Merten}, J. and {Bellagamba}, F. and {Ettori}, S. and {Mazzotta}, P. and {Dolag}, K. and {Marri}, S.},
        title = "{Weighing simulated galaxy clusters using lensing and X-ray}",
      journal = {\aap},
         year = 2010,
        month = may,
       volume = {514},
          eid = {A93},
        pages = {A93},
          doi = {10.1051/0004-6361/200913222},
archivePrefix = {arXiv},
       eprint = {0912.1343},
 primaryClass = {astro-ph.CO},
       adsurl = {https://ui.adsabs.harvard.edu/abs/2010A&A...514A..93M}
}

@ARTICLE{Fabjan2010MNRAS_AGNfeedback,
       author = {{Fabjan}, D. and {Borgani}, S. and {Tornatore}, L. and {Saro}, A. and {Murante}, G. and {Dolag}, K.},
        title = "{Simulating the effect of active galactic nuclei feedback on the metal enrichment of galaxy clusters}",
      journal = {\mnras},
         year = 2010,
        month = jan,
       volume = {401},
       number = {3},
        pages = {1670-1690},
          doi = {10.1111/j.1365-2966.2009.15794.x},
archivePrefix = {arXiv},
       eprint = {0909.0664},
 primaryClass = {astro-ph.CO},
       adsurl = {https://ui.adsabs.harvard.edu/abs/2010MNRAS.401.1670F}
}

@ARTICLE{Schaye2010MNRAS_owls,
       author = {{Schaye}, Joop and {Dalla Vecchia}, Claudio and {Booth}, C.~M. and {Wiersma}, Robert P.~C. and {Theuns}, Tom and {Haas}, Marcel R. and {Bertone}, Serena and {Duffy}, Alan R. and {McCarthy}, I.~G. and {van de Voort}, Freeke},
        title = "{The physics driving the cosmic star formation history}",
      journal = {\mnras},
         year = 2010,
        month = mar,
       volume = {402},
       number = {3},
        pages = {1536-1560},
          doi = {10.1111/j.1365-2966.2009.16029.x},
archivePrefix = {arXiv},
       eprint = {0909.5196},
 primaryClass = {astro-ph.CO},
       adsurl = {https://ui.adsabs.harvard.edu/abs/2010MNRAS.402.1536S}
}

@ARTICLE{Meneghetti2011A&A_cluobssim,
       author = {{Meneghetti}, M. and {Fedeli}, C. and {Zitrin}, A. and {Bartelmann}, M. and {Broadhurst}, T. and {Gottl{\"o}ber}, S. and {Moscardini}, L. and {Yepes}, G.},
        title = "{Comparison of an X-ray-selected sample of massive lensing clusters with the MareNostrum Universe {\ensuremath{\Lambda}}CDM simulation}",
      journal = {\aap},
         year = 2011,
        month = jun,
       volume = {530},
          eid = {A17},
        pages = {A17},
          doi = {10.1051/0004-6361/201016040},
archivePrefix = {arXiv},
       eprint = {1103.0044},
 primaryClass = {astro-ph.CO},
       adsurl = {https://ui.adsabs.harvard.edu/abs/2011A&A...530A..17M}
}

@ARTICLE{Komatsu2011ApJS..192...18K,
       author = {{Komatsu}, E. and {Smith}, K.~M. and {Dunkley}, J. and {Bennett}, C.~L. and {Gold}, B. and {Hinshaw}, G. and {Jarosik}, N. and {Larson}, D. and {Nolta}, M.~R. and {Page}, L. and {Spergel}, D.~N. and {Halpern}, M. and {Hill}, R.~S. and {Kogut}, A. and {Limon}, M. and {Meyer}, S.~S. and {Odegard}, N. and {Tucker}, G.~S. and {Weiland}, J.~L. and {Wollack}, E. and {Wright}, E.~L.},
        title = "{Seven-year Wilkinson Microwave Anisotropy Probe (WMAP) Observations: Cosmological Interpretation}",
      journal = {\apjs},
         year = 2011,
        month = feb,
       volume = {192},
       number = {2},
          eid = {18},
        pages = {18},
          doi = {10.1088/0067-0049/192/2/18},
archivePrefix = {arXiv},
       eprint = {1001.4538},
 primaryClass = {astro-ph.CO},
       adsurl = {https://ui.adsabs.harvard.edu/abs/2011ApJS..192...18K}
}

@ARTICLE{Biffi2012MNRASphox,
       author = {{Biffi}, V. and {Dolag}, K. and {B{\"o}hringer}, H. and {Lemson}, G.},
        title = "{Observing simulated galaxy clusters with PHOX: a novel X-ray photon simulator}",
      journal = {\mnras},
         year = 2012,
        month = mar,
       volume = {420},
       number = {4},
        pages = {3545-3556},
          doi = {10.1111/j.1365-2966.2011.20278.x},
archivePrefix = {arXiv},
       eprint = {1112.0314},
 primaryClass = {astro-ph.CO},
       adsurl = {https://ui.adsabs.harvard.edu/abs/2012MNRAS.420.3545B}
}

@ARTICLE{Lehmer2012ApJ_logNlogSChandra,
       author = {{Lehmer}, B.~D. and {Xue}, Y.~Q. and {Brandt}, W.~N. and {Alexander}, D.~M. and {Bauer}, F.~E. and {Brusa}, M. and {Comastri}, A. and {Gilli}, R. and {Hornschemeier}, A.~E. and {Luo}, B. and {Paolillo}, M. and {Ptak}, A. and {Shemmer}, O. and {Schneider}, D.~P. and {Tozzi}, P. and {Vignali}, C.},
        title = "{The 4 Ms Chandra Deep Field-South Number Counts Apportioned by Source Class: Pervasive Active Galactic Nuclei and the Ascent of Normal Galaxies}",
      journal = {\apj},
         year = 2012,
        month = jun,
       volume = {752},
       number = {1},
          eid = {46},
        pages = {46},
          doi = {10.1088/0004-637X/752/1/46},
archivePrefix = {arXiv},
       eprint = {1204.1977},
 primaryClass = {astro-ph.CO},
       adsurl = {https://ui.adsabs.harvard.edu/abs/2012ApJ...752...46L}
}

@ARTICLE{Kravtsov2012ARA&ABorgani,
       author = {{Kravtsov}, Andrey V. and {Borgani}, Stefano},
        title = "{Formation of Galaxy Clusters}",
      journal = {\araa},
         year = 2012,
        month = sep,
       volume = {50},
        pages = {353-409},
          doi = {10.1146/annurev-astro-081811-125502},
archivePrefix = {arXiv},
       eprint = {1205.5556},
 primaryClass = {astro-ph.CO},
       adsurl = {https://ui.adsabs.harvard.edu/abs/2012ARA&A..50..353K}
}

@ARTICLE{Rasia2012NJPh_xraylensing,
       author = {{Rasia}, E. and {Meneghetti}, M. and {Martino}, R. and {Borgani}, S. and {Bonafede}, A. and {Dolag}, K. and {Ettori}, S. and {Fabjan}, D. and {Giocoli}, C. and {Mazzotta}, P. and {Merten}, J. and {Radovich}, M. and {Tornatore}, L.},
        title = "{Lensing and x-ray mass estimates of clusters (simulations)}",
      journal = {New Journal of Physics},
         year = 2012,
        month = may,
       volume = {14},
       number = {5},
          eid = {055018},
        pages = {055018},
          doi = {10.1088/1367-2630/14/5/055018},
archivePrefix = {arXiv},
       eprint = {1201.1569},
 primaryClass = {astro-ph.CO},
       adsurl = {https://ui.adsabs.harvard.edu/abs/2012NJPh...14e5018R}
}

@ARTICLE{Biffi2013MNRAS,
       author = {{Biffi}, V. and {Dolag}, K. and {B{\"o}hringer}, H.},
        title = "{Investigating the velocity structure and X-ray observable properties of simulated galaxy clusters with PHOX}",
      journal = {\mnras},
         year = 2013,
        month = jan,
       volume = {428},
       number = {2},
        pages = {1395-1409},
          doi = {10.1093/mnras/sts120},
archivePrefix = {arXiv},
       eprint = {1210.4158},
 primaryClass = {astro-ph.CO},
       adsurl = {https://ui.adsabs.harvard.edu/abs/2013MNRAS.428.1395B}
}

@ARTICLE{Behroozi2013,
   author = {{Behroozi}, P.S. and {Wechsler}, R.H. and {Wu}, H.-Y.},
    title = "{The ROCKSTAR Phase-space Temporal Halo Finder and the Velocity Offsets of Cluster Cores}",
  journal = {\apj},
archivePrefix = "arXiv",
   eprint = {1110.4372},
 primaryClass = "astro-ph.CO",
     year = 2013,
    month = jan,
   volume = 762,
      eid = {109},
    pages = {109},
      doi = {10.1088/0004-637X/762/2/109},
   adsurl = {http://adsabs.harvard.edu/abs/2013ApJ...762..109B}
}

@ARTICLE{Battaglia2013ApJ_fgassims,
       author = {{Battaglia}, N. and {Bond}, J.~R. and {Pfrommer}, C. and {Sievers}, J.~L.},
        title = "{On the Cluster Physics of Sunyaev-Zel'dovich and X-Ray Surveys. III. Measurement Biases and Cosmological Evolution of Gas and Stellar Mass Fractions}",
      journal = {\apj},
         year = 2013,
        month = nov,
       volume = {777},
       number = {2},
          eid = {123},
        pages = {123},
          doi = {10.1088/0004-637X/777/2/123},
archivePrefix = {arXiv},
       eprint = {1209.4082},
 primaryClass = {astro-ph.CO},
       adsurl = {https://ui.adsabs.harvard.edu/abs/2013ApJ...777..123B}
}

@ARTICLE{Rasia2014ApJ_Tstruc,
       author = {{Rasia}, Elena and {Lau}, Erwin T. and {Borgani}, Stefano and {Nagai}, Daisuke and {Dolag}, Klaus and {Avestruz}, Camille and {Granato}, Gian Luigi and {Mazzotta}, Pasquale and {Murante}, Giuseppe and {Nelson}, Kaylea and {Ragone-Figueroa}, Cinthia},
        title = "{Temperature Structure of the Intracluster Medium from Smoothed-particle Hydrodynamics and Adaptive-mesh Refinement Simulations}",
      journal = {\apj},
         year = 2014,
        month = aug,
       volume = {791},
       number = {2},
          eid = {96},
        pages = {96},
          doi = {10.1088/0004-637X/791/2/96},
archivePrefix = {arXiv},
       eprint = {1406.4410},
 primaryClass = {astro-ph.CO},
       adsurl = {https://ui.adsabs.harvard.edu/abs/2014ApJ...791...96R}
}

@ARTICLE{Ueda2014ApJ_AGN,
       author = {{Ueda}, Yoshihiro and {Akiyama}, Masayuki and {Hasinger}, G{\"u}nther and {Miyaji}, Takamitsu and {Watson}, Michael G.},
        title = "{Toward the Standard Population Synthesis Model of the X-Ray Background: Evolution of X-Ray Luminosity and Absorption Functions of Active Galactic Nuclei Including Compton-thick Populations}",
      journal = {\apj},
         year = 2014,
        month = may,
       volume = {786},
       number = {2},
          eid = {104},
        pages = {104},
          doi = {10.1088/0004-637X/786/2/104},
archivePrefix = {arXiv},
       eprint = {1402.1836},
 primaryClass = {astro-ph.CO},
       adsurl = {https://ui.adsabs.harvard.edu/abs/2014ApJ...786..104U}
}

@ARTICLE{Nelson2014ApJ_hydrob,
       author = {{Nelson}, Kaylea and {Lau}, Erwin T. and {Nagai}, Daisuke and {Rudd}, Douglas H. and {Yu}, Liang},
        title = "{Weighing Galaxy Clusters with Gas. II. On the Origin of Hydrostatic Mass Bias in {\ensuremath{\Lambda}}CDM Galaxy Clusters}",
      journal = {\apj},
         year = 2014,
        month = feb,
       volume = {782},
       number = {2},
          eid = {107},
        pages = {107},
          doi = {10.1088/0004-637X/782/2/107},
archivePrefix = {arXiv},
       eprint = {1308.6589},
 primaryClass = {astro-ph.CO},
       adsurl = {https://ui.adsabs.harvard.edu/abs/2014ApJ...782..107N}
}

@ARTICLE{Hirschmann2014MNRAS.442.2304H,
       author = {{Hirschmann}, Michaela and {Dolag}, Klaus and {Saro}, Alexandro and {Bachmann}, Lisa and {Borgani}, Stefano and {Burkert}, Andreas},
        title = "{Cosmological simulations of black hole growth: AGN luminosities and downsizing}",
      journal = {\mnras},
         year = 2014,
        month = aug,
       volume = {442},
       number = {3},
        pages = {2304-2324},
          doi = {10.1093/mnras/stu1023},
archivePrefix = {arXiv},
       eprint = {1308.0333},
 primaryClass = {astro-ph.CO},
       adsurl = {https://ui.adsabs.harvard.edu/abs/2014MNRAS.442.2304H}
}

@ARTICLE{Planck2014A&A...571A..20P,
       author = {{Planck Collaboration} and {Ade}, P.~A.~R. and {Aghanim}, N. and {Armitage-Caplan}, C. and {Arnaud}, M. and {Ashdown}, M. and {Atrio-Barandela}, F. and {Aumont}, J. and {Baccigalupi}, C. and {Banday}, A.~J. and et al.},
        title = "{Planck 2013 results. XX. Cosmology from Sunyaev-Zeldovich cluster counts}",
      journal = {\aap},
         year = 2014,
        month = nov,
       volume = {571},
          eid = {A20},
        pages = {A20},
          doi = {10.1051/0004-6361/201321521},
archivePrefix = {arXiv},
       eprint = {1303.5080},
 primaryClass = {astro-ph.CO},
       adsurl = {https://ui.adsabs.harvard.edu/abs/2014A&A...571A..20P}
}

@ARTICLE{Rykoff2014redmapper,
       author = {{Rykoff}, E.~S. and {Rozo}, E. and {Busha}, M.~T. and {Cunha}, C.~E. and
         {Finoguenov}, A. and {Evrard}, A. and {Hao}, J. and {Koester}, B.~P. and
         {Leauthaud}, A. and {Nord}, B. and {Pierre}, M. and {Reddick}, R. and
         {Sadibekova}, T. and {Sheldon}, E.~S. and {Wechsler}, R.~H.},
        title = "{redMaPPer. I. Algorithm and SDSS DR8 Catalog}",
      journal = {\apj},
         year = 2014,
        month = apr,
       volume = {785},
       number = {2},
          eid = {104},
        pages = {104},
          doi = {10.1088/0004-637X/785/2/104},
archivePrefix = {arXiv},
       eprint = {1303.3562},
 primaryClass = {astro-ph.CO},
       adsurl = {https://ui.adsabs.harvard.edu/abs/2014ApJ...785..104R}
}

@ARTICLE{Planck2014,
   author = {{Planck Collaboration} and {Ade}, P.~A.~R. and {Aghanim}, N. and 
	{Armitage-Caplan}, C. and {Arnaud}, M. and {Ashdown}, M. and 
	{Atrio-Barandela}, F. and {Aumont}, J. and {Baccigalupi}, C. and 
	{Banday}, A.~J. and et al.},
    title = "{Planck 2013 results. XVI. Cosmological parameters}",
  journal = {\aap},
archivePrefix = "arXiv",
   eprint = {1303.5076},
     year = 2014,
    month = nov,
   volume = 571,
      eid = {A16},
    pages = {A16},
      doi = {10.1051/0004-6361/201321591},
   adsurl = {http://adsabs.harvard.edu/abs/2014A%26A...571A..16P}
}

@ARTICLE{Rasia2015ApJCCNCCsims,
       author = {{Rasia}, E. and {Borgani}, S. and {Murante}, G. and {Planelles}, S. and {Beck}, A.~M. and {Biffi}, V. and {Ragone-Figueroa}, C. and {Granato}, G.~L. and {Steinborn}, L.~K. and {Dolag}, K.},
        title = "{Cool Core Clusters from Cosmological Simulations}",
      journal = {\apjl},
         year = 2015,
        month = nov,
       volume = {813},
       number = {1},
          eid = {L17},
        pages = {L17},
          doi = {10.1088/2041-8205/813/1/L17},
archivePrefix = {arXiv},
       eprint = {1509.04247},
 primaryClass = {astro-ph.CO},
       adsurl = {https://ui.adsabs.harvard.edu/abs/2015ApJ...813L..17R}
}

@ARTICLE{Steinborn2015MNRAS.448.1504S,
       author = {{Steinborn}, Lisa K. and {Dolag}, Klaus and {Hirschmann}, Michaela and {Prieto}, M. Almudena and {Remus}, Rhea-Silvia},
        title = "{A refined sub-grid model for black hole accretion and AGN feedback in large cosmological simulations}",
      journal = {\mnras},
         year = 2015,
        month = apr,
       volume = {448},
       number = {2},
        pages = {1504-1525},
          doi = {10.1093/mnras/stv072},
archivePrefix = {arXiv},
       eprint = {1409.3221},
 primaryClass = {astro-ph.GA},
       adsurl = {https://ui.adsabs.harvard.edu/abs/2015MNRAS.448.1504S}
}

@ARTICLE{Teklu2015ApJ...812...29T,
       author = {{Teklu}, Adelheid F. and {Remus}, Rhea-Silvia and {Dolag}, Klaus and {Beck}, Alexander M. and {Burkert}, Andreas and {Schmidt}, Andreas S. and {Schulze}, Felix and {Steinborn}, Lisa K.},
        title = "{Connecting Angular Momentum and Galactic Dynamics: The Complex Interplay between Spin, Mass, and Morphology}",
      journal = {\apj},
         year = 2015,
        month = oct,
       volume = {812},
       number = {1},
          eid = {29},
        pages = {29},
          doi = {10.1088/0004-637X/812/1/29},
archivePrefix = {arXiv},
       eprint = {1503.03501},
 primaryClass = {astro-ph.GA},
       adsurl = {https://ui.adsabs.harvard.edu/abs/2015ApJ...812...29T}
}

@INPROCEEDINGS{Dolag2015IAUGA..2250156D,
       author = {{Dolag}, Klaus},
        title = "{The Magneticum Simulations, from Galaxies to Galaxy Clusters}",
    booktitle = {IAU General Assembly},
         year = 2015,
       volume = {29},
        month = aug,
          eid = {2250156},
        pages = {2250156},
       adsurl = {https://ui.adsabs.harvard.edu/abs/2015IAUGA..2250156D}
}

@ARTICLE{Beck2016MNRAS_gadgetX,
       author = {{Beck}, A.~M. and {Murante}, G. and {Arth}, A. and {Remus}, R. -S. and {Teklu}, A.~F. and {Donnert}, J.~M.~F. and {Planelles}, S. and {Beck}, M.~C. and {F{\"o}rster}, P. and {Imgrund}, M. and {Dolag}, K. and {Borgani}, S.},
        title = "{An improved SPH scheme for cosmological simulations}",
      journal = {\mnras},
         year = 2016,
        month = jan,
       volume = {455},
       number = {2},
        pages = {2110-2130},
          doi = {10.1093/mnras/stv2443},
archivePrefix = {arXiv},
       eprint = {1502.07358},
 primaryClass = {astro-ph.CO},
       adsurl = {https://ui.adsabs.harvard.edu/abs/2016MNRAS.455.2110B}
}

@ARTICLE{Biffi2016HE,
       author = {{Biffi}, V. and {Borgani}, S. and {Murante}, G. and {Rasia}, E. and {Planelles}, S. and {Granato}, G.~L. and {Ragone-Figueroa}, C. and {Beck}, A.~M. and {Gaspari}, M. and {Dolag}, K.},
        title = "{On the Nature of Hydrostatic Equilibrium in Galaxy Clusters}",
      journal = {\apj},
         year = 2016,
        month = aug,
       volume = {827},
       number = {2},
          eid = {112},
        pages = {112},
          doi = {10.3847/0004-637X/827/2/112},
archivePrefix = {arXiv},
       eprint = {1606.02293},
 primaryClass = {astro-ph.CO},
       adsurl = {https://ui.adsabs.harvard.edu/abs/2016ApJ...827..112B}
}

@ARTICLE{Planck2016A&A..cosmopars,
       author = {{Planck Collaboration} and {Ade}, P.~A.~R. and {Aghanim}, N. and {Arnaud}, M. and {Ashdown}, M. and {Aumont}, J. and {Baccigalupi}, C. and {Banday}, A.~J. and {Barreiro}, R.~B. and {Bartlett}, J.~G. and et al.},
        title = "{Planck 2015 results. XIII. Cosmological parameters}",
      journal = {\aap},
         year = 2016,
        month = sep,
       volume = {594},
          eid = {A13},
        pages = {A13},
          doi = {10.1051/0004-6361/201525830},
archivePrefix = {arXiv},
       eprint = {1502.01589},
 primaryClass = {astro-ph.CO},
       adsurl = {https://ui.adsabs.harvard.edu/abs/2016A&A...594A..13P}
}

@ARTICLE{Klypin2016,
   author = {{Klypin}, A. and {Yepes}, G. and {Gottl{\"o}ber}, S. and {Prada}, F. and 
	{He{\ss}}, S.},
    title = "{MultiDark simulations: the story of dark matter halo concentrations and density profiles}",
  journal = {\mnras},
     year = 2016,
    month = apr,
   volume = 457,
    pages = {4340-4359},
      doi = {10.1093/mnras/stw248},
   adsurl = {http://adsabs.harvard.edu/abs/2016MNRAS.457.4340K}
}

@ARTICLE{Pierre2016XLL,
       author = {{Pierre}, M. and {Pacaud}, F. and {Adami}, C. and {Alis}, S. and
         {Altieri}, B. and {Baran}, N. and {Benoist}, C. and {Birkinshaw}, M. and
         {Bongiorno}, A. and {Bremer}, M.~N. and {Brusa}, M. and {Butler}, A. and
         {Ciliegi}, P. and {Chiappetti}, L. and {Clerc}, N. and
         {Corasaniti}, P.~S. and {Coupon}, J. and {De Breuck}, C. and
         {Democles}, J. and {Desai}, S. and {Delhaize}, J. and {Devriendt}, J. and
         {Dubois}, Y. and {Eckert}, D. and {Elyiv}, A. and {Ettori}, S. and
         {Evrard}, A. and {Faccioli}, L. and {Farahi}, A. and {Ferrari}, C. and
         {Finet}, F. and {Fotopoulou}, S. and {Fourmanoit}, N. and {Gandhi}, P. and
         {Gastaldello}, F. and {Gastaud}, R. and {Georgantopoulos}, I. and
         {Giles}, P. and {Guennou}, L. and {Guglielmo}, V. and {Horellou}, C. and
         {Husband}, K. and {Huynh}, M. and {Iovino}, A. and {Kilbinger}, M. and
         {Koulouridis}, E. and {Lavoie}, S. and {Le Brun}, A.~M.~C. and
         {Le Fevre}, J.~P. and {Lidman}, C. and {Lieu}, M. and {Lin}, C.~A. and
         {Mantz}, A. and {Maughan}, B.~J. and {Maurogordato}, S. and
         {McCarthy}, I.~G. and {McGee}, S. and {Melin}, J.~B. and {Melnyk}, O. and
         {Menanteau}, F. and {Novak}, M. and {Paltani}, S. and {Plionis}, M. and
         {Poggianti}, B.~M. and {Pomarede}, D. and {Pompei}, E. and
         {Ponman}, T.~J. and {Ramos-Ceja}, M.~E. and {Ranalli}, P. and
         {Rapetti}, D. and {Raychaudury}, S. and {Reiprich}, T.~H. and
         {Rottgering}, H. and {Rozo}, E. and {Rykoff}, E. and {Sadibekova}, T. and
         {Santos}, J. and {Sauvageot}, J.~L. and {Schimd}, C. and {Sereno}, M. and
         {Smith}, G.~P. and {Smol{\v{c}}i{\'c}}, V. and {Snowden}, S. and
         {Spergel}, D. and {Stanford}, S. and {Surdej}, J. and {Valageas}, P. and
         {Valotti}, A. and {Valtchanov}, I. and {Vignali}, C. and {Willis}, J. and
         {Ziparo}, F.},
        title = "{The XXL Survey. I. Scientific motivations - XMM-Newton observing plan - Follow-up observations and simulation programme}",
      journal = {\aap},
         year = 2016,
        month = jun,
       volume = {592},
          eid = {A1},
        pages = {A1},
          doi = {10.1051/0004-6361/201526766},
archivePrefix = {arXiv},
       eprint = {1512.04317},
 primaryClass = {astro-ph.CO},
       adsurl = {https://ui.adsabs.harvard.edu/abs/2016A&A...592A...1P}
}

@ARTICLE{McCarthy2017MNRAS_bahamas,
       author = {{McCarthy}, Ian G. and {Schaye}, Joop and {Bird}, Simeon and {Le Brun}, Amandine M.~C.},
        title = "{The BAHAMAS project: calibrated hydrodynamical simulations for large-scale structure cosmology}",
      journal = {\mnras},
         year = 2017,
        month = mar,
       volume = {465},
       number = {3},
        pages = {2936-2965},
          doi = {10.1093/mnras/stw2792},
archivePrefix = {arXiv},
       eprint = {1603.02702},
 primaryClass = {astro-ph.CO},
       adsurl = {https://ui.adsabs.harvard.edu/abs/2017MNRAS.465.2936M}
}

@ARTICLE{Dolag2017_Magneticum_metals,
       author = {{Dolag}, Klaus and {Mevius}, Emilio and {Remus}, Rhea-Silvia},
        title = "{Distribution and Evolution of Metals in the Magneticum Simulations}",
      journal = {Galaxies},
         year = 2017,
        month = aug,
       volume = {5},
       number = {3},
        pages = {35},
          doi = {10.3390/galaxies5030035},
archivePrefix = {arXiv},
       eprint = {1708.00027},
 primaryClass = {astro-ph.GA},
       adsurl = {https://ui.adsabs.harvard.edu/abs/2017Galax...5...35D}
}

@ARTICLE{Remus2017MNRAS_Magneticum,
       author = {{Remus}, Rhea-Silvia and {Dolag}, Klaus and {Naab}, Thorsten and {Burkert}, Andreas and {Hirschmann}, Michaela and {Hoffmann}, Tadziu L. and {Johansson}, Peter H.},
        title = "{The co-evolution of total density profiles and central dark matter fractions in simulated early-type galaxies}",
      journal = {\mnras},
         year = 2017,
        month = jan,
       volume = {464},
       number = {3},
        pages = {3742-3756},
          doi = {10.1093/mnras/stw2594},
archivePrefix = {arXiv},
       eprint = {1603.01619},
 primaryClass = {astro-ph.GA},
       adsurl = {https://ui.adsabs.harvard.edu/abs/2017MNRAS.464.3742R}
}

@ARTICLE{Gupta2017_pressureprof_magneticum,
       author = {{Gupta}, N. and {Saro}, A. and {Mohr}, J.~J. and {Dolag}, K. and {Liu}, J.},
        title = "{SZE observables, pressure profiles and centre offsets in Magneticum simulation galaxy clusters}",
      journal = {\mnras},
         year = 2017,
        month = aug,
       volume = {469},
       number = {3},
        pages = {3069-3087},
          doi = {10.1093/mnras/stx715},
archivePrefix = {arXiv},
       eprint = {1612.05266},
 primaryClass = {astro-ph.CO},
       adsurl = {https://ui.adsabs.harvard.edu/abs/2017MNRAS.469.3069G}
}

@ARTICLE{Henson2017MNRAS.465.3361H,
       author = {{Henson}, Monique A. and {Barnes}, David J. and {Kay}, Scott T. and
         {McCarthy}, Ian G. and {Schaye}, Joop},
        title = "{The impact of baryons on massive galaxy clusters: halo structure and cluster mass estimates}",
      journal = {\mnras},
         year = 2017,
        month = mar,
       volume = {465},
       number = {3},
        pages = {3361-3378},
          doi = {10.1093/mnras/stw2899},
archivePrefix = {arXiv},
       eprint = {1607.08550},
 primaryClass = {astro-ph.CO},
       adsurl = {https://ui.adsabs.harvard.edu/abs/2017MNRAS.465.3361H}
}

@ARTICLE{Barnes2017MNRAS_macsis,
       author = {{Barnes}, David J. and {Kay}, Scott T. and {Henson}, Monique A. and {McCarthy}, Ian G. and {Schaye}, Joop and {Jenkins}, Adrian},
        title = "{The redshift evolution of massive galaxy clusters in the MACSIS simulations}",
      journal = {\mnras},
         year = 2017,
        month = feb,
       volume = {465},
       number = {1},
        pages = {213-233},
          doi = {10.1093/mnras/stw2722},
archivePrefix = {arXiv},
       eprint = {1607.04569},
 primaryClass = {astro-ph.CO},
       adsurl = {https://ui.adsabs.harvard.edu/abs/2017MNRAS.465..213B}
}

@ARTICLE{Ruppin2018A&A_NIKA2,
       author = {{Ruppin}, F. and {Mayet}, F. and {Pratt}, G.~W. and {Adam}, R. and {Ade}, P. and {Andr{\'e}}, P. and {Arnaud}, M. and {Aussel}, H. and {Bartalucci}, I. and {Beelen}, A. and {Beno{\^\i}t}, A. and {Bideaud}, A. and {Bourrion}, O. and {Calvo}, M. and {Catalano}, A. and {Comis}, B. and {De Petris}, M. and {D{\'e}sert}, F.-X. and {Doyle}, S. and {Driessen}, E.~F.~C. and {Goupy}, J. and {Kramer}, C. and {Lagache}, G. and {Leclercq}, S. and {Lestrade}, J.-F. and {Mac{\'\i}as-P{\'e}rez}, J.~F. and {Mauskopf}, P. and {Monfardini}, A. and {Perotto}, L. and {Pisano}, G. and {Pointecouteau}, E. and {Ponthieu}, N. and {Rev{\'e}ret}, V. and {Ritacco}, A. and {Romero}, C. and {Roussel}, H. and {Schuster}, K. and {Sievers}, A. and {Tucker}, C. and {Zylka}, R.},
        title = "{First Sunyaev-Zel'dovich mapping with the NIKA2 camera: Implication of cluster substructures for the pressure profile and mass estimate}",
      journal = {\aap},
         year = 2018,
        month = jul,
       volume = {615},
          eid = {A112},
        pages = {A112},
          doi = {10.1051/0004-6361/201732558},
archivePrefix = {arXiv},
       eprint = {1712.09587},
 primaryClass = {astro-ph.CO},
       adsurl = {https://ui.adsabs.harvard.edu/abs/2018A&A...615A.112R}
}

@ARTICLE{Cui2018MNRAS_The300,
       author = {{Cui}, Weiguang and {Knebe}, Alexander and {Yepes}, Gustavo and {Pearce}, Frazer and {Power}, Chris and {Dave}, Romeel and {Arth}, Alexander and {Borgani}, Stefano and {Dolag}, Klaus and {Elahi}, Pascal and {Mostoghiu}, Robert and {Murante}, Giuseppe and {Rasia}, Elena and {Stoppacher}, Doris and {Vega-Ferrero}, Jesus and {Wang}, Yang and {Yang}, Xiaohu and {Benson}, Andrew and {Cora}, Sof{\'\i}a A. and {Croton}, Darren J. and {Sinha}, Manodeep and {Stevens}, Adam R.~H. and {Vega-Mart{\'\i}nez}, Cristian A. and {Arthur}, Jake and {Baldi}, Anna S. and {Ca{\~n}as}, Rodrigo and {Cialone}, Giammarco and {Cunnama}, Daniel and {De Petris}, Marco and {Durando}, Giacomo and {Ettori}, Stefano and {Gottl{\"o}ber}, Stefan and {Nuza}, Sebasti{\'a}n E. and {Old}, Lyndsay J. and {Pilipenko}, Sergey and {Sorce}, Jenny G. and {Welker}, Charlotte},
        title = "{The Three Hundred project: a large catalogue of theoretically modelled galaxy clusters for cosmological and astrophysical applications}",
      journal = {\mnras},
         year = 2018,
        month = nov,
       volume = {480},
       number = {3},
        pages = {2898-2915},
          doi = {10.1093/mnras/sty2111},
archivePrefix = {arXiv},
       eprint = {1809.04622},
 primaryClass = {astro-ph.GA},
       adsurl = {https://ui.adsabs.harvard.edu/abs/2018MNRAS.480.2898C}
}

@ARTICLE{Miyazaki2018PASJ_WLclu,
       author = {{Miyazaki}, Satoshi and {Oguri}, Masamune and {Hamana}, Takashi and {Shirasaki}, Masato and {Koike}, Michitaro and {Komiyama}, Yutaka and {Umetsu}, Keiichi and {Utsumi}, Yousuke and {Okabe}, Nobuhiro and {More}, Surhud and {Medezinski}, Elinor and {Lin}, Yen-Ting and {Miyatake}, Hironao and {Murayama}, Hitoshi and {Ota}, Naomi and {Mitsuishi}, Ikuyuki},
        title = "{A large sample of shear-selected clusters from the Hyper Suprime-Cam Subaru Strategic Program S16A Wide field mass maps}",
      journal = {\pasj},
         year = 2018,
        month = jan,
       volume = {70},
          eid = {S27},
        pages = {S27},
          doi = {10.1093/pasj/psx120},
archivePrefix = {arXiv},
       eprint = {1802.10290},
 primaryClass = {astro-ph.CO},
       adsurl = {https://ui.adsabs.harvard.edu/abs/2018PASJ...70S..27M}
}

@ARTICLE{Ghirardini2019A&A...621A..41G,
       author = {{Ghirardini}, V. and {Eckert}, D. and {Ettori}, S. and {Pointecouteau}, E. and {Molendi}, S. and {Gaspari}, M. and {Rossetti}, M. and {De Grandi}, S. and {Roncarelli}, M. and {Bourdin}, H. and {Mazzotta}, P. and {Rasia}, E. and {Vazza}, F.},
        title = "{Universal thermodynamic properties of the intracluster medium over two decades in radius in the X-COP sample}",
      journal = {\aap},
         year = 2019,
        month = jan,
       volume = {621},
          eid = {A41},
        pages = {A41},
          doi = {10.1051/0004-6361/201833325},
archivePrefix = {arXiv},
       eprint = {1805.00042},
 primaryClass = {astro-ph.CO},
       adsurl = {https://ui.adsabs.harvard.edu/abs/2019A&A...621A..41G}
}

@ARTICLE{Eckert2019xcop,
       author = {{Eckert}, D. and {Ghirardini}, V. and {Ettori}, S. and {Rasia}, E. and {Biffi}, V. and {Pointecouteau}, E. and {Rossetti}, M. and {Molendi}, S. and {Vazza}, F. and {Gastaldello}, F. and {Gaspari}, M. and {De Grandi}, S. and {Ghizzardi}, S. and {Bourdin}, H. and {Tchernin}, C. and {Roncarelli}, M.},
        title = "{Non-thermal pressure support in X-COP galaxy clusters}",
      journal = {\aap},
         year = 2019,
        month = jan,
       volume = {621},
          eid = {A40},
        pages = {A40},
          doi = {10.1051/0004-6361/201833324},
archivePrefix = {arXiv},
       eprint = {1805.00034},
 primaryClass = {astro-ph.CO},
       adsurl = {https://ui.adsabs.harvard.edu/abs/2019A&A...621A..40E}
}

@ARTICLE{Pratt2019SSRv..215...25P,
       author = {{Pratt}, G.~W. and {Arnaud}, M. and {Biviano}, A. and {Eckert}, D. and {Ettori}, S. and {Nagai}, D. and {Okabe}, N. and {Reiprich}, T.~H.},
        title = "{The Galaxy Cluster Mass Scale and Its Impact on Cosmological Constraints from the Cluster Population}",
      journal = {\ssr},
         year = 2019,
        month = feb,
       volume = {215},
       number = {2},
          eid = {25},
        pages = {25},
          doi = {10.1007/s11214-019-0591-0},
archivePrefix = {arXiv},
       eprint = {1902.10837},
 primaryClass = {astro-ph.CO},
       adsurl = {https://ui.adsabs.harvard.edu/abs/2019SSRv..215...25P}
}

@ARTICLE{Ansarifard2020A&A_300bhse,
       author = {{Ansarifard}, S. and {Rasia}, E. and {Biffi}, V. and {Borgani}, S. and {Cui}, W. and {De Petris}, M. and {Dolag}, K. and {Ettori}, S. and {Movahed}, S.~M.~S. and {Murante}, G. and {Yepes}, G.},
        title = "{The Three Hundred Project: Correcting for the hydrostatic-equilibrium mass bias in X-ray and SZ surveys}",
      journal = {\aap},
         year = 2020,
        month = feb,
       volume = {634},
          eid = {A113},
        pages = {A113},
          doi = {10.1051/0004-6361/201936742},
archivePrefix = {arXiv},
       eprint = {1911.07878},
 primaryClass = {astro-ph.CO},
       adsurl = {https://ui.adsabs.harvard.edu/abs/2020A&A...634A.113A}
}

@ARTICLE{Gaspari2020multiscale,
       author = {{Gaspari}, Massimo and {Tombesi}, Francesco and {Cappi}, Massimo},
        title = "{Linking macro-, meso- and microscales in multiphase AGN feeding and feedback}",
      journal = {Nature Astronomy},
         year = 2020,
        month = jan,
       volume = {4},
        pages = {10-13},
          doi = {10.1038/s41550-019-0970-1},
archivePrefix = {arXiv},
       eprint = {2001.04985},
 primaryClass = {astro-ph.GA},
       adsurl = {https://ui.adsabs.harvard.edu/abs/2020NatAs...4...10G}
}

@ARTICLE{Pearce2020MNRAS.491.1622P,
       author = {{Pearce}, Francesca A. and {Kay}, Scott T. and {Barnes}, David J. and {Bower}, Richard G. and {Schaller}, Matthieu},
        title = "{Hydrostatic mass estimates of massive galaxy clusters: a study with varying hydrodynamics flavours and non-thermal pressure support}",
      journal = {\mnras},
         year = 2020,
        month = jan,
       volume = {491},
       number = {2},
        pages = {1622-1642},
          doi = {10.1093/mnras/stz3003},
archivePrefix = {arXiv},
       eprint = {1910.10217},
 primaryClass = {astro-ph.CO},
       adsurl = {https://ui.adsabs.harvard.edu/abs/2020MNRAS.491.1622P}
}

@ARTICLE{Bleem2020ApJS_sptEXT,
       author = {{Bleem}, L.~E. and {Bocquet}, S. and {Stalder}, B. and {Gladders}, M.~D. and {Ade}, P.~A.~R. and {Allen}, S.~W. and {Anderson}, A.~J. and {Annis}, J. and {Ashby}, M.~L.~N. and {Austermann}, J.~E. and {Avila}, S. and {Avva}, J.~S. and {Bayliss}, M. and {Beall}, J.~A. and {Bechtol}, K. and {Bender}, A.~N. and {Benson}, B.~A. and {Bertin}, E. and {Bianchini}, F. and {Blake}, C. and {Brodwin}, M. and {Brooks}, D. and {Buckley-Geer}, E. and {Burke}, D.~L. and {Carlstrom}, J.~E. and {Rosell}, A. Carnero and {Carrasco Kind}, M. and {Carretero}, J. and {Chang}, C.~L. and {Chiang}, H.~C. and {Citron}, R. and {Moran}, C. Corbett and {Costanzi}, M. and {Crawford}, T.~M. and {Crites}, A.~T. and {da Costa}, L.~N. and {de Haan}, T. and {De Vicente}, J. and {Desai}, S. and {Diehl}, H.~T. and {Dietrich}, J.~P. and {Dobbs}, M.~A. and {Eifler}, T.~F. and {Everett}, W. and {Flaugher}, B. and {Floyd}, B. and {Frieman}, J. and {Gallicchio}, J. and {Garc{\'\i}a-Bellido}, J. and {George}, E.~M. and {Gerdes}, D.~W. and {Gilbert}, A. and {Gruen}, D. and {Gruendl}, R.~A. and {Gschwend}, J. and {Gupta}, N. and {Gutierrez}, G. and {Halverson}, N.~W. and {Harrington}, N. and {Henning}, J.~W. and {Heymans}, C. and {Holder}, G.~P. and {Hollowood}, D.~L. and {Holzapfel}, W.~L. and {Honscheid}, K. and {Hrubes}, J.~D. and {Huang}, N. and {Hubmayr}, J. and {Irwin}, K.~D. and {James}, D.~J. and {Jeltema}, T. and {Joudaki}, S. and {Khullar}, G. and {Klein}, M. and {Knox}, L. and {Kuropatkin}, N. and {Lee}, A.~T. and {Li}, D. and {Lidman}, C. and {Lowitz}, A. and {MacCrann}, N. and {Mahler}, G. and {Maia}, M.~A.~G. and {Marshall}, J.~L. and {McDonald}, M. and {McMahon}, J.~J. and {Melchior}, P. and {Menanteau}, F. and {Meyer}, S.~S. and {Miquel}, R. and {Mocanu}, L.~M. and {Mohr}, J.~J. and {Montgomery}, J. and {Nadolski}, A. and {Natoli}, T. and {Nibarger}, J.~P. and {Noble}, G. and {Novosad}, V. and {Padin}, S. and {Palmese}, A. and {Parkinson}, D. and {Patil}, S. and {Paz-Chinch{\'o}n}, F. and {Plazas}, A.~A. and {Pryke}, C. and {Ramachandra}, N.~S. and {Reichardt}, C.~L. and {Remolina Gonz{\'a}lez}, J.~D. and {Romer}, A.~K. and {Roodman}, A. and {Ruhl}, J.~E. and {Rykoff}, E.~S. and {Saliwanchik}, B.~R. and {Sanchez}, E. and {Saro}, A. and {Sayre}, J.~T. and {Schaffer}, K.~K. and {Schrabback}, T. and {Serrano}, S. and {Sharon}, K. and {Sievers}, C. and {Smecher}, G. and {Smith}, M. and {Soares-Santos}, M. and {Stark}, A.~A. and {Story}, K.~T. and {Suchyta}, E. and {Tarle}, G. and {Tucker}, C. and {Vanderlinde}, K. and {Veach}, T. and {Vieira}, J.~D. and {Wang}, G. and {Weller}, J. and {Whitehorn}, N. and {Wu}, W.~L.~K. and {Yefremenko}, V. and {Zhang}, Y.},
        title = "{The SPTpol Extended Cluster Survey}",
      journal = {\apjs},
         year = 2020,
        month = mar,
       volume = {247},
       number = {1},
          eid = {25},
        pages = {25},
          doi = {10.3847/1538-4365/ab6993},
archivePrefix = {arXiv},
       eprint = {1910.04121},
 primaryClass = {astro-ph.CO},
       adsurl = {https://ui.adsabs.harvard.edu/abs/2020ApJS..247...25B}
}

@ARTICLE{Virtanen2020SciPy-NMeth,
  author  = {Virtanen, Pauli and Gommers, Ralf and Oliphant, Travis E. and
            Haberland, Matt and Reddy, Tyler and Cournapeau, David and
            Burovski, Evgeni and Peterson, Pearu and Weckesser, Warren and
            Bright, Jonathan and {van der Walt}, St{\'e}fan J. and
            Brett, Matthew and Wilson, Joshua and Millman, K. Jarrod and
            Mayorov, Nikolay and Nelson, Andrew R. J. and Jones, Eric and
            Kern, Robert and Larson, Eric and Carey, C J and
            Polat, {\.I}lhan and Feng, Yu and Moore, Eric W. and
            {VanderPlas}, Jake and Laxalde, Denis and Perktold, Josef and
            Cimrman, Robert and Henriksen, Ian and Quintero, E. A. and
            Harris, Charles R. and Archibald, Anne M. and
            Ribeiro, Ant{\^o}nio H. and Pedregosa, Fabian and
            {van Mulbregt}, Paul and {SciPy 1.0 Contributors}},
  title   = {{{SciPy} 1.0: Fundamental Algorithms for Scientific
            Computing in Python}},
  journal = {Nature Methods},
  year    = {2020},
  volume  = {17},
  pages   = {261--272},
  adsurl  = {https://rdcu.be/b08Wh},
  doi     = {10.1038/s41592-019-0686-2},
}

@ARTICLE{Abbott2020DESY1_clusters,
       author = {{Abbott}, T.~M.~C. and {Aguena}, M. and {Alarcon}, A. and {Allam}, S. and {Allen}, S. and {Annis}, J. and {Avila}, S. and {Bacon}, D. and {Bechtol}, K. and {Bermeo}, A. and {Bernstein}, G.~M. and {Bertin}, E. and {Bhargava}, S. and {Bocquet}, S. and {Brooks}, D. and {Brout}, D. and {Buckley-Geer}, E. and {Burke}, D.~L. and {Carnero Rosell}, A. and {Carrasco Kind}, M. and {Carretero}, J. and {Castander}, F.~J. and {Cawthon}, R. and {Chang}, C. and {Chen}, X. and {Choi}, A. and {Costanzi}, M. and {Crocce}, M. and {da Costa}, L.~N. and {Davis}, T.~M. and {De Vicente}, J. and {DeRose}, J. and {Desai}, S. and {Diehl}, H.~T. and {Dietrich}, J.~P. and {Dodelson}, S. and {Doel}, P. and {Drlica-Wagner}, A. and {Eckert}, K. and {Eifler}, T.~F. and {Elvin-Poole}, J. and {Estrada}, J. and {Everett}, S. and {Evrard}, A.~E. and {Farahi}, A. and {Ferrero}, I. and {Flaugher}, B. and {Fosalba}, P. and {Frieman}, J. and {Garc{\'\i}a-Bellido}, J. and {Gatti}, M. and {Gaztanaga}, E. and {Gerdes}, D.~W. and {Giannantonio}, T. and {Giles}, P. and {Grandis}, S. and {Gruen}, D. and {Gruendl}, R.~A. and {Gschwend}, J. and {Gutierrez}, G. and {Hartley}, W.~G. and {Hinton}, S.~R. and {Hollowood}, D.~L. and {Honscheid}, K. and {Hoyle}, B. and {Huterer}, D. and {James}, D.~J. and {Jarvis}, M. and {Jeltema}, T. and {Johnson}, M.~W.~G. and {Johnson}, M.~D. and {Kent}, S. and {Krause}, E. and {Kron}, R. and {Kuehn}, K. and {Kuropatkin}, N. and {Lahav}, O. and {Li}, T.~S. and {Lidman}, C. and {Lima}, M. and {Lin}, H. and {MacCrann}, N. and {Maia}, M.~A.~G. and {Mantz}, A. and {Marshall}, J.~L. and {Martini}, P. and {Mayers}, J. and {Melchior}, P. and {Mena-Fern{\'a}ndez}, J. and {Menanteau}, F. and {Miquel}, R. and {Mohr}, J.~J. and {Nichol}, R.~C. and {Nord}, B. and {Ogando}, R.~L.~C. and {Palmese}, A. and {Paz-Chinch{\'o}n}, F. and {Plazas}, A.~A. and {Prat}, J. and {Rau}, M.~M. and {Romer}, A.~K. and {Roodman}, A. and {Rooney}, P. and {Rozo}, E. and {Rykoff}, E.~S. and {Sako}, M. and {Samuroff}, S. and {S{\'a}nchez}, C. and {Sanchez}, E. and {Saro}, A. and {Scarpine}, V. and {Schubnell}, M. and {Scolnic}, D. and {Serrano}, S. and {Sevilla-Noarbe}, I. and {Sheldon}, E. and {Smith}, J. Allyn. and {Smith}, M. and {Suchyta}, E. and {Swanson}, M.~E.~C. and {Tarle}, G. and {Thomas}, D. and {To}, C. and {Troxel}, M.~A. and {Tucker}, D.~L. and {Varga}, T.~N. and {von der Linden}, A. and {Walker}, A.~R. and {Wechsler}, R.~H. and {Weller}, J. and {Wilkinson}, R.~D. and {Wu}, H. and {Yanny}, B. and {Zhang}, Y. and {Zhang}, Z. and {Zuntz}, J. and {DES Collaboration}},
        title = "{Dark Energy Survey Year 1 Results: Cosmological constraints from cluster abundances and weak lensing}",
      journal = {\prd},
         year = 2020,
        month = jul,
       volume = {102},
       number = {2},
          eid = {023509},
        pages = {023509},
          doi = {10.1103/PhysRevD.102.023509},
archivePrefix = {arXiv},
       eprint = {2002.11124},
 primaryClass = {astro-ph.CO},
       adsurl = {https://ui.adsabs.harvard.edu/abs/2020PhRvD.102b3509A}
}

@ARTICLE{Eckert2020_pyproffit,
       author = {{Eckert}, Dominique and {Finoguenov}, Alexis and {Ghirardini}, Vittorio and {Grandis}, Sebastian and {Kaefer}, Florian and {Sanders}, Jeremy and {Ramos-Ceja}, Miriam},
        title = "{Low-scatter galaxy cluster mass proxies for the eROSITA all-sky survey}",
      journal = {The Open Journal of Astrophysics},
         year = 2020,
        month = sep,
       volume = {3},
          eid = {12},
        pages = {12},
          doi = {10.21105/astro.2009.13944},
archivePrefix = {arXiv},
       eprint = {2009.03944},
 primaryClass = {astro-ph.CO},
       adsurl = {https://ui.adsabs.harvard.edu/abs/2020OJAp....3E..12E}
}

@ARTICLE{Barnes2021MNRAS_bhe,
       author = {{Barnes}, David J. and {Vogelsberger}, Mark and {Pearce}, Francesca A. and {Pop}, Ana-Roxana and {Kannan}, Rahul and {Cao}, Kaili and {Kay}, Scott T. and {Hernquist}, Lars},
        title = "{Characterizing hydrostatic mass bias with MOCK-X}",
      journal = {\mnras},
         year = 2021,
        month = sep,
       volume = {506},
       number = {2},
        pages = {2533-2550},
          doi = {10.1093/mnras/stab1276},
archivePrefix = {arXiv},
       eprint = {2001.11508},
 primaryClass = {astro-ph.CO},
       adsurl = {https://ui.adsabs.harvard.edu/abs/2021MNRAS.506.2533B}
}

@ARTICLE{Deluca2021MNRAS_300MAPS,
       author = {{De Luca}, Federico and {De Petris}, Marco and {Yepes}, Gustavo and {Cui}, Weiguang and {Knebe}, Alexander and {Rasia}, Elena},
        title = "{The Three Hundred project: dynamical state of galaxy clusters and morphology from multiwavelength synthetic maps}",
      journal = {\mnras},
         year = 2021,
        month = jul,
       volume = {504},
       number = {4},
        pages = {5383-5400},
          doi = {10.1093/mnras/stab1073},
archivePrefix = {arXiv},
       eprint = {2011.09002},
 primaryClass = {astro-ph.CO},
       adsurl = {https://ui.adsabs.harvard.edu/abs/2021MNRAS.504.5383D}
}

@ARTICLE{Gianfagna2021MNRAS_hydrob,
       author = {{Gianfagna}, Giulia and {De Petris}, Marco and {Yepes}, Gustavo and {De Luca}, Federico and {Sembolini}, Federico and {Cui}, Weiguang and {Biffi}, Veronica and {K{\'e}ruzor{\'e}}, Florian and {Mac{\'\i}as-P{\'e}rez}, Juan and {Mayet}, Fr{\'e}d{\'e}ric and {Perotto}, Laurence and {Rasia}, Elena and {Ruppin}, Florian},
        title = "{Exploring the hydrostatic mass bias in MUSIC clusters: application to the NIKA2 mock sample}",
      journal = {\mnras},
         year = 2021,
        month = apr,
       volume = {502},
       number = {4},
        pages = {5115-5133},
          doi = {10.1093/mnras/stab308},
archivePrefix = {arXiv},
       eprint = {2010.03634},
 primaryClass = {astro-ph.CO},
       adsurl = {https://ui.adsabs.harvard.edu/abs/2021MNRAS.502.5115G}
}

@ARTICLE{CHEX-MATE2021A&A_intro,
       author = {{CHEX-MATE Collaboration} and {Arnaud}, M. and {Ettori}, S. and {Pratt}, G.~W. and {Rossetti}, M. and {Eckert}, D. and {Gastaldello}, F. and {Gavazzi}, R. and {Kay}, S.~T. and {Lovisari}, L. and {Maughan}, B.~J. and {Pointecouteau}, E. and {Sereno}, M. and {Bartalucci}, I. and {Bonafede}, A. and {Bourdin}, H. and {Cassano}, R. and {Duffy}, R.~T. and {Iqbal}, A. and {Maurogordato}, S. and {Rasia}, E. and {Sayers}, J. and {Andrade-Santos}, F. and {Aussel}, H. and {Barnes}, D.~J. and {Barrena}, R. and {Borgani}, S. and {Burkutean}, S. and {Clerc}, N. and {Corasaniti}, P. -S. and {Cuillandre}, J. -C. and {De Grandi}, S. and {De Petris}, M. and {Dolag}, K. and {Donahue}, M. and {Ferragamo}, A. and {Gaspari}, M. and {Ghizzardi}, S. and {Gitti}, M. and {Haines}, C.~P. and {Jauzac}, M. and {Johnston-Hollitt}, M. and {Jones}, C. and {K{\'e}ruzor{\'e}}, F. and {Le Brun}, A.~M.~C. and {Mayet}, F. and {Mazzotta}, P. and {Melin}, J. -B. and {Molendi}, S. and {Nonino}, M. and {Okabe}, N. and {Paltani}, S. and {Perotto}, L. and {Pires}, S. and {Radovich}, M. and {Rubino-Martin}, J. -A. and {Salvati}, L. and {Saro}, A. and {Sartoris}, B. and {Schellenberger}, G. and {Streblyanska}, A. and {Tarr{\'\i}o}, P. and {Tozzi}, P. and {Umetsu}, K. and {van der Burg}, R.~F.~J. and {Vazza}, F. and {Venturi}, T. and {Yepes}, G. and {Zarattini}, S.},
        title = "{The Cluster HEritage project with XMM-Newton: Mass Assembly and Thermodynamics at the Endpoint of structure formation. I. Programme overview}",
      journal = {\aap},
         year = 2021,
        month = jun,
       volume = {650},
          eid = {A104},
        pages = {A104},
          doi = {10.1051/0004-6361/202039632},
archivePrefix = {arXiv},
       eprint = {2010.11972},
 primaryClass = {astro-ph.CO},
       adsurl = {https://ui.adsabs.harvard.edu/abs/2021A&A...650A.104C}
}

@ARTICLE{Pratt2022A&A_density_profs,
       author = {{Pratt}, G.~W. and {Arnaud}, M. and {Maughan}, B.~J. and {Melin}, J.-B.},
        title = "{Linking a universal gas density profile to the core-excised X-ray luminosity in galaxy clusters up to z {\ensuremath{\sim}} 1.1}",
      journal = {\aap},
         year = 2022,
        month = sep,
       volume = {665},
          eid = {A24},
        pages = {A24},
          doi = {10.1051/0004-6361/202243074},
archivePrefix = {arXiv},
       eprint = {2206.06656},
 primaryClass = {astro-ph.CO},
       adsurl = {https://ui.adsabs.harvard.edu/abs/2022A&A...665A..24P}
}

@ARTICLE{Eckert2022_hydromass,
       author = {{Eckert}, D. and {Ettori}, S. and {Pointecouteau}, E. and {van der Burg}, R.~F.~J. and {Loubser}, S.~I.},
        title = "{The gravitational field of X-COP galaxy clusters}",
      journal = {\aap},
         year = 2022,
        month = jun,
       volume = {662},
          eid = {A123},
        pages = {A123},
          doi = {10.1051/0004-6361/202142507},
archivePrefix = {arXiv},
       eprint = {2205.01110},
 primaryClass = {astro-ph.CO},
       adsurl = {https://ui.adsabs.harvard.edu/abs/2022A&A...662A.123E}
}

@ARTICLE{Hilton2021ApJACT,
       author = {{Hilton}, M. and {Sif{\'o}n}, C. and {Naess}, S. and {Madhavacheril}, M. and {Oguri}, M. and {Rozo}, E. and {Rykoff}, E. and {Abbott}, T.~M.~C. and {Adhikari}, S. and {Aguena}, M. and {Aiola}, S. and {Allam}, S. and {Amodeo}, S. and {Amon}, A. and {Annis}, J. and {Ansarinejad}, B. and {Aros-Bunster}, C. and {Austermann}, J.~E. and {Avila}, S. and {Bacon}, D. and {Battaglia}, N. and {Beall}, J.~A. and {Becker}, D.~T. and {Bernstein}, G.~M. and {Bertin}, E. and {Bhandarkar}, T. and {Bhargava}, S. and {Bond}, J.~R. and {Brooks}, D. and {Burke}, D.~L. and {Calabrese}, E. and {Carrasco Kind}, M. and {Carretero}, J. and {Choi}, S.~K. and {Choi}, A. and {Conselice}, C. and {da Costa}, L.~N. and {Costanzi}, M. and {Crichton}, D. and {Crowley}, K.~T. and {D{\"u}nner}, R. and {Denison}, E.~V. and {Devlin}, M.~J. and {Dicker}, S.~R. and {Diehl}, H.~T. and {Dietrich}, J.~P. and {Doel}, P. and {Duff}, S.~M. and {Duivenvoorden}, A.~J. and {Dunkley}, J. and {Everett}, S. and {Ferraro}, S. and {Ferrero}, I. and {Fert{\'e}}, A. and {Flaugher}, B. and {Frieman}, J. and {Gallardo}, P.~A. and {Garc{\'\i}a-Bellido}, J. and {Gaztanaga}, E. and {Gerdes}, D.~W. and {Giles}, P. and {Golec}, J.~E. and {Gralla}, M.~B. and {Grandis}, S. and {Gruen}, D. and {Gruendl}, R.~A. and {Gschwend}, J. and {Gutierrez}, G. and {Han}, D. and {Hartley}, W.~G. and {Hasselfield}, M. and {Hill}, J.~C. and {Hilton}, G.~C. and {Hincks}, A.~D. and {Hinton}, S.~R. and {Ho}, S. -P.~P. and {Honscheid}, K. and {Hoyle}, B. and {Hubmayr}, J. and {Huffenberger}, K.~M. and {Hughes}, J.~P. and {Jaelani}, A.~T. and {Jain}, B. and {James}, D.~J. and {Jeltema}, T. and {Kent}, S. and {Knowles}, K. and {Koopman}, B.~J. and {Kuehn}, K. and {Lahav}, O. and {Lima}, M. and {Lin}, Y. -T. and {Lokken}, M. and {Loubser}, S.~I. and {MacCrann}, N. and {Maia}, M.~A.~G. and {Marriage}, T.~A. and {Martin}, J. and {McMahon}, J. and {Melchior}, P. and {Menanteau}, F. and {Miquel}, R. and {Miyatake}, H. and {Moodley}, K. and {Morgan}, R. and {Mroczkowski}, T. and {Nati}, F. and {Newburgh}, L.~B. and {Niemack}, M.~D. and {Nishizawa}, A.~J. and {Ogando}, R.~L.~C. and {Orlowski-Scherer}, J. and {Page}, L.~A. and {Palmese}, A. and {Partridge}, B. and {Paz-Chinch{\'o}n}, F. and {Phakathi}, P. and {Plazas}, A.~A. and {Robertson}, N.~C. and {Romer}, A.~K. and {Carnero Rosell}, A. and {Salatino}, M. and {Sanchez}, E. and {Schaan}, E. and {Schillaci}, A. and {Sehgal}, N. and {Serrano}, S. and {Shin}, T. and {Simon}, S.~M. and {Smith}, M. and {Soares-Santos}, M. and {Spergel}, D.~N. and {Staggs}, S.~T. and {Storer}, E.~R. and {Suchyta}, E. and {Swanson}, M.~E.~C. and {Tarle}, G. and {Thomas}, D. and {To}, C. and {Trac}, H. and {Ullom}, J.~N. and {Vale}, L.~R. and {Van Lanen}, J. and {Vavagiakis}, E.~M. and {De Vicente}, J. and {Wilkinson}, R.~D. and {Wollack}, E.~J. and {Xu}, Z. and {Zhang}, Y.},
        title = "{The Atacama Cosmology Telescope: A Catalog of >4000 Sunyaev-Zel{\textquoteright}dovich Galaxy Clusters}",
      journal = {\apjs},
         year = 2021,
        month = mar,
       volume = {253},
       number = {1},
          eid = {3},
        pages = {3},
          doi = {10.3847/1538-4365/abd023},
archivePrefix = {arXiv},
       eprint = {2009.11043},
 primaryClass = {astro-ph.CO},
       adsurl = {https://ui.adsabs.harvard.edu/abs/2021ApJS..253....3H}
}

@ARTICLE{Predehl2021A&Aerosita,
       author = {{Predehl}, P. and {Andritschke}, R. and {Arefiev}, V. and {Babyshkin}, V. and {Batanov}, O. and {Becker}, W. and {B{\"o}hringer}, H. and {Bogomolov}, A. and {Boller}, T. and {Borm}, K. and {Bornemann}, W. and {Br{\"a}uninger}, H. and {Br{\"u}ggen}, M. and {Brunner}, H. and {Brusa}, M. and {Bulbul}, E. and {Buntov}, M. and {Burwitz}, V. and {Burkert}, W. and {Clerc}, N. and {Churazov}, E. and {Coutinho}, D. and {Dauser}, T. and {Dennerl}, K. and {Doroshenko}, V. and {Eder}, J. and {Emberger}, V. and {Eraerds}, T. and {Finoguenov}, A. and {Freyberg}, M. and {Friedrich}, P. and {Friedrich}, S. and {F{\"u}rmetz}, M. and {Georgakakis}, A. and {Gilfanov}, M. and {Granato}, S. and {Grossberger}, C. and {Gueguen}, A. and {Gureev}, P. and {Haberl}, F. and {H{\"a}lker}, O. and {Hartner}, G. and {Hasinger}, G. and {Huber}, H. and {Ji}, L. and {Kienlin}, A. v. and {Kink}, W. and {Korotkov}, F. and {Kreykenbohm}, I. and {Lamer}, G. and {Lomakin}, I. and {Lapshov}, I. and {Liu}, T. and {Maitra}, C. and {Meidinger}, N. and {Menz}, B. and {Merloni}, A. and {Mernik}, T. and {Mican}, B. and {Mohr}, J. and {M{\"u}ller}, S. and {Nandra}, K. and {Nazarov}, V. and {Pacaud}, F. and {Pavlinsky}, M. and {Perinati}, E. and {Pfeffermann}, E. and {Pietschner}, D. and {Ramos-Ceja}, M.~E. and {Rau}, A. and {Reiffers}, J. and {Reiprich}, T.~H. and {Robrade}, J. and {Salvato}, M. and {Sanders}, J. and {Santangelo}, A. and {Sasaki}, M. and {Scheuerle}, H. and {Schmid}, C. and {Schmitt}, J. and {Schwope}, A. and {Shirshakov}, A. and {Steinmetz}, M. and {Stewart}, I. and {Str{\"u}der}, L. and {Sunyaev}, R. and {Tenzer}, C. and {Tiedemann}, L. and {Tr{\"u}mper}, J. and {Voron}, V. and {Weber}, P. and {Wilms}, J. and {Yaroshenko}, V.},
        title = "{The eROSITA X-ray telescope on SRG}",
      journal = {\aap},
         year = 2021,
        month = mar,
       volume = {647},
          eid = {A1},
        pages = {A1},
          doi = {10.1051/0004-6361/202039313},
archivePrefix = {arXiv},
       eprint = {2010.03477},
 primaryClass = {astro-ph.HE},
       adsurl = {https://ui.adsabs.harvard.edu/abs/2021A&A...647A...1P}
}

@ARTICLE{Campitiello2022A&A_chexmate_morph,
       author = {{Campitiello}, M.~G. and {Ettori}, S. and {Lovisari}, L. and {Bartalucci}, I. and {Eckert}, D. and {Rasia}, E. and {Rossetti}, M. and {Gastaldello}, F. and {Pratt}, G.~W. and {Maughan}, B. and {Pointecouteau}, E. and {Sereno}, M. and {Biffi}, V. and {Borgani}, S. and {De Luca}, F. and {De Petris}, M. and {Gaspari}, M. and {Ghizzardi}, S. and {Mazzotta}, P. and {Molendi}, S.},
        title = "{CHEX-MATE: Morphological analysis of the sample}",
      journal = {\aap},
         year = 2022,
        month = sep,
       volume = {665},
          eid = {A117},
        pages = {A117},
          doi = {10.1051/0004-6361/202243470},
archivePrefix = {arXiv},
       eprint = {2205.11326},
 primaryClass = {astro-ph.CO},
       adsurl = {https://ui.adsabs.harvard.edu/abs/2022A&A...665A.117C}
}

@ARTICLE{Clerc2022arXiv220311906C_review,
       author = {{Clerc}, Nicolas and {Finoguenov}, Alexis},
        title = "{X-ray cluster cosmology}",
      journal = {arXiv e-prints},
         year = 2022,
        month = mar,
          eid = {arXiv:2203.11906},
        pages = {arXiv:2203.11906},
archivePrefix = {arXiv},
       eprint = {2203.11906},
 primaryClass = {astro-ph.CO},
       adsurl = {https://ui.adsabs.harvard.edu/abs/2022arXiv220311906C}
}

@ARTICLE{ZuHone2023A&A_LXMAGerosita,
       author = {{ZuHone}, J. and {Bahar}, Y.~E. and {Biffi}, V. and {Dolag}, K. and {Sanders}, J. and {Bulbul}, E. and {Liu}, T. and {Dauser}, T. and {K{\"o}nig}, O. and {Zhang}, X. and {Ghirardini}, V.},
        title = "{Effects of multiphase gas and projection on X-ray observables in simulated galaxy clusters as seen by eROSITA}",
      journal = {\aap},
         year = 2023,
        month = jul,
       volume = {675},
          eid = {A150},
        pages = {A150},
          doi = {10.1051/0004-6361/202245749},
archivePrefix = {arXiv},
       eprint = {2212.11028},
 primaryClass = {astro-ph.CO},
       adsurl = {https://ui.adsabs.harvard.edu/abs/2023A&A...675A.150Z}
}

@ARTICLE{Scheck2023A&A_hydrobiasMAG,
       author = {{Scheck}, Dominik and {Sanders}, Jeremy S. and {Biffi}, Veronica and {Dolag}, Klaus and {Bulbul}, Esra and {Liu}, Ang},
        title = "{Hydrostatic mass profiles of galaxy clusters in the eROSITA survey}",
      journal = {\aap},
         year = 2023,
        month = feb,
       volume = {670},
          eid = {A33},
        pages = {A33},
          doi = {10.1051/0004-6361/202244582},
archivePrefix = {arXiv},
       eprint = {2211.12146},
 primaryClass = {astro-ph.CO},
       adsurl = {https://ui.adsabs.harvard.edu/abs/2023A&A...670A..33S}
}

@ARTICLE{Wicker2023A&A_hydrobOBS,
       author = {{Wicker}, R. and {Douspis}, M. and {Salvati}, L. and {Aghanim}, N.},
        title = "{Constraining the mass and redshift evolution of the hydrostatic mass bias using the gas mass fraction in galaxy clusters}",
      journal = {\aap},
         year = 2023,
        month = jun,
       volume = {674},
          eid = {A48},
        pages = {A48},
          doi = {10.1051/0004-6361/202243922},
archivePrefix = {arXiv},
       eprint = {2204.12823},
 primaryClass = {astro-ph.CO},
       adsurl = {https://ui.adsabs.harvard.edu/abs/2023A&A...674A..48W}
}

@ARTICLE{Gianfagna2023MNRAS_hydrob_300,
       author = {{Gianfagna}, Giulia and {Rasia}, Elena and {Cui}, Weiguang and {De Petris}, Marco and {Yepes}, Gustavo and {Contreras-Santos}, Ana and {Knebe}, Alexander},
        title = "{A study of the hydrostatic mass bias dependence and evolution within The Three Hundred clusters}",
      journal = {\mnras},
         year = 2023,
        month = jan,
       volume = {518},
       number = {3},
        pages = {4238-4248},
          doi = {10.1093/mnras/stac3364},
archivePrefix = {arXiv},
       eprint = {2211.08372},
 primaryClass = {astro-ph.CO},
       adsurl = {https://ui.adsabs.harvard.edu/abs/2023MNRAS.518.4238G}
}

@ARTICLE{Bartalucci2023A&A_chexmateSB,
       author = {{Bartalucci}, I. and {Molendi}, S. and {Rasia}, E. and {Pratt}, G.~W. and {Arnaud}, M. and {Rossetti}, M. and {Gastaldello}, F. and {Eckert}, D. and {Balboni}, M. and {Borgani}, S. and {Bourdin}, H. and {Campitiello}, M.~G. and {De Grandi}, S. and {De Petris}, M. and {Duffy}, R.~T. and {Ettori}, S. and {Ferragamo}, A. and {Gaspari}, M. and {Gavazzi}, R. and {Ghizzardi}, S. and {Iqbal}, A. and {Kay}, S.~T. and {Lovisari}, L. and {Mazzotta}, P. and {Maughan}, B.~J. and {Pointecouteau}, E. and {Riva}, G. and {Sereno}, M.},
        title = "{CHEX-MATE: Constraining the origin of the scatter in galaxy cluster radial X-ray surface brightness profiles}",
      journal = {\aap},
         year = 2023,
        month = jun,
       volume = {674},
          eid = {A179},
        pages = {A179},
          doi = {10.1051/0004-6361/202346189},
archivePrefix = {arXiv},
       eprint = {2305.03082},
 primaryClass = {astro-ph.CO},
       adsurl = {https://ui.adsabs.harvard.edu/abs/2023A&A...674A.179B}
}

@ARTICLE{Jennings2023MNRAS_simbahydrob,
       author = {{Jennings}, Fred and {Dav{\'e}}, Romeel},
        title = "{Halo scaling relations and hydrostatic mass bias in the SIMBA simulation from realistic mock X-ray catalogues}",
      journal = {\mnras},
         year = 2023,
        month = nov,
       volume = {526},
       number = {1},
        pages = {1367-1387},
          doi = {10.1093/mnras/stad2666},
archivePrefix = {arXiv},
       eprint = {2306.01397},
 primaryClass = {astro-ph.GA},
       adsurl = {https://ui.adsabs.harvard.edu/abs/2023MNRAS.526.1367J}
}

@ARTICLE{Beauchesne2024MNRAS_lensingXraykine,
       author = {{Beauchesne}, Benjamin and {Cl{\'e}ment}, Benjamin and {Hibon}, Pascale and {Limousin}, Marceau and {Eckert}, Dominique and {Kneib}, Jean-Paul and {Richard}, Johan and {Natarajan}, Priyamvada and {Jauzac}, Mathilde and {Montes}, Mireia and {Mahler}, Guillaume and {Claeyssens}, Ad{\'e}la{\"\i}de and {Jeanneau}, Alexandre and {Koekemoer}, Anton M. and {Lagattuta}, David and {Pagul}, Amanda and {S{\'a}nchez}, Javier},
        title = "{A new step forward in realistic cluster lens mass modelling: analysis of Hubble Frontier Field Cluster Abell S1063 from joint lensing, X-ray, and galaxy kinematics data}",
      journal = {\mnras},
         year = 2024,
        month = jan,
       volume = {527},
       number = {2},
        pages = {3246-3275},
          doi = {10.1093/mnras/stad3308},
archivePrefix = {arXiv},
       eprint = {2301.10907},
 primaryClass = {astro-ph.CO},
       adsurl = {https://ui.adsabs.harvard.edu/abs/2024MNRAS.527.3246B}
}

@ARTICLE{Giocolieuclid2024A&A_WL,
       author = {{Euclid Collaboration} and {Giocoli}, C. and {Meneghetti}, M. and {Rasia}, E. and {Borgani}, S. and {Despali}, G. and {Lesci}, G.~F. and {Marulli}, F. and {Moscardini}, L. and {Sereno}, M. and {Cui}, W. and {Knebe}, A. and {Yepes}, G. and {Castro}, T. and {Corasaniti}, P. -S. and {Pires}, S. and {Castignani}, G. and {Schrabback}, T. and {Pratt}, G.~W. and {Le Brun}, A.~M.~C. and {Aghanim}, N. and {Amendola}, L. and {Auricchio}, N. and {Baldi}, M. and {Bodendorf}, C. and {Bonino}, D. and {Branchini}, E. and {Brescia}, M. and {Brinchmann}, J. and {Camera}, S. and {Capobianco}, V. and {Carbone}, C. and {Carretero}, J. and {Castander}, F.~J. and {Castellano}, M. and {Cavuoti}, S. and {Cledassou}, R. and {Congedo}, G. and {Conselice}, C.~J. and {Conversi}, L. and {Copin}, Y. and {Corcione}, L. and {Courbin}, F. and {Cropper}, M. and {Da Silva}, A. and {Degaudenzi}, H. and {Dinis}, J. and {Dubath}, F. and {Dupac}, X. and {Dusini}, S. and {Farrens}, S. and {Ferriol}, S. and {Fosalba}, P. and {Frailis}, M. and {Franceschi}, E. and {Fumana}, M. and {Galeotta}, S. and {Garilli}, B. and {Gillis}, B. and {Grazian}, A. and {Grupp}, F. and {Haugan}, S.~V.~H. and {Holmes}, W. and {Hornstrup}, A. and {Jahnke}, K. and {K{\"u}mmel}, M. and {Kermiche}, S. and {Kilbinger}, M. and {Kunz}, M. and {Kurki-Suonio}, H. and {Ligori}, S. and {Lilje}, P.~B. and {Lloro}, I. and {Maiorano}, E. and {Mansutti}, O. and {Marggraf}, O. and {Markovic}, K. and {Massey}, R. and {Maurogordato}, S. and {Mei}, S. and {Merlin}, E. and {Meylan}, G. and {Moresco}, M. and {Munari}, E. and {Niemi}, S. -M. and {Nightingale}, J. and {Nutma}, T. and {Padilla}, C. and {Paltani}, S. and {Pasian}, F. and {Pedersen}, K. and {Pettorino}, V. and {Polenta}, G. and {Poncet}, M. and {Popa}, L.~A. and {Raison}, F. and {Renzi}, A. and {Rhodes}, J. and {Riccio}, G. and {Romelli}, E. and {Roncarelli}, M. and {Rossetti}, E. and {Saglia}, R. and {Sapone}, D. and {Sartoris}, B. and {Schneider}, P. and {Secroun}, A. and {Serrano}, S. and {Sirignano}, C. and {Sirri}, G. and {Stanco}, L. and {Starck}, J. -L. and {Tallada-Cresp{\'\i}}, P. and {Taylor}, A.~N. and {Tereno}, I. and {Toledo-Moreo}, R. and {Torradeflot}, F. and {Tutusaus}, I. and {Valentijn}, E.~A. and {Valenziano}, L. and {Vassallo}, T. and {Wang}, Y. and {Weller}, J. and {Zamorani}, G. and {Zoubian}, J. and {Andreon}, S. and {Bardelli}, S. and {Boucaud}, A. and {Bozzo}, E. and {Colodro-Conde}, C. and {Di Ferdinando}, D. and {Fabbian}, G. and {Farina}, M. and {Israel}, H. and {Keih{\"a}nen}, E. and {Lindholm}, V. and {Mauri}, N. and {Neissner}, C. and {Schirmer}, M. and {Scottez}, V. and {Tenti}, M. and {Zucca}, E. and {Akrami}, Y. and {Baccigalupi}, C. and {Ballardini}, M. and {Bernardeau}, F. and {Biviano}, A. and {Borlaff}, A.~S. and {Burigana}, C. and {Cabanac}, R. and {Cappi}, A. and {Carvalho}, C.~S. and {Casas}, S. and {Chambers}, K.~C. and {Cooray}, A.~R. and {Courtois}, H.~M. and {Davini}, S. and {de la Torre}, S. and {De Lucia}, G. and {Desprez}, G. and {Dole}, H. and {Escartin}, J.~A. and {Escoffier}, S. and {Ferrero}, I. and {Finelli}, F. and {Gabarra}, L. and {Ganga}, K. and {Garcia-Bellido}, J. and {George}, K. and {Giacomini}, F. and {Gozaliasl}, G. and {Hildebrandt}, H. and {Hook}, I. and {Jimenez Mu{\~n}oz}, A. and {Joachimi}, B. and {Kajava}, J.~J.~E. and {Kansal}, V. and {Kirkpatrick}, C.~C. and {Legrand}, L. and {Loureiro}, A. and {Macias-Perez}, J. and {Magliocchetti}, M. and {Mainetti}, G. and {Maoli}, R. and {Marcin}, S. and {Martinelli}, M. and {Martinet}, N. and {Martins}, C.~J.~A.~P. and {Matthew}, S. and {Maurin}, L. and {Metcalf}, R.~B. and {Monaco}, P. and {Morgante}, G. and {Nadathur}, S. and {Nucita}, A.~A. and {Patrizii}, L. and {Peel}, A. and {Pollack}, J. and {Popa}, V. and {Porciani}, C.},
        title = "{Euclid preparation. XXXII. Evaluating the weak-lensing cluster mass biases using the Three Hundred Project hydrodynamical simulations}",
      journal = {\aap},
         year = 2024,
        month = jan,
       volume = {681},
          eid = {A67},
        pages = {A67},
          doi = {10.1051/0004-6361/202346058},
archivePrefix = {arXiv},
       eprint = {2302.00687},
 primaryClass = {astro-ph.CO},
       adsurl = {https://ui.adsabs.harvard.edu/abs/2024A&A...681A..67E}
}

@ARTICLE{Kim2024A&A_clump3d,
       author = {{Kim}, Junhan and {Sayers}, Jack and {Sereno}, Mauro and {Bartalucci}, Iacopo and {Chappuis}, Loris and {De Grandi}, Sabrina and {De Luca}, Federico and {De Petris}, Marco and {Donahue}, Megan E. and {Eckert}, Dominique and {Ettori}, Stefano and {Gaspari}, Massimo and {Gastaldello}, Fabio and {Gavazzi}, Raphael and {Gavidia}, Adriana and {Ghizzardi}, Simona and {Iqbal}, Asif and {Kay}, Scott T. and {Lovisari}, Lorenzo and {Maughan}, Ben J. and {Mazzotta}, Pasquale and {Okabe}, Nobuhiro and {Pointecouteau}, Etienne and {Pratt}, Gabriel W. and {Rossetti}, Mariachiara and {Umetsu}, Keiichi},
        title = "{CHEX-MATE: CLUster Multi-Probes in Three Dimensions (CLUMP-3D). I. Gas analysis method using X-ray and Sunyaev-Zel'dovich effect data}",
      journal = {\aap},
         year = 2024,
        month = jun,
       volume = {686},
          eid = {A97},
        pages = {A97},
          doi = {10.1051/0004-6361/202347399},
archivePrefix = {arXiv},
       eprint = {2307.04794},
 primaryClass = {astro-ph.CO},
       adsurl = {https://ui.adsabs.harvard.edu/abs/2024A&A...686A..97K}
}

@ARTICLE{Bulbul2024,
       author = {{Bulbul}, E. and {Liu}, A. and {Kluge}, M. and {Zhang}, X. and {Sanders}, J.~S. and {Bahar}, Y.~E. and {Ghirardini}, V. and {Artis}, E. and {Seppi}, R. and {Garrel}, C. and {Ramos-Ceja}, M.~E. and {Comparat}, J. and {Balzer}, F. and {B{\"o}ckmann}, K. and {Br{\"u}ggen}, M. and {Clerc}, N. and {Dennerl}, K. and {Dolag}, K. and {Freyberg}, M. and {Grandis}, S. and {Gruen}, D. and {Kleinebreil}, F. and {Krippendorf}, S. and {Lamer}, G. and {Merloni}, A. and {Migkas}, K. and {Nandra}, K. and {Pacaud}, F. and {Predehl}, P. and {Reiprich}, T.~H. and {Schrabback}, T. and {Veronica}, A. and {Weller}, J. and {Zelmer}, S.},
        title = "{The SRG/eROSITA All-Sky Survey: The first catalog of galaxy clusters and groups in the Western Galactic Hemisphere}",
      journal = {arXiv e-prints},
         year = 2024,
        month = feb,
          eid = {arXiv:2402.08452},
        pages = {arXiv:2402.08452},
          doi = {10.48550/arXiv.2402.08452},
archivePrefix = {arXiv},
       eprint = {2402.08452},
 primaryClass = {astro-ph.CO},
       adsurl = {https://ui.adsabs.harvard.edu/abs/2024arXiv240208452B}
}

@ARTICLE{Ghirardini2024_erass1cosmo,
       author = {{Ghirardini}, V. and {Bulbul}, E. and {Artis}, E. and {Clerc}, N. and {Garrel}, C. and {Grandis}, S. and {Kluge}, M. and {Liu}, A. and {Bahar}, Y.~E. and {Balzer}, F. and {Chiu}, I. and {Comparat}, J. and {Gruen}, D. and {Kleinebreil}, F. and {Krippendorf}, S. and {Merloni}, A. and {Nandra}, K. and {Okabe}, N. and {Pacaud}, F. and {Predehl}, P. and {Ramos-Ceja}, M.~E. and {Reiprich}, T.~H. and {Sanders}, J.~S. and {Schrabback}, T. and {Seppi}, R. and {Zelmer}, S. and {Zhang}, X. and {Bornemann}, W. and {Brunner}, H. and {Burwitz}, V. and {Coutinho}, D. and {Dennerl}, K. and {Freyberg}, M. and {Friedrich}, S. and {Gaida}, R. and {Gueguen}, A. and {Haberl}, F. and {Kink}, W. and {Lamer}, G. and {Li}, X. and {Liu}, T. and {Maitra}, C. and {Meidinger}, N. and {Mueller}, S. and {Miyatake}, H. and {Miyazaki}, S. and {Robrade}, J. and {Schwope}, A. and {Stewart}, I.},
        title = "{The SRG/eROSITA all-sky survey: Cosmology constraints from cluster abundances in the western Galactic hemisphere}",
      journal = {\aap},
         year = 2024,
        month = sep,
       volume = {689},
          eid = {A298},
        pages = {A298},
          doi = {10.1051/0004-6361/202348852},
archivePrefix = {arXiv},
       eprint = {2402.08458},
 primaryClass = {astro-ph.CO},
       adsurl = {https://ui.adsabs.harvard.edu/abs/2024A&A...689A.298G}
}

@ARTICLE{Rossetti2024A&A_TxprofCXM,
       author = {{Rossetti}, M. and {Eckert}, D. and {Gastaldello}, F. and {Rasia}, E. and {Pratt}, G.~W. and {Ettori}, S. and {Molendi}, S. and {Arnaud}, M. and {Balboni}, M. and {Bartalucci}, I. and {Batalha}, R.~M. and {Borgani}, S. and {Bourdin}, H. and {De Grandi}, S. and {De Luca}, F. and {De Petris}, M. and {Forman}, W. and {Gaspari}, M. and {Ghizzardi}, S. and {Iqbal}, A. and {Kay}, S. and {Lovisari}, L. and {Maughan}, B.~J. and {Mazzotta}, P. and {Pointecouteau}, E. and {Riva}, G. and {Sayers}, J. and {Sereno}, M.},
        title = "{CHEX-MATE: Robust reconstruction of temperature profiles in galaxy clusters with XMM-Newton}",
      journal = {\aap},
         year = 2024,
        month = jun,
       volume = {686},
          eid = {A68},
        pages = {A68},
          doi = {10.1051/0004-6361/202348853},
       adsurl = {https://ui.adsabs.harvard.edu/abs/2024A&A...686A..68R}
}

@ARTICLE{Bahar2024A&A_grpsFeedback,
       author = {{Bahar}, Y.~E. and {Bulbul}, E. and {Ghirardini}, V. and {Sanders}, J.~S. and {Zhang}, X. and {Liu}, A. and {Clerc}, N. and {Artis}, E. and {Balzer}, F. and {Biffi}, V. and {Bose}, S. and {Comparat}, J. and {Dolag}, K. and {Garrel}, C. and {Hadzhiyska}, B. and {Hern{\'a}ndez-Aguayo}, C. and {Hernquist}, L. and {Kluge}, M. and {Krippendorf}, S. and {Merloni}, A. and {Nandra}, K. and {Pakmor}, R. and {Popesso}, P. and {Ramos-Ceja}, M. and {Seppi}, R. and {Springel}, V. and {Weller}, J. and {Zelmer}, S.},
        title = "{The SRG/eROSITA All-Sky Survey: Constraints on AGN feedback in galaxy groups}",
      journal = {\aap},
         year = 2024,
        month = nov,
       volume = {691},
          eid = {A188},
        pages = {A188},
          doi = {10.1051/0004-6361/202449399},
archivePrefix = {arXiv},
       eprint = {2401.17276},
 primaryClass = {astro-ph.CO},
       adsurl = {https://ui.adsabs.harvard.edu/abs/2024A&A...691A.188B}
}

@ARTICLE{Riva2024A&Achexmate_icm,
       author = {{Riva}, G. and {Pratt}, G.~W. and {Rossetti}, M. and {Bartalucci}, I. and {Kay}, S.~T. and {Rasia}, E. and {Gavazzi}, R. and {Umetsu}, K. and {Arnaud}, M. and {Balboni}, M. and {Bonafede}, A. and {Bourdin}, H. and {De Grandi}, S. and {De Luca}, F. and {Eckert}, D. and {Ettori}, S. and {Gaspari}, M. and {Gastaldello}, F. and {Ghirardini}, V. and {Ghizzardi}, S. and {Gitti}, M. and {Lovisari}, L. and {Maughan}, B.~J. and {Mazzotta}, P. and {Molendi}, S. and {Pointecouteau}, E. and {Sayers}, J. and {Sereno}, M. and {Towler}, I.},
        title = "{CHEX-MATE: The intracluster medium entropy distribution in the gravity-dominated regime}",
      journal = {\aap},
         year = 2024,
        month = nov,
       volume = {691},
          eid = {A340},
        pages = {A340},
          doi = {10.1051/0004-6361/202451455},
archivePrefix = {arXiv},
       eprint = {2410.11947},
 primaryClass = {astro-ph.CO},
       adsurl = {https://ui.adsabs.harvard.edu/abs/2024A&A...691A.340R}
}

@ARTICLE{Lovisari2024A&A_CXMATETX,
       author = {{Lovisari}, L. and {Ettori}, S. and {Rasia}, E. and {Gaspari}, M. and {Bourdin}, H. and {Campitiello}, M.~G. and {Rossetti}, M. and {Bartalucci}, I. and {De Grandi}, S. and {De Luca}, F. and {De Petris}, M. and {Eckert}, D. and {Forman}, W. and {Gastaldello}, F. and {Ghizzardi}, S. and {Jones}, C. and {Kay}, S. and {Kim}, J. and {Maughan}, B.~J. and {Mazzotta}, P. and {Pointecouteau}, E. and {Pratt}, G.~W. and {Sayers}, J. and {Sereno}, M. and {Simonte}, M. and {Tozzi}, P.},
        title = "{CHEX-MATE: Characterization of the intra-cluster medium temperature distribution}",
      journal = {\aap},
         year = 2024,
        month = feb,
       volume = {682},
          eid = {A45},
        pages = {A45},
          doi = {10.1051/0004-6361/202346651},
archivePrefix = {arXiv},
       eprint = {2311.02176},
 primaryClass = {astro-ph.CO},
       adsurl = {https://ui.adsabs.harvard.edu/abs/2024A&A...682A..45L}
}

@ARTICLE{Miren2024A&A_bhe,
       author = {{Mu{\~n}oz-Echeverr{\'\i}a}, M. and {Mac{\'\i}as-P{\'e}rez}, J.~F. and {Pratt}, G.~W. and {Pointecouteau}, E. and {Bartalucci}, I. and {De Petris}, M. and {Ferragamo}, A. and {Hanser}, C. and {K{\'e}ruzor{\'e}}, F. and {Mayet}, F. and {Moyer-Anin}, A. and {Paliwal}, A. and {Perotto}, L. and {Yepes}, G.},
        title = "{The hydrostatic-to-lensing mass bias from resolved X-ray and optical-IR data}",
      journal = {\aap},
         year = 2024,
        month = feb,
       volume = {682},
          eid = {A147},
        pages = {A147},
          doi = {10.1051/0004-6361/202347584},
archivePrefix = {arXiv},
       eprint = {2312.01154},
 primaryClass = {astro-ph.CO},
       adsurl = {https://ui.adsabs.harvard.edu/abs/2024A&A...682A.147M}
}

@ARTICLE{Kay2024MNRAS_flamingo,
       author = {{Kay}, Scott T. and {Braspenning}, Joey and {Chluba}, Jens and {Helly}, John C. and {Kugel}, Roi and {Schaller}, Matthieu and {Schaye}, Joop},
        title = "{Relativistic SZ temperatures and hydrostatic mass bias for massive clusters in the FLAMINGO simulations}",
      journal = {\mnras},
         year = 2024,
        month = oct,
       volume = {534},
       number = {1},
        pages = {251-270},
          doi = {10.1093/mnras/stae1991},
archivePrefix = {arXiv},
       eprint = {2404.08539},
 primaryClass = {astro-ph.CO},
       adsurl = {https://ui.adsabs.harvard.edu/abs/2024MNRAS.534..251K}
}

@ARTICLE{Popesso2024arXiv_grpsprop,
       author = {{Popesso}, P. and {Marini}, I. and {Dolag}, K. and {Lamer}, G. and {Csizi}, B. and {Biffi}, V. and {Robothan}, A. and {Bravo}, M. and {Biviano}, A. and {Vladutesku-Zopp}, S. and {Lovisari}, L. and {Ettori}, S. and {Angelinelli}, M. and {Driver}, S. and {Toptun}, V. and {Dev}, A. and {Mazengo}, D. and {Merloni}, A. and {Zhang}, Y. and {Comparat}, J. and {Ponti}, G. and {Mroczkowski}, T. and {Bulbul}, E.},
        title = "{Average X-ray properties of galaxy groups. From Milky Way-like halos to massive clusters}",
      journal = {arXiv e-prints},
         year = 2024,
        month = nov,
          eid = {arXiv:2411.17120},
        pages = {arXiv:2411.17120},
          doi = {10.48550/arXiv.2411.17120},
archivePrefix = {arXiv},
       eprint = {2411.17120},
 primaryClass = {astro-ph.GA},
       adsurl = {https://ui.adsabs.harvard.edu/abs/2024arXiv241117120P}
}

@ARTICLE{Cruise2025NatAs_newathena,
       author = {{Cruise}, Mike and {Guainazzi}, Matteo and {Aird}, James and {Carrera}, Francisco J. and {Costantini}, Elisa and {Corrales}, Lia and {Dauser}, Thomas and {Eckert}, Dominique and {Gastaldello}, Fabio and {Matsumoto}, Hironori and {Osten}, Rachel and {Petrucci}, Pierre-Olivier and {Porquet}, Delphine and {Pratt}, Gabriel W. and {Rea}, Nanda and {Reiprich}, Thomas H. and {Simionescu}, Aurora and {Spiga}, Daniele and {Troja}, Eleonora},
        title = "{The NewAthena mission concept in the context of the next decade of X-ray astronomy}",
      journal = {Nature Astronomy},
         year = 2025,
        month = jan,
       volume = {9},
        pages = {36-44},
          doi = {10.1038/s41550-024-02416-3},
archivePrefix = {arXiv},
       eprint = {2501.03100},
 primaryClass = {astro-ph.IM},
       adsurl = {https://ui.adsabs.harvard.edu/abs/2025NatAs...9...36C}
}

@ARTICLE{Braspenning2025MNRAS_bhe,
       author = {{Braspenning}, Joey and {Schaye}, Joop and {Schaller}, Matthieu and {Kugel}, Roi and {Kay}, Scott T.},
        title = "{Hydrostatic mass bias for galaxy groups and clusters in the FLAMINGO simulations}",
      journal = {\mnras},
         year = 2025,
        month = feb,
       volume = {536},
       number = {4},
        pages = {3784-3802},
          doi = {10.1093/mnras/stae2798},
archivePrefix = {arXiv},
       eprint = {2409.07849},
 primaryClass = {astro-ph.CO},
       adsurl = {https://ui.adsabs.harvard.edu/abs/2025MNRAS.536.3784B}
}

@ARTICLE{Gavidia2025_clump3d,
       author = {{Gavidia}, Adriana and {Kim}, Junhan and {Sayers}, Jack and {Sereno}, Mauro and {Chappuis}, Loris and {Eckert}, Dominique and {Umetsu}, Keiichi and {Bourdin}, Herve and {De Luca}, Federico and {Ettori}, Stefano and {Gaspari}, Massimo and {Gavazzi}, Raphael and {Kay}, Scott and {Lovisari}, Lorenzo and {Mazzotta}, Pasquale and {Pratt}, Gabriel and {Rasia}, Elena and {Rossetti}, Mariachiara},
        title = "{CHEX-MATE: Cluster Multi-Probes in Three Dimensions (CLUMP-3D) II. Combined Gas and Dark Matter Analysis from X-ray, SZE, and WL}",
      journal = {arXiv e-prints},
         year = 2025,
        month = jul,
          eid = {arXiv:2507.10857},
        pages = {arXiv:2507.10857},
          doi = {10.48550/arXiv.2507.10857},
archivePrefix = {arXiv},
       eprint = {2507.10857},
 primaryClass = {astro-ph.CO},
       adsurl = {https://ui.adsabs.harvard.edu/abs/2025arXiv250710857G}
}

@ARTICLE{Xrism2025ApJ_coma,
       author = {{Xrism Collaboration} and {Audard}, Marc and {Awaki}, Hisamitsu and {Ballhausen}, Ralf and {Bamba}, Aya and {Behar}, Ehud and {Boissay-Malaquin}, Rozenn and {Brenneman}, Laura and {Brown}, Gregory V. and {Corrales}, Lia and {Costantini}, Elisa and {Cumbee}, Renata and {Diaz Trigo}, Maria and {Done}, Chris and {Dotani}, Tadayasu and {Ebisawa}, Ken and {Eckart}, Megan E. and {Eckert}, Dominique and {Eguchi}, Satoshi and {Enoto}, Teruaki and {Ezoe}, Yuichiro and {Foster}, Adam and {Fujimoto}, Ryuichi and {Fujita}, Yutaka and {Fukazawa}, Yasushi and {Fukushima}, Kotaro and {Furuzawa}, Akihiro and {Gallo}, Luigi and {Garc{\'\i}a}, Javier A. and {Gu}, Liyi and {Guainazzi}, Matteo and {Hagino}, Kouichi and {Hamaguchi}, Kenji and {Hatsukade}, Isamu and {Hayashi}, Katsuhiro and {Hayashi}, Takayuki and {Hell}, Natalie and {Hodges-Kluck}, Edmund and {Hornschemeier}, Ann and {Ichinohe}, Yuto and {Ishi}, Daiki and {Ishida}, Manabu and {Ishikawa}, Kumi and {Ishisaki}, Yoshitaka and {Kaastra}, Jelle and {Kallman}, Timothy and {Kara}, Erin and {Katsuda}, Satoru and {Kanemaru}, Yoshiaki and {Kelley}, Richard and {Kilbourne}, Caroline and {Kitamoto}, Shunji and {Kobayashi}, Shogo and {Kohmura}, Takayoshi and {Kubota}, Aya and {Leutenegger}, Maurice and {Loewenstein}, Michael and {Maeda}, Yoshitomo and {Markevitch}, Maxim and {Matsumoto}, Hironori and {Matsushita}, Kyoko and {McCammon}, Dan and {McNamara}, Brian and {Mernier}, Fran{\c{c}}ois and {Miller}, Eric D. and {Miller}, Jon M. and {Mitsuishi}, Ikuyuki and {Mizumoto}, Misaki and {Mizuno}, Tsunefumi and {Mori}, Koji and {Mukai}, Koji and {Murakami}, Hiroshi and {Mushotzky}, Richard and {Nakajima}, Hiroshi and {Nakazawa}, Kazuhiro and {Ness}, Jan-Uwe and {Nobukawa}, Kumiko and {Nobukawa}, Masayoshi and {Noda}, Hirofumi and {Odaka}, Hirokazu and {Ogawa}, Shoji and {Ogorza{\l}ek}, Anna and {Okajima}, Takashi and {Ota}, Naomi and {Paltani}, St{\'e}phane and {Petre}, Robert and {Plucinsky}, Paul and {Porter}, Frederick S. and {Pottschmidt}, Katja and {Sato}, Kosuke and {Sato}, Toshiki and {Sawada}, Makoto and {Seta}, Hiromi and {Shidatsu}, Megumi and {Simionescu}, Aurora and {Smith}, Randall and {Suzuki}, Hiromasa and {Szymkowiak}, Andrew and {Takahashi}, Hiromitsu and {Takeo}, Mai and {Tamagawa}, Toru and {Tamura}, Keisuke and {Tanaka}, Takaaki and {Tanimoto}, Atsushi and {Tashiro}, Makoto and {Terada}, Yukikatsu and {Terashima}, Yuichi and {Tsuboi}, Yohko and {Tsujimoto}, Masahiro and {Tsunemi}, Hiroshi and {Tsuru}, Takeshi and {T{\"u}mer}, Ay{\c{s}}eg{\"u}l and {Uchida}, Hiroyuki and {Uchida}, Nagomi and {Uchida}, Yuusuke and {Uchiyama}, Hideki and {Ueda}, Shutaro and {Ueda}, Yoshihiro and {Uno}, Shinichiro and {Vink}, Jacco and {Watanabe}, Shin and {Williams}, Brian J. and {Yamada}, Satoshi and {Yamada}, Shinya and {Yamaguchi}, Hiroya and {Yamaoka}, Kazutaka and {Yamasaki}, Noriko and {Yamauchi}, Makoto and {Yamauchi}, Shigeo and {Yaqoob}, Tahir and {Yoneyama}, Tomokage and {Yoshida}, Tessei and {Yukita}, Mihoko and {Zhuravleva}, Irina and {Fabian}, Andrew and {Nelson}, Dylan and {Okabe}, Nobuhiro and {Pillepich}, Annalisa and {Potter}, Cicely and {Regamey}, Manon and {Sakai}, Kosei and {Shishido}, Mona and {Truong}, Nhut and {Wik}, Daniel R. and {Zuhone}, John},
        title = "{XRISM Forecast for the Coma Cluster: Stormy, with a Steep Power Spectrum}",
      journal = {\apjl},
         year = 2025,
        month = may,
       volume = {985},
       number = {1},
          eid = {L20},
        pages = {L20},
          doi = {10.3847/2041-8213/add2f6},
archivePrefix = {arXiv},
       eprint = {2504.20928},
 primaryClass = {astro-ph.HE},
       adsurl = {https://ui.adsabs.harvard.edu/abs/2025ApJ...985L..20X}
}

@ARTICLE{XRISM2025Natur_centaurus,
       author = {{XRISM Collaboration} and {Audard}, Marc and {Awaki}, Hisamitsu and {Ballhausen}, Ralf and {Bamba}, Aya and {Behar}, Ehud and {Boissay-Malaquin}, Rozenn and {Brenneman}, Laura and {Brown}, Gregory V. and {Corrales}, Lia and {Costantini}, Elisa and {Cumbee}, Renata and {Done}, Chris and {Dotani}, Tadayasu and {Ebisawa}, Ken and {Eckart}, Megan E. and {Eckert}, Dominique and {Enoto}, Teruaki and {Eguchi}, Satoshi and {Ezoe}, Yuichiro and {Foster}, Adam and {Fujimoto}, Ryuichi and {Fujita}, Yutaka and {Fukazawa}, Yasushi and {Fukushima}, Kotaro and {Furuzawa}, Akihiro and {Gallo}, Luigi and {Garc{\'\i}a}, Javier A. and {Gu}, Liyi and {Guainazzi}, Matteo and {Hagino}, Kouichi and {Hamaguchi}, Kenji and {Hatsukade}, Isamu and {Hayashi}, Katsuhiro and {Hayashi}, Takayuki and {Hell}, Natalie and {Hodges-Kluck}, Edmund and {Hornschemeier}, Ann and {Ichinohe}, Yuto and {Ishida}, Manabu and {Ishikawa}, Kumi and {Ishisaki}, Yoshitaka and {Kaastra}, Jelle and {Kallman}, Timothy and {Kara}, Erin and {Katsuda}, Satoru and {Kanemaru}, Yoshiaki and {Kelley}, Richard and {Kilbourne}, Caroline and {Kitamoto}, Shunji and {Kobayashi}, Shogo and {Kohmura}, Takayoshi and {Kubota}, Aya and {Leutenegger}, Maurice and {Loewenstein}, Michael and {Maeda}, Yoshitomo and {Markevitch}, Maxim and {Matsumoto}, Hironori and {Matsushita}, Kyoko and {McCammon}, Dan and {McNamara}, Brian and {Mernier}, Fran{\c{c}}ois and {Miller}, Eric D. and {Miller}, Jon M. and {Mitsuishi}, Ikuyuki and {Mizumoto}, Misaki and {Mizuno}, Tsunefumi and {Mori}, Koji and {Mukai}, Koji and {Murakami}, Hiroshi and {Mushotzky}, Richard and {Nakajima}, Hiroshi and {Nakazawa}, Kazuhiro and {Ness}, Jan-Uwe and {Nobukawa}, Kumiko and {Nobukawa}, Masayoshi and {Noda}, Hirofumi and {Odaka}, Hirokazu and {Ogawa}, Shoji and {Ogorzalek}, Anna and {Okajima}, Takashi and {Ota}, Naomi and {Paltani}, Stephane and {Petre}, Robert and {Plucinsky}, Paul and {Porter}, Frederick Scott and {Pottschmidt}, Katja and {Sato}, Kosuke and {Sato}, Toshiki and {Sawada}, Makoto and {Seta}, Hiromi and {Shidatsu}, Megumi and {Simionescu}, Aurora and {Smith}, Randall and {Suzuki}, Hiromasa and {Szymkowiak}, Andrew and {Takahashi}, Hiromitsu and {Takeo}, Mai and {Tamagawa}, Toru and {Tamura}, Keisuke and {Tanaka}, Takaaki and {Tanimoto}, Atsushi and {Tashiro}, Makoto and {Terada}, Yukikatsu and {Terashima}, Yuichi and {Trigo}, Mar{\'\i}a D{\'\i}az and {Tsuboi}, Yohko and {Tsujimoto}, Masahiro and {Tsunemi}, Hiroshi and {Tsuru}, Takeshi G. and {Uchida}, Hiroyuki and {Uchida}, Nagomi and {Uchida}, Yuusuke and {Uchiyama}, Hideki and {Ueda}, Yoshihiro and {Uno}, Shinichiro and {Vink}, Jacco and {Watanabe}, Shin and {Williams}, Brian J. and {Yamada}, Satoshi and {Yamada}, Shinya and {Yamaguchi}, Hiroya and {Yamaoka}, Kazutaka and {Yamasaki}, Noriko Y. and {Yamauchi}, Makoto and {Yamauchi}, Shigeo and {Yaqoob}, Tahir and {Yoneyama}, Tomokage and {Yoshida}, Tessei and {Yukita}, Mihoko and {Zhuravleva}, Irina and {Kondo}, Marie and {Werner}, Norbert and {Pl{\v{s}}ek}, Tom{\'a}{\v{s}} and {Sun}, Ming and {Hosogi}, Kokoro and {Majumder}, Anwesh},
        title = "{The bulk motion of gas in the core of the Centaurus galaxy cluster}",
      journal = {\nat},
         year = 2025,
        month = feb,
       volume = {638},
       number = {8050},
        pages = {365-369},
          doi = {10.1038/s41586-024-08561-z},
archivePrefix = {arXiv},
       eprint = {2502.08722},
 primaryClass = {astro-ph.HE},
       adsurl = {https://ui.adsabs.harvard.edu/abs/2025Natur.638..365X}
}

@ARTICLE{Xrism2025ApJ_a2029,
       author = {{Xrism Collaboration} and {Audard}, Marc and {Awaki}, Hisamitsu and {Ballhausen}, Ralf and {Bamba}, Aya and {Behar}, Ehud and {Boissay-Malaquin}, Rozenn and {Brenneman}, Laura and {Brown}, Gregory V. and {Corrales}, Lia and {Costantini}, Elisa and {Cumbee}, Renata and {Diaz Trigo}, Maria and {Done}, Chris and {Dotani}, Tadayasu and {Ebisawa}, Ken and {Eckart}, Megan E. and {Eckert}, Dominique and {Eguchi}, Satoshi and {Enoto}, Teruaki and {Ezoe}, Yuichiro and {Foster}, Adam and {Fujimoto}, Ryuichi and {Fujita}, Yutaka and {Fukazawa}, Yasushi and {Fukushima}, Kotaro and {Furuzawa}, Akihiro and {Gallo}, Luigi and {Garc{\'\i}a}, Javier A. and {Gu}, Liyi and {Guainazzi}, Matteo and {Hagino}, Kouichi and {Hamaguchi}, Kenji and {Hatsukade}, Isamu and {Hayashi}, Katsuhiro and {Hayashi}, Takayuki and {Hell}, Natalie and {Hodges-Kluck}, Edmund and {Hornschemeier}, Ann and {Ichinohe}, Yuto and {Ishida}, Manabu and {Ishikawa}, Kumi and {Ishisaki}, Yoshitaka and {Kaastra}, Jelle and {Kallman}, Timothy and {Kara}, Erin and {Katsuda}, Satoru and {Kanemaru}, Yoshiaki and {Kelley}, Richard and {Kilbourne}, Caroline and {Kitamoto}, Shunji and {Kobayashi}, Shogo and {Kohmura}, Takayoshi and {Kubota}, Aya and {Leutenegger}, Maurice and {Loewenstein}, Michael and {Maeda}, Yoshitomo and {Markevitch}, Maxim and {Matsumoto}, Hironori and {Matsushita}, Kyoko and {McCammon}, Dan and {McNamara}, Brian and {Mernier}, Fran{\c{c}}ois and {Miller}, Eric D. and {Miller}, Jon M. and {Mitsuishi}, Ikuyuki and {Mizumoto}, Misaki and {Mizuno}, Tsunefumi and {Mori}, Koji and {Mukai}, Koji and {Murakami}, Hiroshi and {Mushotzky}, Richard and {Nakajima}, Hiroshi and {Nakazawa}, Kazuhiro and {Ness}, Jan-Uwe and {Nobukawa}, Kumiko and {Nobukawa}, Masayoshi and {Noda}, Hirofumi and {Odaka}, Hirokazu and {Ogawa}, Shoji and {Ogorzalek}, Anna and {Okajima}, Takashi and {Ota}, Naomi and {Paltani}, Stephane and {Petre}, Robert and {Plucinsky}, Paul and {Porter}, Frederick S. and {Pottschmidt}, Katja and {Sato}, Kosuke and {Sato}, Toshiki and {Sawada}, Makoto and {Seta}, Hiromi and {Shidatsu}, Megumi and {Simionescu}, Aurora and {Smith}, Randall and {Suzuki}, Hiromasa and {Szymkowiak}, Andrew and {Takahashi}, Hiromitsu and {Takeo}, Mai and {Tamagawa}, Toru and {Tamura}, Keisuke and {Tanaka}, Takaaki and {Tanimoto}, Atsushi and {Tashiro}, Makoto and {Terada}, Yukikatsu and {Terashima}, Yuichi and {Tsuboi}, Yohko and {Tsujimoto}, Masahiro and {Tsunemi}, Hiroshi and {Tsuru}, Takeshi and {Uchida}, Hiroyuki and {Uchida}, Nagomi and {Uchida}, Yuusuke and {Uchiyama}, Hideki and {Ueda}, Yoshihiro and {Uno}, Shinichiro and {Vink}, Jacco and {Watanabe}, Shin and {Williams}, Brian J. and {Yamada}, Satoshi and {Yamada}, Shinya and {Yamaguchi}, Hiroya and {Yamaoka}, Kazutaka and {Yamasaki}, Noriko and {Yamauchi}, Makoto and {Yamauchi}, Shigeo and {Yaqoob}, Tahir and {Yoneyama}, Tomokage and {Yoshida}, Tessei and {Yukita}, Mihoko and {Zhuravleva}, Irina and {Bartalesi}, Tommaso and {Ettori}, Stefano and {Kosarzycki}, Roman and {Lovisari}, Lorenzo and {Rose}, Tom and {Sarkar}, Arnab and {Sun}, Ming and {Tamhane}, Prathamesh},
        title = "{XRISM Reveals Low Nonthermal Pressure in the Core of the Hot, Relaxed Galaxy Cluster A2029}",
      journal = {\apjl},
         year = 2025,
        month = mar,
       volume = {982},
       number = {1},
          eid = {L5},
        pages = {L5},
          doi = {10.3847/2041-8213/ada7cd},
archivePrefix = {arXiv},
       eprint = {2501.05514},
 primaryClass = {astro-ph.HE},
       adsurl = {https://ui.adsabs.harvard.edu/abs/2025ApJ...982L...5X}
}

@ARTICLE{XRISM2025arXiv_A2029II,
       author = {{XRISM Collaboration} and {Audard}, Marc and {Awaki}, Hisamitsu and {Ballhausen}, Ralf and {Bamba}, Aya and {Behar}, Ehud and {Boissay-Malaquin}, Rozenn and {Brenneman}, Laura and {Brown}, Gregory and {Corrales}, Lia and {Costantini}, Elisa and {Cumbee}, Renata and {Diaz Trigo}, Maria and {Done}, Chris and {Dotani}, Tadayasu and {Ebisawa}, Ken and {Eckart}, Megan and {Eckert}, Dominique and {Eguchi}, Satoshi and {Enoto}, Teruaki and {Ezoe}, Yuichiro and {Foster}, Adam and {Fujimoto}, Ryuichi and {Fujita}, Yutaka and {Fukazawa}, Yasushi and {Fukushima}, Kotaro and {Furuzawa}, Akihiro and {Gallo}, Luigi and {Garc{\'\i}a}, Javier and {Gu}, Liyi and {Guainazzi}, Matteo and {Hagino}, Kouichi and {Hamaguchi}, Kenji and {Hatsukade}, Isamu and {Hayashi}, Katsuhiro and {Hayashi}, Takayuki and {Hell}, Natalie and {Hodges-Kluck}, Edmund and {Hornschemeier}, Ann and {Ichinohe}, Yuto and {Ishi}, Daiki and {Ishida}, Manabu and {Ishikawa}, Kumi and {Ishisaki}, Yoshitaka and {Kaastra}, Jelle and {Kallman}, Timothy and {Kara}, Erin and {Katsuda}, Satoru and {Kanemaru}, Yoshiaki and {Kelley}, Richard and {Kilbourne}, Caroline and {Kitamoto}, Shunji and {Kobayashi}, Shogo and {Kohmura}, Takayoshi and {Kubota}, Aya and {Leutenegger}, Maurice and {Loewenstein}, Michael and {Maeda}, Yoshitomo and {Markevitch}, Maxim and {Matsumoto}, Hironori and {Matsushita}, Kyoko and {McCammon}, Dan and {McNamara}, Brian and {Mernier}, Francois and {Miller}, Eric and {Miller}, Jon and {Mitsuishi}, Ikuyuki and {Mizumoto}, Misaki and {Mizuno}, Tsunefumi and {Mori}, Koji and {Mukai}, Koji and {Murakami}, Hiroshi and {Mushotzky}, Richard and {Nakajima}, Hiroshi and {Nakazawa}, Kazuhiro and {Ness}, Jan-Uwe and {Nobukawa}, Kumiko and {Nobukawa}, Masayoshi and {Noda}, Hirofumi and {Odaka}, Hirokazu and {Ogawa}, Shoji and {Ogorzalek}, Anna and {Okajima}, Takashi and {Ota}, Naomi and {Paltani}, Stephane and {Petre}, Robert and {Plucinsky}, Paul and {Porter}, Frederick and {Pottschmidt}, Katja and {Sato}, Kosuke and {Sato}, Toshiki and {Sawada}, Makoto and {Seta}, Hiromi and {Shidatsu}, Megumi and {Simionescu}, Aurora and {Smith}, Randall and {Suzuki}, Hiromasa and {Szymkowiak}, Andrew and {Takahashi}, Hiromitsu and {Takeo}, Mai and {Tamagawa}, Toru and {Tamura}, Keisuke and {Tanaka}, Takaaki and {Tanimoto}, Atsushi and {Tashiro}, Makoto and {Terada}, Yukikatsu and {Terashima}, Yuichi and {Tsuboi}, Yohko and {Tsujimoto}, Masahiro and {Tsunemi}, Hiroshi and {Tsuru}, Takeshi and {Uchida}, Hiroyuki and {Uchida}, Nagomi and {Uchida}, Yuusuke and {Uchiyama}, Hideki and {Ueda}, Yoshihiro and {Uno}, Shinichiro and {Vink}, Jacco and {Watanabe}, Shin and {Williams}, Brian J. and {Yamada}, Satoshi and {Yamada}, Shinya and {Yamaguchi}, Hiroya and {Yamaoka}, Kazutaka and {Yamasaki}, Noriko and {Yamauchi}, Makoto and {Yamauchi}, Shigeo and {Yaqoob}, Tahir and {Yoneyama}, Tomokage and {Yoshida}, Tessei and {Yukita}, Mihoko and {Zhuravleva}, Irina and {Bartalesi}, Tommaso and {Ettori}, Stefano and {Kosarzycki}, Roman and {Lovisari}, Lorenzo and {Rose}, Tom and {Sarkar}, Arnab and {Sun}, Ming and {Tamhane}, Prathamesh},
        title = "{Constraining gas motion and non-thermal pressure beyond the core of the Abell 2029 galaxy cluster with XRISM}",
      journal = {arXiv e-prints},
         year = 2025,
        month = may,
          eid = {arXiv:2505.06533},
        pages = {arXiv:2505.06533},
          doi = {10.48550/arXiv.2505.06533},
archivePrefix = {arXiv},
       eprint = {2505.06533},
 primaryClass = {astro-ph.CO},
       adsurl = {https://ui.adsabs.harvard.edu/abs/2025arXiv250506533X}
}

@ARTICLE{Fujita2025arXiv_xrismOphiucus,
       author = {{Fujita}, Yutaka and {Fukushima}, Kotaro and {Sato}, Kosuke and {Fukazawa}, Yasushi and {Kondo}, Marie},
        title = "{XRISM Observation of the Ophiuchus Galaxy Cluster: Quiescent Velocity Structure in the Dynamically Disturbed Core}",
      journal = {arXiv e-prints},
         year = 2025,
        month = jun,
          eid = {arXiv:2507.00126},
        pages = {arXiv:2507.00126},
          doi = {10.48550/arXiv.2507.00126},
archivePrefix = {arXiv},
       eprint = {2507.00126},
 primaryClass = {astro-ph.HE},
       adsurl = {https://ui.adsabs.harvard.edu/abs/2025arXiv250700126F}
}

@ARTICLE{Chappuis2025A&A_1689,
       author = {{Chappuis}, L. and {Eckert}, D. and {Sereno}, M. and {Gavidia}, A. and {Sayers}, J. and {Kim}, J. and {Rossetti}, M. and {Umetsu}, K. and {Saxena}, H. and {Bartalucci}, I. and {Gavazzi}, R. and {Rowlands Doblas}, A. and {Pointecouteau}, E. and {Ettori}, S. and {Pratt}, G.~W. and {Bourdin}, H. and {Cassano}, R. and {De Luca}, F. and {De Petris}, M. and {Donahue}, M. and {Gaspari}, M. and {Gastaldello}, F. and {Ghirardini}, V. and {Gitti}, M. and {Maughan}, B. and {Mazzotta}, P. and {Oppizzi}, F. and {Rasia}, E. and {Radovich}, M.},
        title = "{CHEX-MATE: Multiprobe analysis of Abell 1689}",
      journal = {\aap},
         year = 2025,
        month = jul,
       volume = {699},
          eid = {A141},
        pages = {A141},
          doi = {10.1051/0004-6361/202554743},
archivePrefix = {arXiv},
       eprint = {2503.22316},
 primaryClass = {astro-ph.CO},
       adsurl = {https://ui.adsabs.harvard.edu/abs/2025A&A...699A.141C}
}

@ARTICLE{Saxena2025arXiv_triaxial,
       author = {{Saxena}, H. and {Sayers}, J. and {Gavidia}, A. and {Melin}, J.~B. and {Lau}, E.~T. and {Kim}, J. and {Chappuis}, L. and {Eckert}, D. and {Ettori}, S. and {Gaspari}, M. and {Gastaldello}, F. and {Kay}, S. and {Lovisari}, L. and {Oppizzi}, F. and {Petris}, M.~D. and {Pratt}, G.~W. and {Pointecouteau}, E. and {Rasia}, E. and {Rossetti}, M. and {Sereno}, M.},
        title = "{CHEX-MATE: The Impact of Triaxiality and Orientation on Planck SZ Cluster Selection and Weak Lensing Mass Measurements}",
      journal = {arXiv e-prints},
         year = 2025,
        month = may,
          eid = {arXiv:2505.23005},
        pages = {arXiv:2505.23005},
          doi = {10.48550/arXiv.2505.23005},
archivePrefix = {arXiv},
       eprint = {2505.23005},
 primaryClass = {astro-ph.CO},
       adsurl = {https://ui.adsabs.harvard.edu/abs/2025arXiv250523005S}
}

@ARTICLE{Rasia2025arXiv_300fbar,
       author = {{Rasia}, Elena and {Tripodi}, Roberta and {Borgani}, Stefano and {Biffi}, Veronica and {Avestruz}, Camille and {Cui}, Weiguang and {De Petris}, Marco and {Dolag}, Klaus and {Eckert}, Dominique and {Ettori}, Stefano and {Gaspari}, Massimo},
        title = "{The Three Hundred Project: Modeling Baryon and Hot-Gas Fraction Evolution in Simulated Clusters}",
      journal = {arXiv e-prints},
         year = 2025,
        month = may,
          eid = {arXiv:2505.21624},
        pages = {arXiv:2505.21624},
          doi = {10.48550/arXiv.2505.21624},
archivePrefix = {arXiv},
       eprint = {2505.21624},
 primaryClass = {astro-ph.CO},
       adsurl = {https://ui.adsabs.harvard.edu/abs/2025arXiv250521624R}
}

@ARTICLE{Sereno2025A&A_chexmate_masses,
       author = {{Sereno}, Mauro and {Maurogordato}, Sophie and {Cappi}, Alberto and {Barrena}, Rafael and {Benoist}, Christophe and {Haines}, Christopher P. and {Radovich}, Mario and {Nonino}, Mario and {Ettori}, Stefano and {Ferragamo}, Antonio and {Gavazzi}, Rapha{\"e}l and {Huot}, Sophie and {Pizzuti}, Lorenzo and {Pratt}, Gabriel W. and {Streblyanska}, Alina and {Zarattini}, Stefano and {Castignani}, Gianluca and {Eckert}, Dominique and {Gastaldello}, Fabio and {Kay}, Scott T. and {Lovisari}, Lorenzo and {Maughan}, Ben J. and {Pointecouteau}, Etienne and {Rasia}, Elena and {Rossetti}, Mariachiara and {Sayers}, Jack},
        title = "{CHEX-MATE: Dynamical masses for a sample of 101 Planck Sunyaev-Zeldovich-selected galaxy clusters}",
      journal = {\aap},
         year = 2025,
        month = jan,
       volume = {693},
          eid = {A2},
        pages = {A2},
          doi = {10.1051/0004-6361/202451610},
archivePrefix = {arXiv},
       eprint = {2410.18165},
 primaryClass = {astro-ph.CO},
       adsurl = {https://ui.adsabs.harvard.edu/abs/2025A&A...693A...2S}
}

@article{giocoli2025_300,
      title={The Three Hundred Project hydrodynamical simulations: Hydrodynamical Weak-Lensing Cluster Mass Biases and Richnesses using different hydro models}, 
      author={Carlo Giocoli and Giulia Despali and Massimo Meneghetti and Elena Rasia and Lauro Moscardini and Stefano Borgani and Giorgio. F. Lesci and Federico Marulli and Weiguang Cui and Gustavo Yepes},
      year={2025},
      eprint={2501.14019},
      archivePrefix={arXiv},
      primaryClass={astro-ph.CO},
      url={https://arxiv.org/abs/2501.14019}, 
}

@ARTICLE{Lesci2025arXiv250714285L,
       author = {{Lesci}, G.~F. and {Marulli}, F. and {Moscardini}, L. and {Maturi}, M. and {Sereno}, M. and {Radovich}, M. and {Romanello}, M. and {Giocoli}, C. and {Wright}, A.~H. and {Bardelli}, S. and {Bilicki}, M. and {Castignani}, G. and {Hildebrandt}, H. and {Kannawadi}, A. and {Ingoglia}, L. and {Joudaki}, S. and {Puddu}, E.},
        title = "{AMICO galaxy clusters in KiDS-1000: cosmological constraints and mass calibration from counts and weak lensing}",
      journal = {arXiv e-prints},
         year = 2025,
        month = jul,
          eid = {arXiv:2507.14285},
        pages = {arXiv:2507.14285},
archivePrefix = {arXiv},
       eprint = {2507.14285},
 primaryClass = {astro-ph.CO},
       adsurl = {https://ui.adsabs.harvard.edu/abs/2025arXiv250714285L}
}

@ARTICLE{Dolag2025arXiv_magneticum,
       author = {{Dolag}, Klaus and {Remus}, Rhea-Silvia and {Valenzuela}, Lucas M. and {Kimmig}, Lucas C. and {Seidel}, Benjamin and {Fortune}, Silvio and {Stoiber}, Johannes and {Ivleva}, Anna and {Hoffmann}, Tadziu and {Biffi}, Veronica and {Marini}, Ilaria and {Popesso}, Paola and {Vladutescu-Zopp}, Stephan},
        title = "{Encyclopedia Magneticum: Scaling Relations from Cosmic Dawn to Present Day}",
      journal = {arXiv e-prints},
         year = 2025,
        month = apr,
          eid = {arXiv:2504.01061},
        pages = {arXiv:2504.01061},
          doi = {10.48550/arXiv.2504.01061},
archivePrefix = {arXiv},
       eprint = {2504.01061},
 primaryClass = {astro-ph.CO},
       adsurl = {https://ui.adsabs.harvard.edu/abs/2025arXiv250401061D}
}

@ARTICLE{gardini.etal.2004,
       author = {{Gardini}, A. and {Rasia}, E. and {Mazzotta}, P. and {Tormen}, G. and {De Grandi}, S. and {Moscardini}, L.},
        title = "{Simulating Chandra observations of galaxy clusters}",
      journal = {\mnras},
         year = 2004,
        month = jun,
       volume = {351},
       number = {2},
        pages = {505-514},
          doi = {10.1111/j.1365-2966.2004.07800.x},
archivePrefix = {arXiv},
       eprint = {astro-ph/0310844},
 primaryClass = {astro-ph},
       adsurl = {https://ui.adsabs.harvard.edu/abs/2004MNRAS.351..505G}
}

@ARTICLE{Miren2025_pressureprof,
       author = {{Mu{\~n}oz-Echeverr{\'\i}a}, M. and {Pointecouteau}, E. and {Pratt}, G.~W. and {Mac{\'\i}as-P{\'e}rez}, J.-F. and {Douspis}, M. and {Salvati}, L. and {Bartalucci}, I. and {Bourdin}, H. and {Clerc}, N. and {De Luca}, F. and {De Petris}, M. and {Donahue}, M. and {Dupourqu{\'e}}, S. and {Eckert}, D. and {Ettori}, S. and {Gaspari}, M. and {Gastaldello}, F. and {Gitti}, M. and {Gorce}, A. and {Ili{\'c}}, S. and {Kay}, S.~T. and {Kim}, J. and {Lovisari}, L. and {Maughan}, B.~J. and {Mazzotta}, P. and {McBride}, L. and {Melin}, J.-B. and {Oppizzi}, F. and {Rasia}, E. and {Rossetti}, M. and {Saxena}, H. and {Sayers}, J. and {Sereno}, M. and {Tristram}, M.},
        title = "{CHEX-MATE: towards a consistent universal pressure profile and cluster mass reconstruction}",
      journal = {arXiv e-prints},
         year = 2025,
        month = oct,
          eid = {arXiv:2510.18578},
        pages = {arXiv:2510.18578},
          doi = {10.48550/arXiv.2510.18578},
archivePrefix = {arXiv},
       eprint = {2510.18578},
 primaryClass = {astro-ph.CO},
       adsurl = {https://ui.adsabs.harvard.edu/abs/2025arXiv251018578M}
}

\appendix

\section{Gas density and gas mass}
\label{appendix:gas_mass}

We further investigate the reconstruction of gas density profiles and gas mass in this Appendix. Fig. \ref{fig:ne_model_300} shows the gas density reconstruction for one of \texttt{The300} clusters. The true profile is shown in black, the reconstructed ones, including the voronoi binning technique, are shown in light blue and red. The bottom panel shows the ratio to the true profile, highlighting the better precision of the profile reconstructed from the voronoi binned image, with local deviations of at most 5$\%$.

A detailed comparison is presented in Fig. \ref{fig:ne_profiles}, where each of the three main panels corresponds to one simulation set. In each case, the top panel displays the gas density profiles reconstructed using the NFW, NP, and FM methods. The middle panel shows the ratio between the azimuthal median reconstructed and true profiles for each method, as derived from the voronoi-binned images.  The bottom panel presents the ratio between the profiles measured directly from the mock EPIC images and the true profiles. Overall, we find excellent agreement among the density profiles recovered with the three methods. The median surface brightness profiles extracted from the voronoi images closely follow the one-to-one relation with the true profiles. In contrast, the direct analysis of the mock EPIC images tends to overestimate the gas density by approximately 5$\%$. This result confirms that the voronoi-based approach enables an unbiased recovery of the gas density profile, owing to its reduced sensitivity to surface brightness fluctuations caused by substructures or colder gas clumps whose emission is not prominent enough to be masked during preprocessing. The profiles in \texttt{Magneticum} tend to have lower normalisation and a flatter slope. This is expected given the slightly different mass selection (see Sect. \ref{subsec:clusample}), because more massive clusters typically show steeper profiles \citep[see e.g.][]{Croston2008A&A_gasProfs_rexcess, Pratt2022A&A_density_profs}. We conclude that despite the intrinsic differences between individual simulations, our methods are able to properly reconstruct the gas density profiles independently of their specific shape.

We estimate the total gas mass content in our mock clusters and compare it to the true value in the hydrodynamical simulations.
The gas mass is computed by integrating the electron number density, following:
\begin{equation}
    M_{\rm gas} = 4\pi \mu_e m_p \int_0 ^{\rm R_{\rm 500c}} n_e(r)r^2 dr,
\end{equation}
where $\mu_e$ is the mean electron molecular weight and $m_p$ is the proton mass. In particular, we convert the gas particle number density to total gas mass in units of solar masses by computing the mean molecular weight per electron in a flexible way based on the chosen abundance table with the latest version of \texttt{hydromass}. For consistency with the generated boxes (see Sect. \ref{sec:mock_gen}) we use the abundances from \citet{Asplund2009ARA&A..47..481A}, which gives $\mu_e=1.146$.
%\begin{align}
%    1 M_\odot &= 1.988\times10^{33} g \nonumber \\
%    \mu_e &= 1.15 \nonumber \\
%    m_p & = 1.67 \times10^{-24} g \nonumber \\
%    1 kpc &= 3.086\times10^{21} cm \nonumber \\
%    n_e [cm^{-3}] &= \rho_{\rm gas}[M_\odot /kpc^3] \frac{1 M_\odot}{\mu_e m_p \times 1 \text{kpc}^3}. 
%\end{align}
Fig. \ref{fig:Mgas} shows the comparison between the measured and true total gas mass within the true R$_{\rm 500c}$ in all three simulations. The panel refers to the gas mass estimated from the integral of the voronoi azimuthal median (mean surface brightness) density profile. When combining the three simulation, the median ratio between measured and true gas mass is 0.993, with 16th and 84th percentile points at 0.916, and 1.067, meaning that the gas mass reconstruction is precise within 1$\%$ and a cluster to cluster scatter variation of about 7.5$\%$. We run a linear regression algorithm using \texttt{scipy} on the logarithm of the gas masses, accounting for a zero point of 14.0 on both axis. Overall, we find excellent agreement with the one to one relation. The best fit slope is closer to one especially for \texttt{The300} and \texttt{Magneticum} using the median voronoi profiles. The fit is less accurate for \texttt{MACSIS}, which lacking the Tier1 mass objects has a limited mass range. We find the residuals to scatter around the mean relation by 0.027 (0.021, 0.022) dex in \texttt{The300} (\texttt{Magneticum}, \texttt{MACSIS}). The gas mass is overestimated by about 5$\%$ using the standard count rate images, this is expected as we also showed that the gas density is overestimated at similar levels in the previous section. We report all parameters in Table \ref{tab:gasmass_pars}. We conclude that our methods are able to properly reconstruct the total gas mass in CHEX-MATE-like cluster samples.

\begin{figure}
    \centering
    \includegraphics[width=\columnwidth]{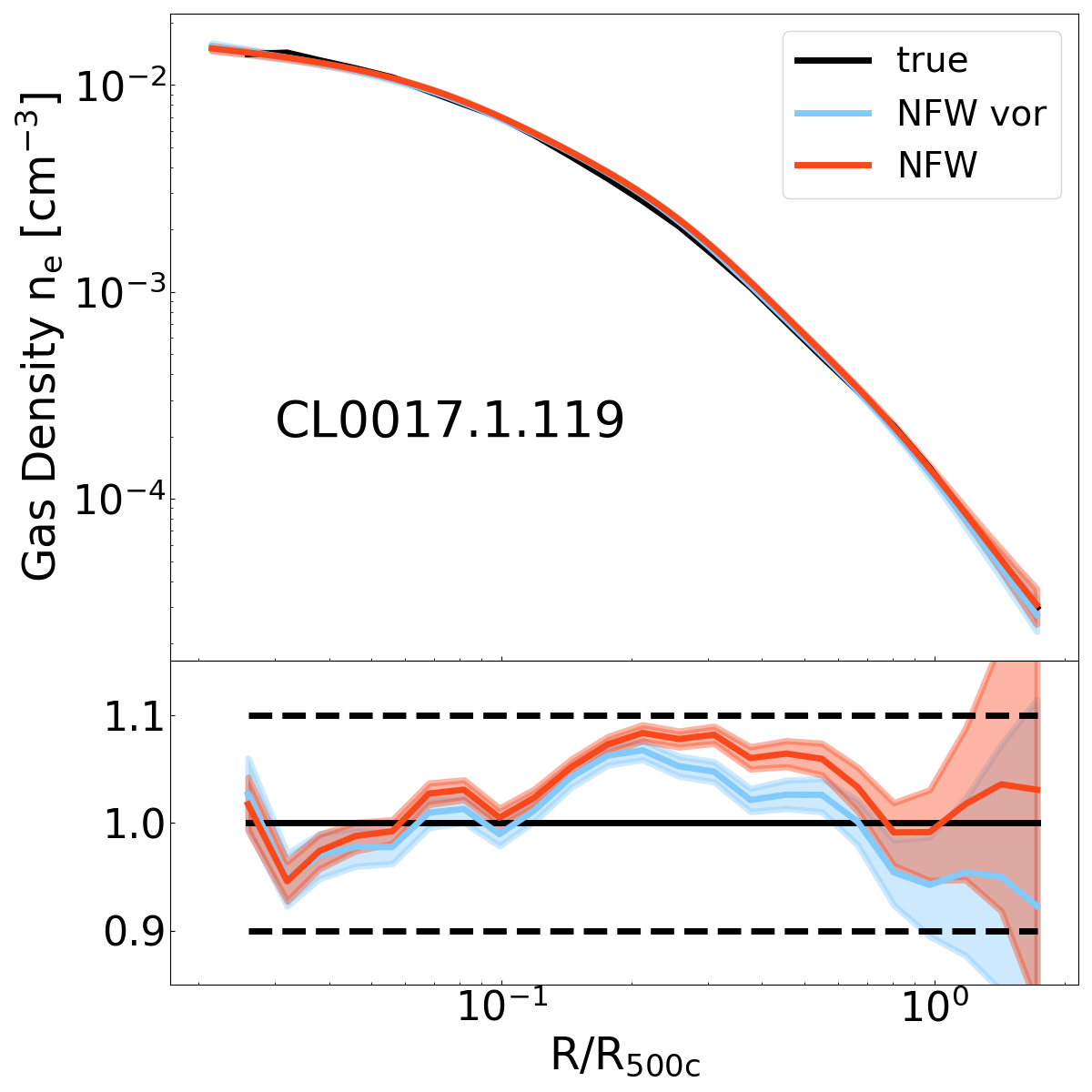}   
    \caption{Density profile reconstruction of CL0017.1.119 from \texttt{The300} simulation.}
    \label{fig:ne_model_300}
\end{figure}

\begin{figure*}
    \centering
    \includegraphics[width=0.66\columnwidth]{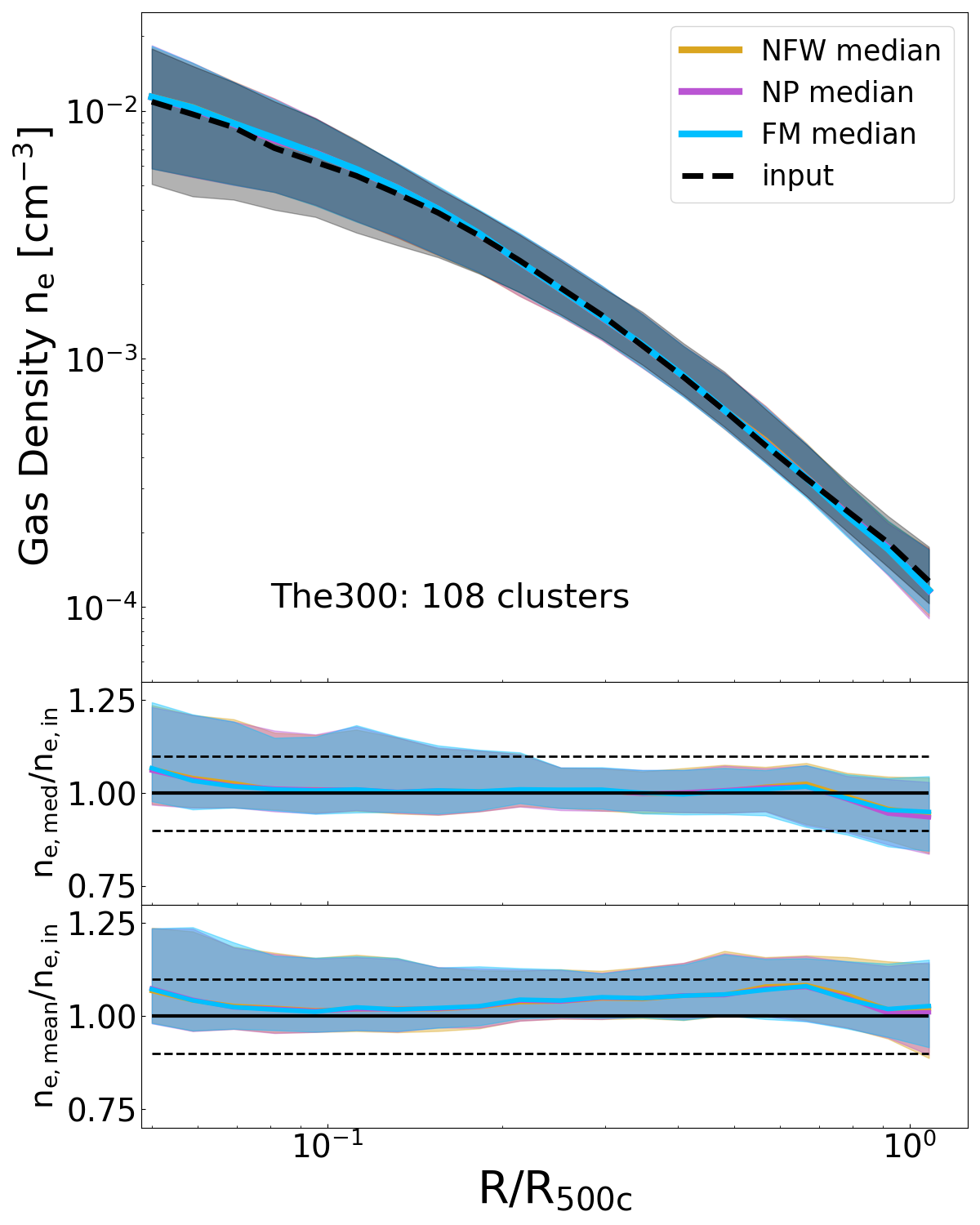}
    \includegraphics[width=0.66\columnwidth]{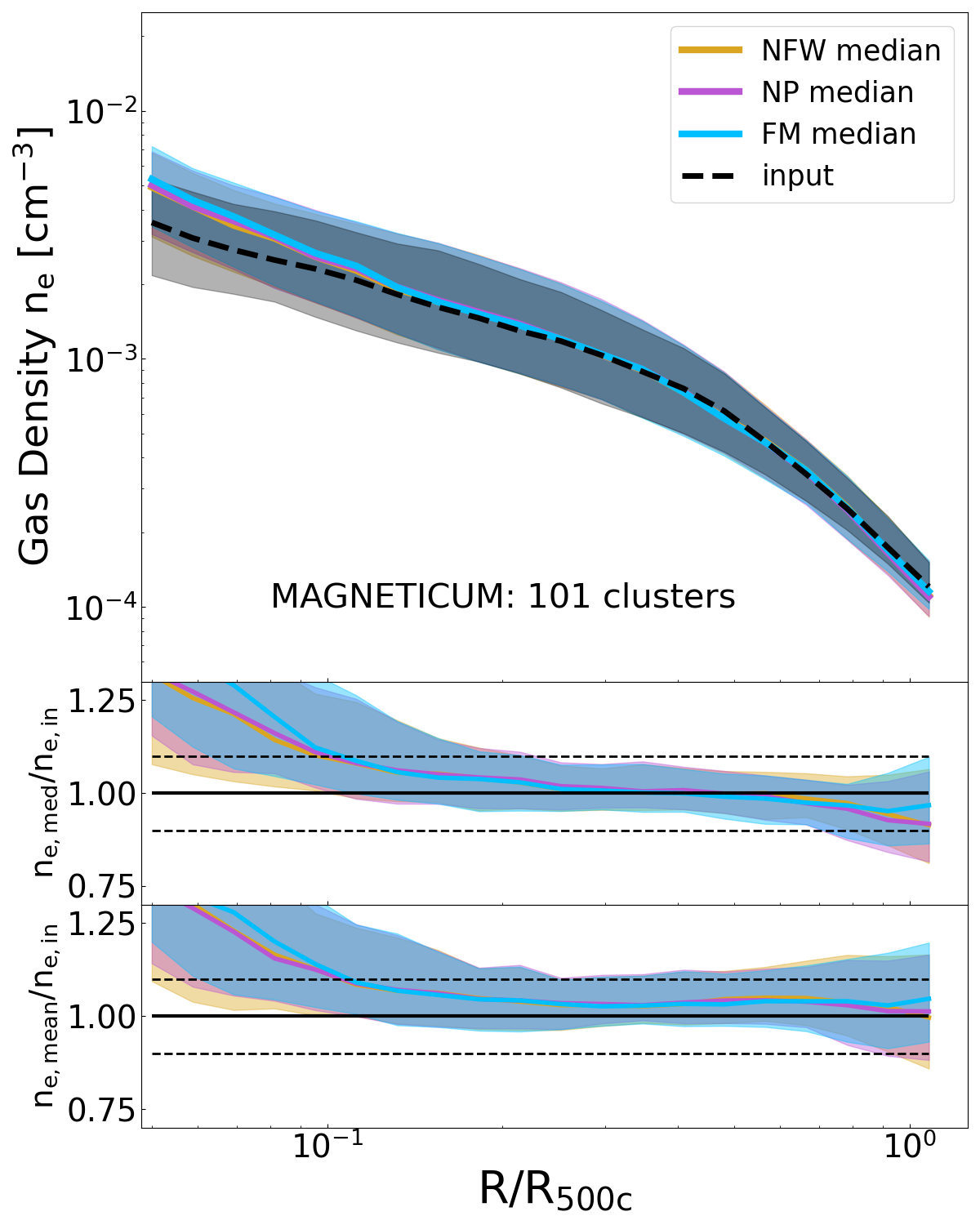}    
    \includegraphics[width=0.66\columnwidth]{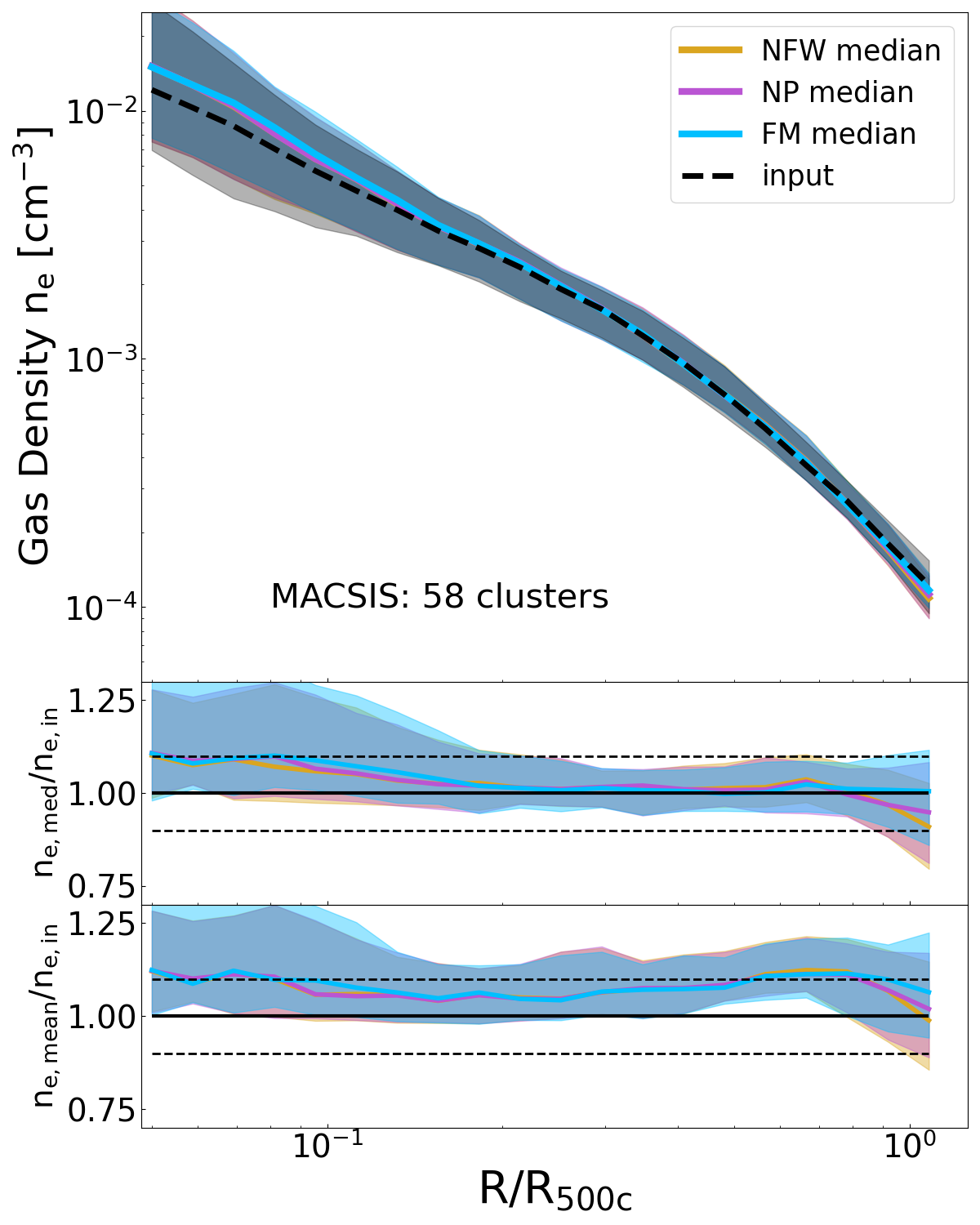}
  
    \caption{Deprojected 3D gas density profiles. Each panel corresponds to one simulation: \texttt{The300} on the left, \texttt{Magneticum} in the centre, and \texttt{MACSIS} on the right. The colours denote the three reconstruction methods. The smaller central (bottom) panels show the ratio between the median voronoi (the standard mean) profile and the true input one.}
    \label{fig:ne_profiles}
\end{figure*}

\begin{table}[]
    \centering
    \caption{Best fit beta model parameters of the gas density profiles in our simulations.}
    \begin{tabular}{|c|c|c|}
        \hline
        \hline
        \textbf{Simulation}   & \textbf{$\beta$} & \textbf{r$_c$} [R$_{\rm 500c}$] \\
        \hline
        \rule{0pt}{2ex}  
        \texttt{The300}  & 0.63$\pm$0.03 & 0.10$\pm$0.02 \\
        \texttt{The300} true & 0.67$\pm$0.03 & 0.13$\pm$0.02 \\
        \texttt{Magneticum} & 0.66$\pm$0.03 & 0.29$\pm$0.02 \\
        \texttt{Magneticum} true & 0.69$\pm$0.03 & 0.33$\pm$0.02 \\
        \texttt{MACSIS} & 0.70$\pm$0.03 & 0.20$\pm$0.02 \\
        \texttt{MACSIS} true & 0.72$\pm$0.03 & 0.21$\pm$0.02 \\
        \hline
    \end{tabular}
    \label{tab:betamodel_pars}
\end{table}

\begin{figure}
    \centering
    \includegraphics[width=\columnwidth]{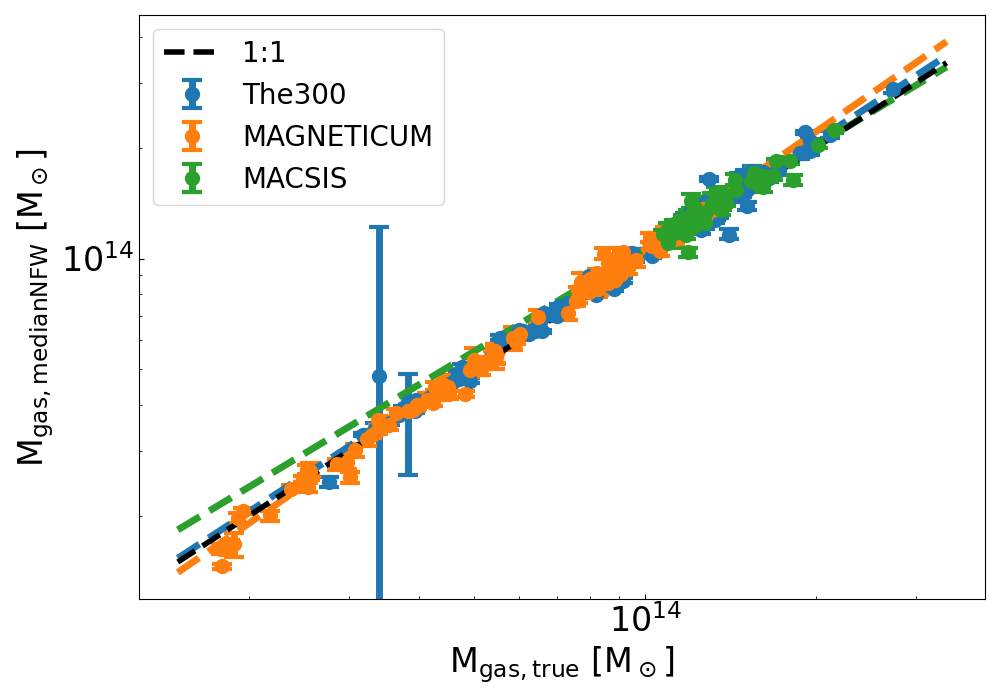}

    \caption{Comparison between the measured and true gas mass within R$_{\rm 500c}$. The panel refers to the masses inferred from the voronoi tessellated images. The dashed coloured lines denote the best linear fit to each simulation.}
    \label{fig:Mgas}
\end{figure}

\begin{table}[]
    \centering
    \caption{Best linear fit parameters of the gas mass recovery in our simulations.}
    \begin{tabular}{|c|c|c|}
        \hline
        \hline
        \textbf{Simulation}   & \textbf{Slope} & \textbf{Intercept} \\
        \hline
        \rule{0pt}{2ex}  
        \texttt{The300} median  & 1.01$\pm$0.01 & 0.01$\pm$0.01 \\
        \texttt{The300} mean & 1.02$\pm$0.02 & 0.04$\pm$0.01 \\
        \texttt{Magneticum} median & 1.05$\pm$0.02 & 0.01$\pm$0.01 \\
        \texttt{Magneticum} mean & 1.07$\pm$0.02 & 0.03$\pm$0.02 \\
        \texttt{MACSIS} median & 0.94$\pm$0.05 & 0.05$\pm$0.01 \\
        \texttt{MACSIS} mean & 0.96$\pm$0.07 & 0.06$\pm$0.01 \\
        \hline
    \end{tabular}
    \vskip.1cm
    \footnotesize{\textbf{Notes.} The relation is fitted on the base 10 logarithm of the gas masses, and the mass on the x-axis was normalised by 10$^{14}$ to have a better handle on the correlation between slope and intercept. }
    \label{tab:gasmass_pars}
\end{table}

\section{Thermodynamical profiles}
\label{appendix:K_P_profs}

\begin{figure}
    \centering
    \includegraphics[width=\columnwidth]{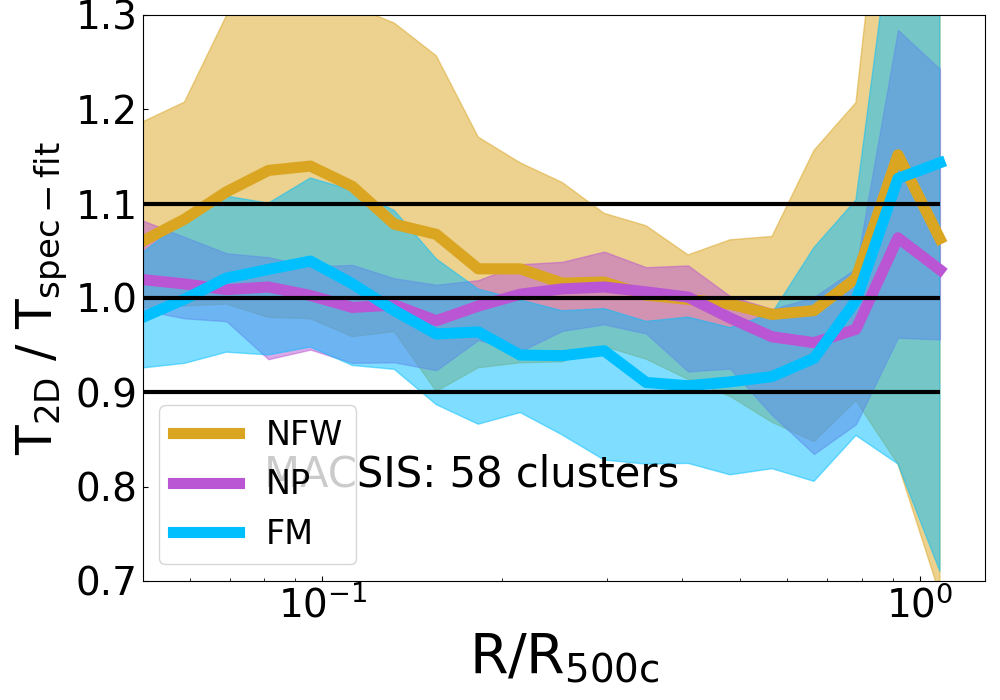}  
    \caption{Ratio between the reconstructed 2D temperature profile with NFW, NP, and FM models to the result of the X-ray spectral fitting in \texttt{MACSIS}.}
    \label{fig:Tspecfit_Tsl_macsis}
\end{figure}

\begin{figure}
    \centering
    \includegraphics[width=\columnwidth]{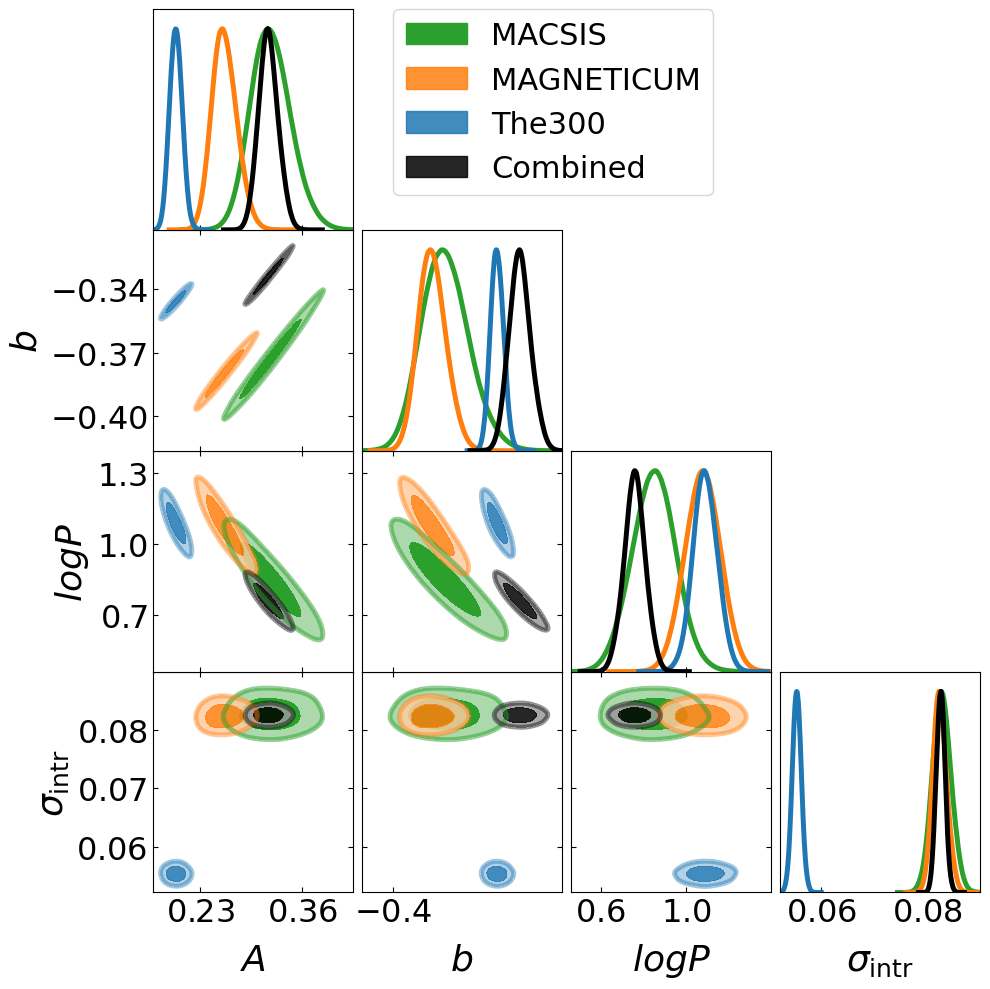}   
    \caption{Corner plot with the marginalised 1d posterior distribution of the parameters describing the ratio between spectroscopic-like and mass weighted temperature as a function of median and width of the temperature fluctuation distribution.}
    \label{fig:cornerTSLTMW_sigma_mu}
\end{figure}

In this Appendix we collect additional results about X-ray observables, starting with temperature. Fig. \ref{fig:Tspecfit_Tsl_macsis} is about the temperature profile reconstruction in \texttt{MACSIS} (Sect. \ref{subsec:Temperature_profiles}): it is the ratio between the model of the temperature profile and the measured one, for different reconstruction methods. It shows how the modelling is able to reproduce the measurement, and confirms that the temperature bias in the core of \texttt{MACSIS} (Fig. \ref{fig:Tx_profiles}) is due to the measurement (see also Fig. \ref{fig:Tspecfit_Tsl}) and not due to the modelling itself.
Fig. \ref{fig:cornerTSLTMW_sigma_mu} is about corner plot showing the best fit parameters of the model describing the ratio between mass weighted and spectroscopic-like temperatures in Sect. \ref{subsec:multiT} (see Eq. \ref{eq:Tratio_model}).

We then focus on derived observables, i.e. the ones that are not directly measured on the X-ray data, such as surface brightness and temperature. We focus on gas pressure and entropy. They are respectively reported in Fig. \ref{fig:Ptot_profiles} and \ref{fig:K_profiles}. 

\begin{figure*}
    \centering
    \includegraphics[width=0.66\columnwidth]{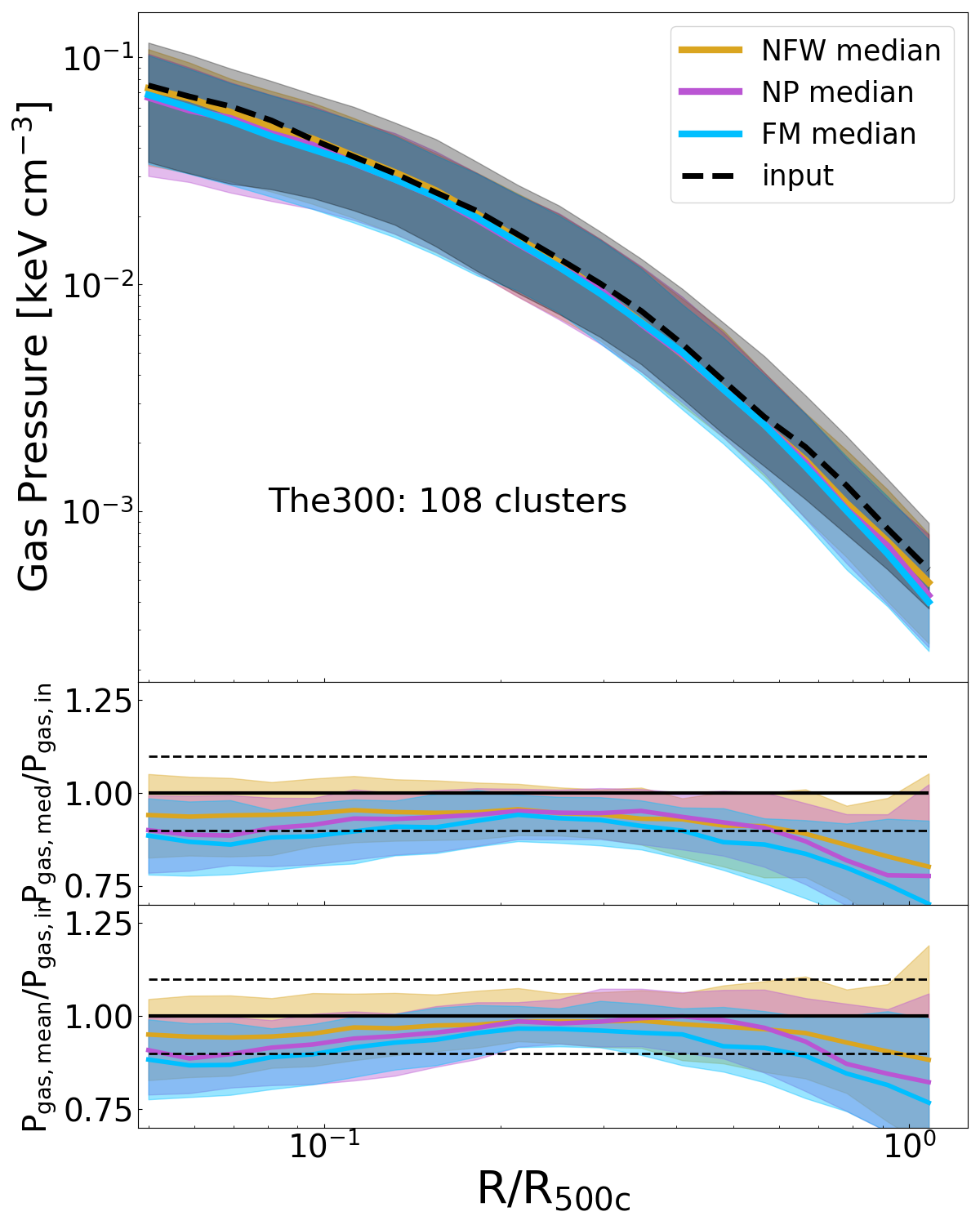}
    \includegraphics[width=0.66\columnwidth]{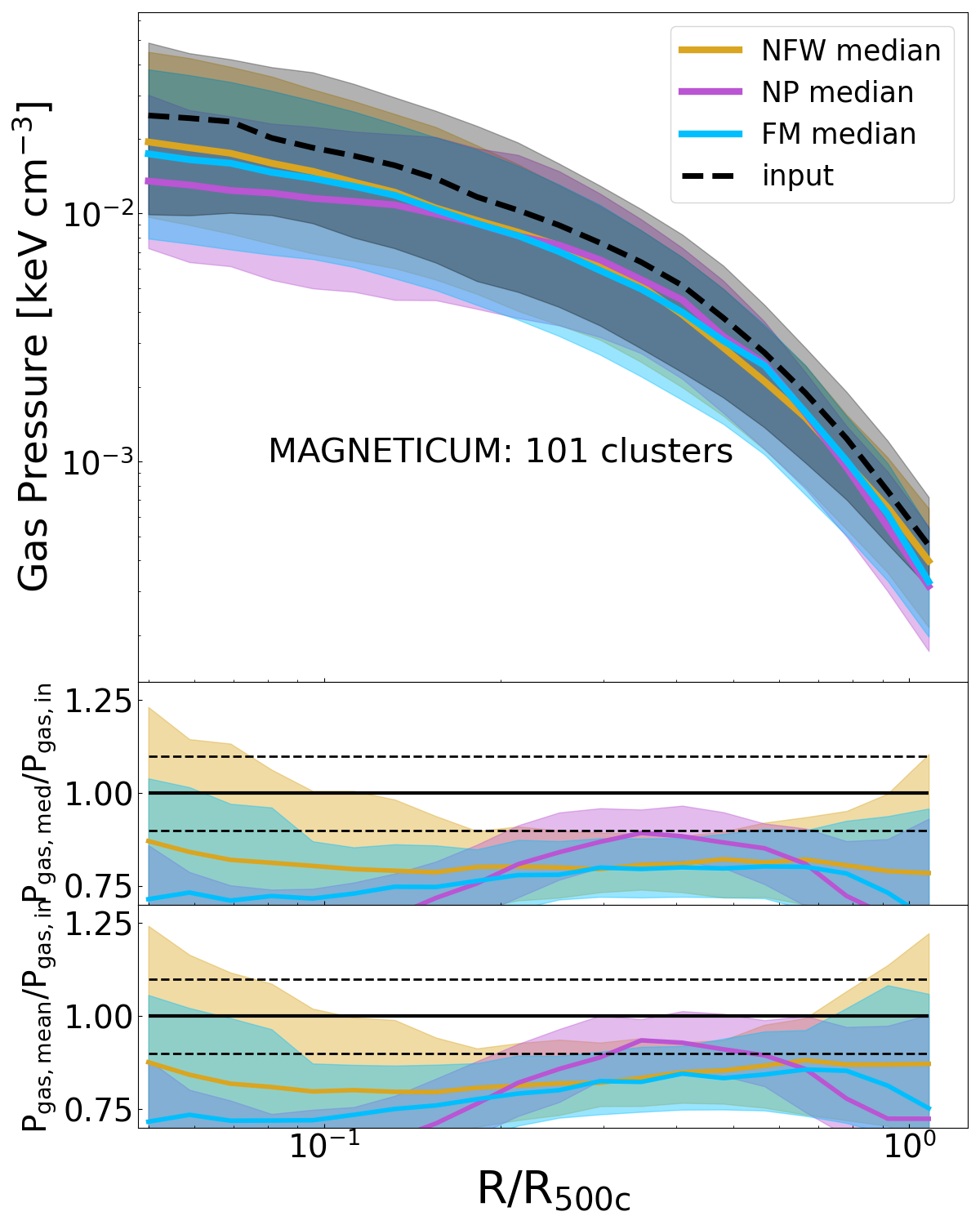}    
    \includegraphics[width=0.66\columnwidth]{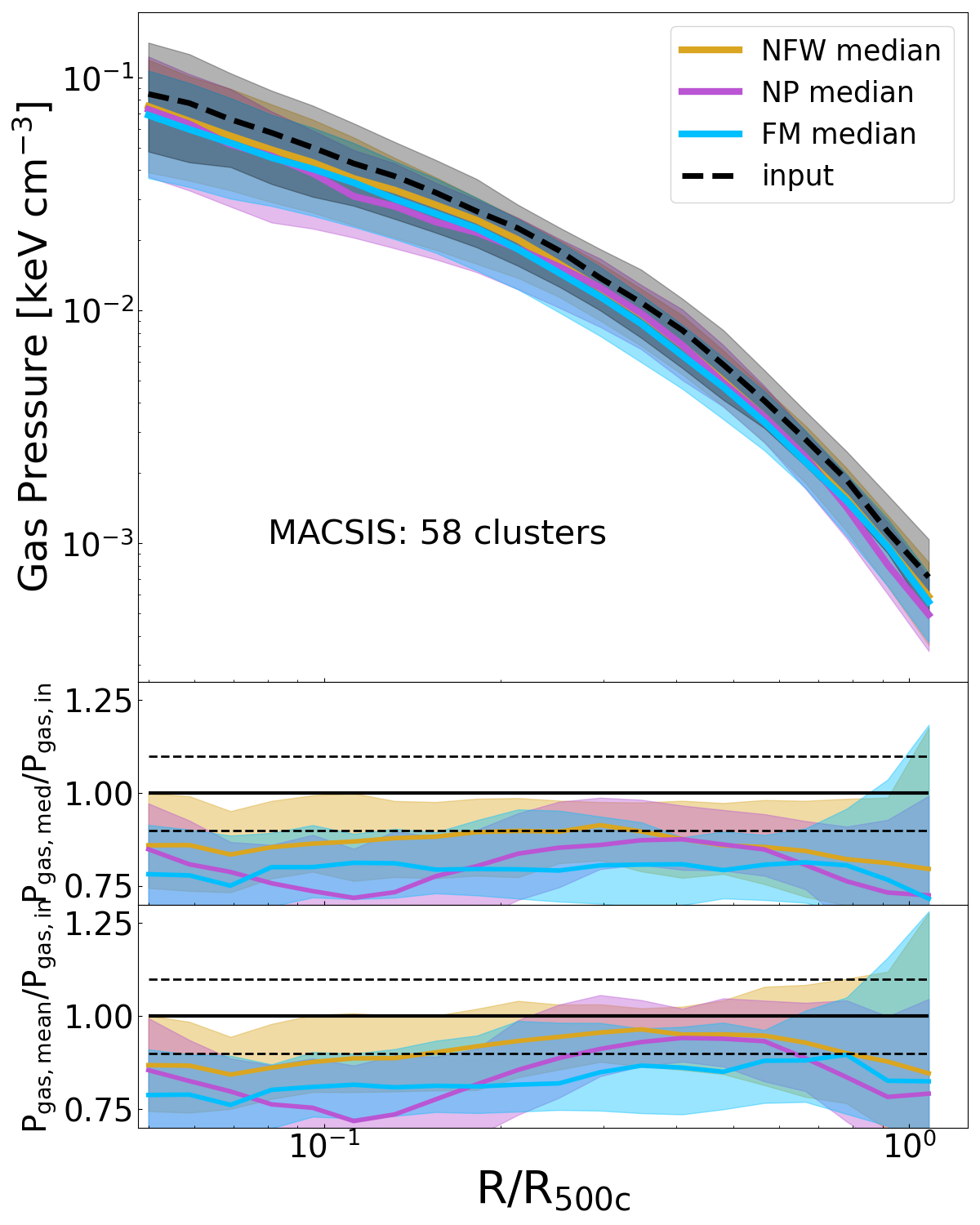}
  
    \caption{Deprojected 3D electron pressure profiles. Each panel corresponds to one simulation: \texttt{The300} on the left, \texttt{Magneticum} in the centre, and \texttt{MACSIS} on the right. The colours denote the three reconstruction methods. The smaller central (bottom) panels show the ratio between the median voronoi (the standard mean) profile and the true input one.}
    \label{fig:Ptot_profiles}
\end{figure*}

\begin{figure*}
    \centering
    \includegraphics[width=0.66\columnwidth]{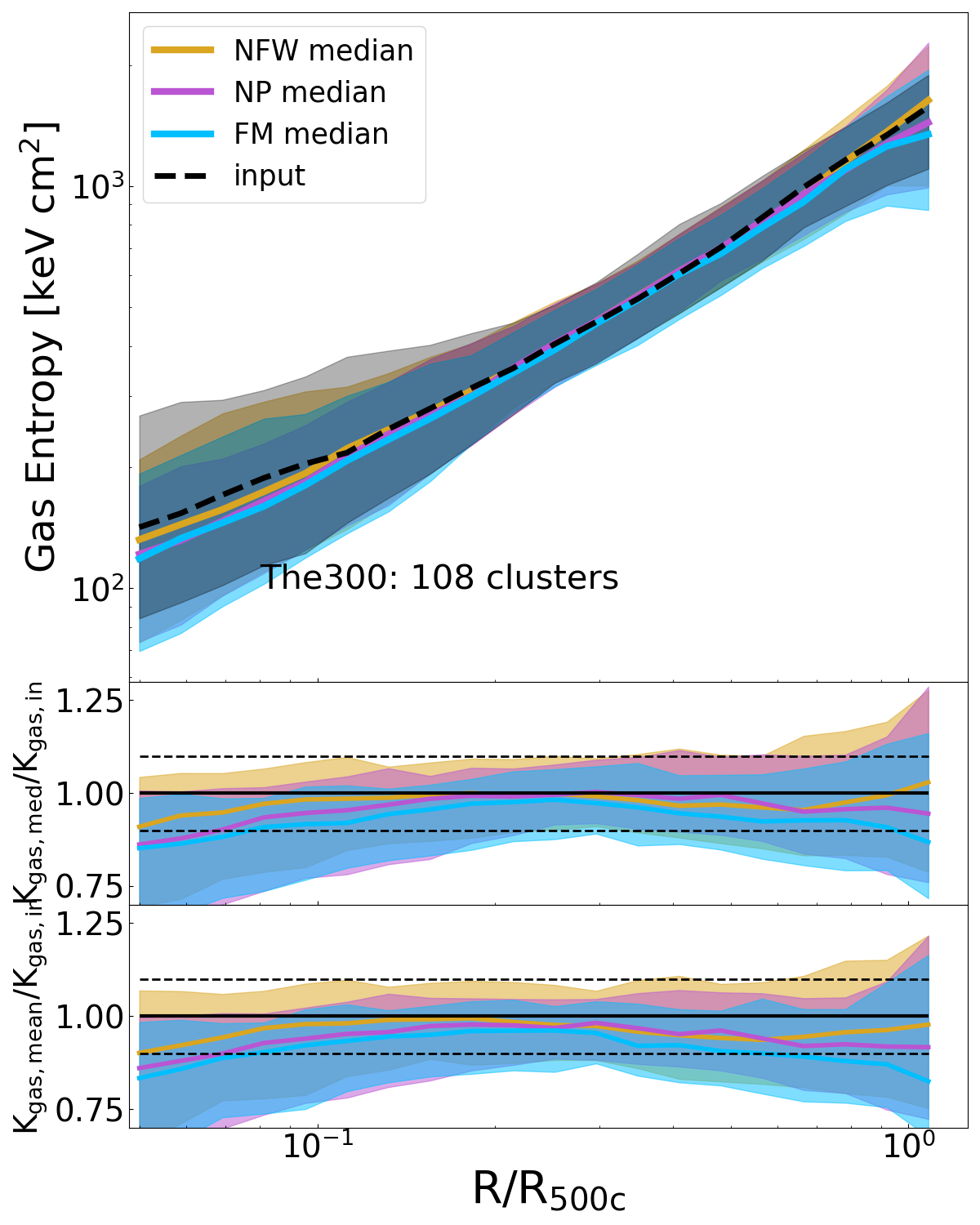}
    \includegraphics[width=0.66\columnwidth]{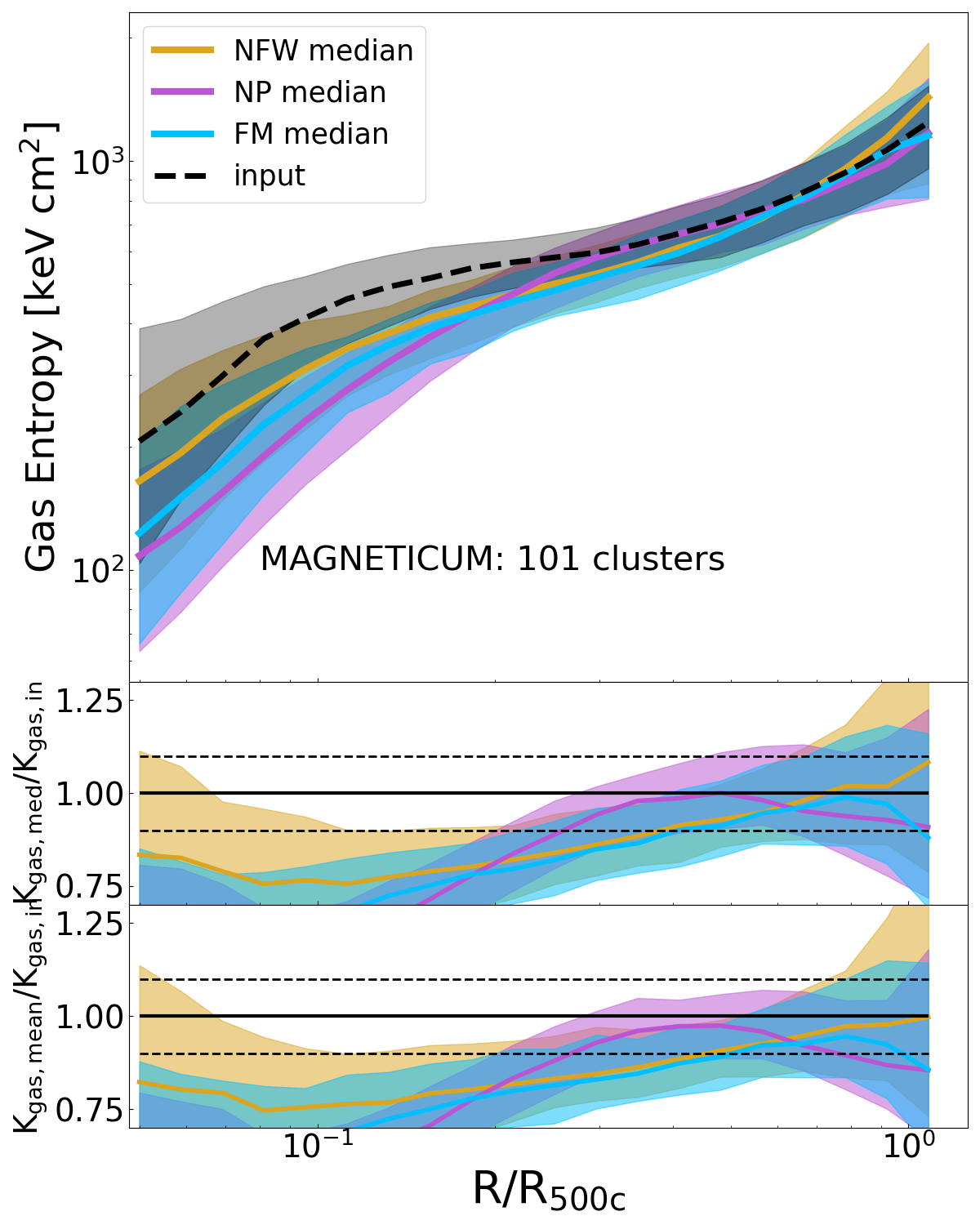}    
    \includegraphics[width=0.66\columnwidth]{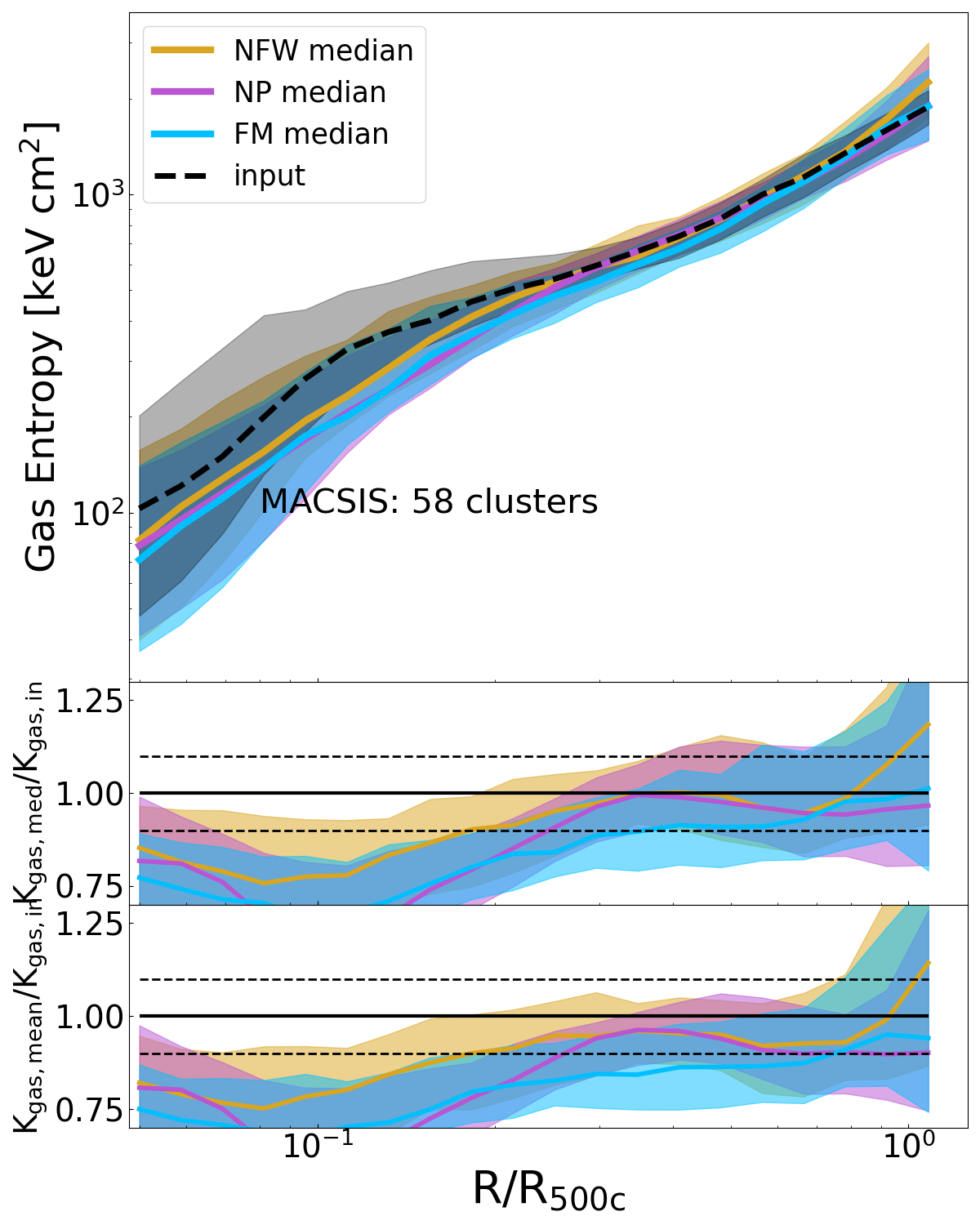}
  
    \caption{Deprojected 3D gas entropy profiles. Each panel corresponds to one simulation: \texttt{The300} on the left, \texttt{Magneticum} in the centre, and \texttt{MACSIS} on the right. The colours denote the three reconstruction methods. The smaller central (bottom) panels show the ratio between the median voronoi (the standard mean) profile and the true input one.}
    \label{fig:K_profiles}
\end{figure*}

We find that both the thermal pressure and entropy profiles recovered from the X-ray analysis are systematically underestimated relative to the true values from the simulations. Among the three simulation suites, \texttt{The300} exhibits the smallest discrepancies: the thermal pressure is underestimated by approximately 10$\%$ across the entire radial range, while the entropy profile closely matches the true values beyond 0.2$\times$R$\_{\rm 500c}$ and is underestimated by about 10$\%$ in the core. Both quantities remain consistent with the one-to-one relation within 1$\sigma$ uncertainties.
In \texttt{Magneticum}, the discrepancies are more pronounced. Thermal pressure is systematically underestimated by about 20$\%$ across the full profile when using the NFW and FM reconstruction methods, while the NP approach shows a smaller discrepancy (about 10$\%$) around 0.4R$_{\rm 500c}$ but exhibits even larger deviations in both the core and outskirts. A similar pattern is observed for entropy, although the reconstructed values converge more closely to the true values near R$_{\rm 500c}$.
In \texttt{MACSIS}, the thermal pressure profile is moderately biased low by about 15$\%$, while the entropy profile shows a level of disagreement similar to that seen in \texttt{Magneticum}. However, the three reconstruction methods yield more consistent results.
In summary, the degree of agreement between the inferred and true thermodynamic profiles varies across simulations and partially depends on the reconstruction method employed. These results highlight the sensitivity of X-ray-derived pressure and entropy profiles to both physical modelling in simulations and methodological assumptions in the analysis pipeline. Understanding the origin of these discrepancies is essential, as both quantities play a central role in characterising the thermodynamic state of the ICM and in deriving cluster masses under the assumption of hydrostatic equilibrium.

The thermal pressure profile is proportional to the electron number density and the gas temperature, through the Boltzmann constant under the assumption of the ideal gas law. Since our density profiles are accurately recovered, as shown in Section \ref{sec:results}, the pressure bias is more likely to be driven by either a bias in the reconstructed temperature, or by simplistic assumptions in the whole deprojection strategy and modelling. The hydrostatic mass modelling used in the temperature reconstruction assumes that the ICM is in equilibrium with the gravitational potential, neglecting any non-thermal pressure support from turbulence, bulk motions, or cosmic rays. If non-thermal pressure contributes significantly to the total pressure budget, the inferred temperature profile from a purely hydrostatic model will be suppressed to compensate, leading to an underestimation of thermal pressure. However, in this case the input pressure is only thermal, i.e. the one computed from the ideal gas law starting from density and temperature. Therefore, any inconsistency in the recovery of thermal pressure is not due to the non-thermal component in this work. In addition, the assumption of spherical symmetry in the hydrostatic analysis may not hold in detail: triaxiality of the halo and projection along preferential axes can distort both the density and temperature gradient estimates and thus affect the pressure reconstruction.

Entropy is directly (inversely) proportional to temperature (density to the two-thirds power). It is a derived quantity that inherits biases from both temperature and density. Entropy is particularly sensitive to the thermal history of the ICM and is therefore more susceptible to localised structures, cooling clumps, or feedback-induced perturbations that may not be properly accounted for in the deprojection process under the assumption of single-temperature ICM within the framework of hydrostatic mass modelling.

Future work will focus on dissecting these contributions in more detail by comparing hydrostatic mass reconstructions to the true simulation masses, investigating the impact of different temperature reconstruction schemes, and exploring projection effects due to triaxiality and substructure. Such an analysis is essential to calibrate cluster thermodynamics for cosmological applications and to refine our interpretation of the hydrostatic mass bias.

\section{Multi-T gas: a toy model}
\label{appendix:toy_model}
\begin{figure}
    \centering
    \includegraphics[width=\columnwidth]{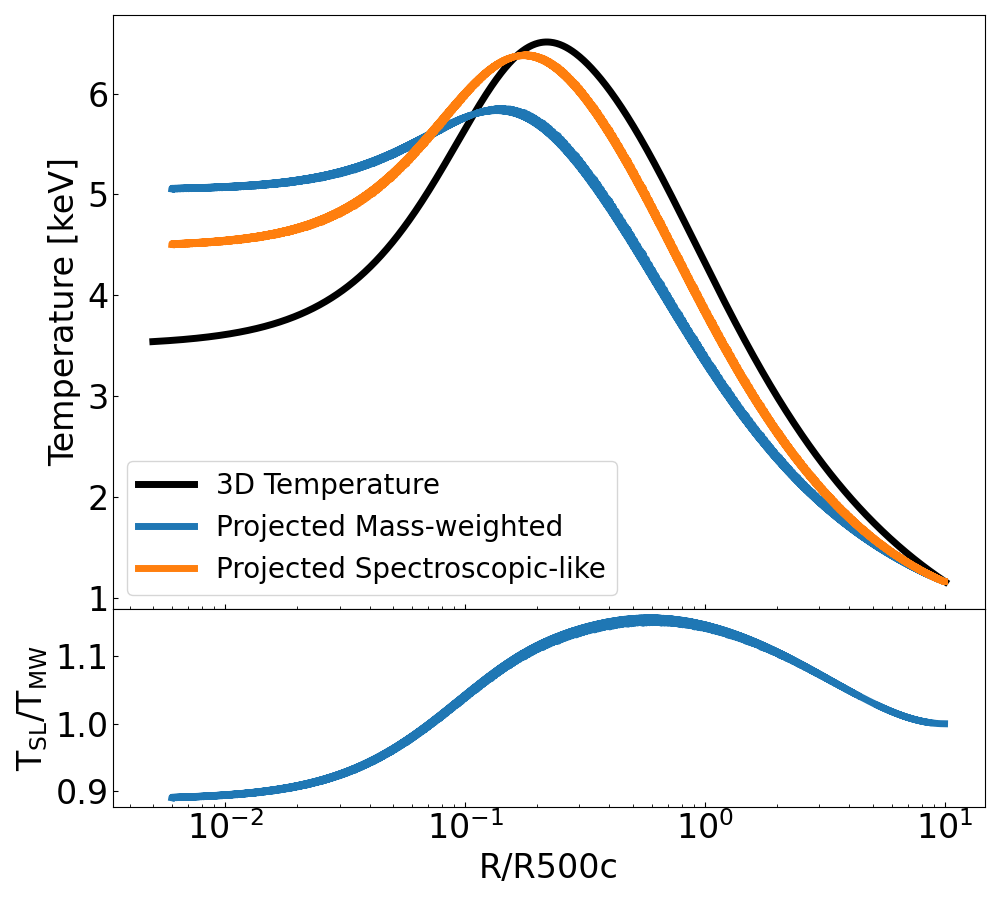}
    \caption{Spherical isothermal toy model of the gas temperature profile, including its projections along the line of sight assuming the spectroscopic-like and mass-weighted temperature schemes.}
    \label{fig:Tx_toymodel}
\end{figure}

To explore projection effects on the cluster temperature profile in a controlled setting, we construct a toy model under the assumption of a spherically symmetric, single-temperature structure. This does not imply that the temperature profile is flat, but rather that the gas follows a single, smooth, and radially dependent profile without azimuthal or multi-phase complexity. We then project this 3D temperature distribution along the line of sight using both the SL and MW weighting schemes (see Eq. \ref{eq:SL_MW}). The gas density profile is modelled with a standard $\beta$-model assuming $\beta = 2/3$ and a core radius of $r_c = 0.1\times R_{\rm 500c}$.  Our toy model includes a cool core and an outer decline and is parametrised by:
\begin{equation}
    T(r) = T_0 \times (1-A\exp[-(r/r_{c1})^{1.5}]) \times (1+(r/r_{c2})^2)^{-0.3},
    \label{eq:Tx_toymodel}
\end{equation}
with $T_0$=7 keV, A=0.5, $r_{c1}$=0.1, and $r_{c2}$=0.5. The projection is performed by looping over the radial grid, computing the line-of-sight depth corresponding to each projected radius, and integrating the temperature profile in Eq. \ref{eq:Tx_toymodel} with the appropriate SL or MW weights within each shell, so that the difference between the weighting schemes only applies to the projection.

The result is shown in Fig. \ref{fig:Tx_toymodel}. We find that in the central region within 0.1$\times$R$_{\rm 500c}$, the MW projection yields higher temperatures than the SL projection. This behaviour arises because the SL weighting scheme gives more emphasis to lower-temperature gas, which is prevalent in the core, thereby biasing the projected temperature downward. Moving toward the outskirts, the density profile steepens and the line-of-sight integration length shortens, reducing the total amount of gas contributing to the projection. In this regime, the SL weighting becomes increasingly sensitive to the denser, hotter gas located in the inner portion of the shell due to its scaling with gas density squared. In contrast, the MW scheme continues to weight gas linearly by mass, giving more equal importance across the shell. As a result, the SL temperature can exceed the MW value beyond the core.
At large radii beyond 3$\times$R$_{\rm 500c}$, the line-of-sight depth becomes small and the integrated signal is dominated by local gas, making both weighting schemes converge and the MW/SL ratio approach unity. These trends arise purely from projection effects applied to a spherically symmetric, single-temperature model.

\section{Voronoi temperature maps}
\label{appendix:2Ms}

\begin{figure}
    \centering
    \includegraphics[width=\columnwidth]{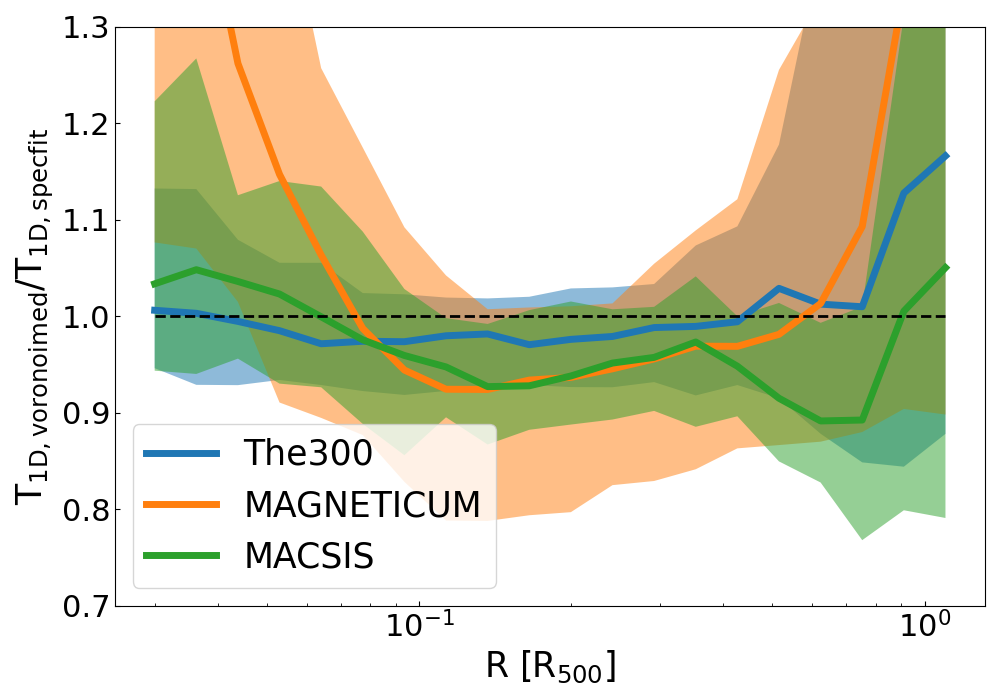}
    \caption{Comparison between the temperature profiles from X-ray spectral fitting and azimuthal median profiles obtained from voronoi tessellated temperature maps.}
    \label{fig:voronoiTxprofs}
\end{figure}

To further investigate on the 2D temperature distribution in our simulations, we assess the intrinsic azimuthal variations in the gas temperature profiles within each radial shell \citep[][]{Rasia2014ApJ_Tstruc, Lovisari2024A&A_CXMATETX}. To this end, we generate voronoi-tessellated images for each cluster, adopting a signal-to-noise ratio threshold of 30 per bin. We then extract and fit the X-ray spectrum in each voronoi region within R$_{\rm 500c}$, following the procedure outlined in Sect. \ref{sec:xray_analysis}. This yields temperature maps based on voronoi tessellation, which we use to characterize the azimuthal median temperature distribution at each radius. By comparing this to the results from direct X-ray spectral fitting, we can identify potential biases in the temperature measurement analogous to the analysis performed for surface brightness and gas density. The comparison is shown in Fig. \ref{fig:voronoiTxprofs}, where we plot the median ratio between the temperature profiles obtained from the azimuthal median of the voronoi temperature maps in concentric annuli and those derived from direct X-ray spectral fitting. Shaded regions represent the 16th to 84th percentile spread of the distribution.
In the innermost region within 0.1$\times$R$_{\rm 500c}$, the voronoi-derived temperature profile is very close or systematically higher than the one obtained from spectral fitting in annuli. This is expected, as the annular spectral fit is particularly sensitive to bright, dense, and cool gas components that dominate the X-ray emission in the core, biasing the temperature low. In contrast, the azimuthal median computed from the voronoi tessellation gives a more equal statistical weight to each spatial bin and is thus less affected by localized cool substructures. At such small radii the voronoi maps do not contain enough resolution elements to provide a meaningful comparison. Since they are on average less massive, reaching a given signal to noise cut requires integrating larger regions. In the voronoi temperature maps, we resolve the innermost 0.05$\times$R$_{\rm 500c}$ with a median of 4 resolution elements, compared to 12 in \texttt{MACSIS} and 25 in \texttt{The300}.
Between 0.1 and about 0.6$\times$R$_{\rm 500c}$, the trend reverses, with the voronoi temperatures falling below the annular spectral fits. Similarly to the projection effects discussed for the SL and MW schemes, this may reflect a stronger sensitivity of the annular spectral fitting to high-emissivity, denser, and hotter gas that dominates the X-ray signal outside the core, where both the temperature and density profiles decrease. The voronoi-based method, which downplays such emissivity-driven biases by construction, results in lower average temperatures.
Beyond about 0.7–1$\times$R$_{\rm 500c}$, the scatter in the temperature ratio increases significantly. This is largely driven by the larger voronoi cells required to maintain the desired signal-to-noise threshold in the outskirts, where the surface brightness is low. The reduced spatial resolution in these regions weakens the reliability of the comparison and introduces larger uncertainties in the azimuthal statistics.

To assess the robustness of our technique and rule out potential systematics, such as the limited spatial resolution of the voronoi bins, low signal-to-noise ratio, or contamination from AGN, we generate a dedicated high-exposure simulation (as described in Sect. \ref{subsec:xmm_simulator}) for a single cluster. This mock observation features a much deeper exposure time of 2 Ms and excludes AGN emission. We apply the same X-ray analysis pipeline and find that all thermodynamic profiles are in excellent agreement with those recovered from the standard 25 ks simulation well within the uncertainties. We also produce the voronoi-tessellated temperature maps for this deep mock observation. Given the computational expense of this process, which involves extracting and fitting spectra in 6002 voronoi bins, we perform it for only one system. This serves as a baseline for comparing the median temperature profile derived from the 2 Ms voronoi map to that obtained from the standard 25 ks exposure. The full map extracted within R$_{\rm 500c}$ is shown in Fig. \ref{fig:2Ms_Txmap}.

\begin{figure}
    \centering
    \includegraphics[width=\columnwidth]{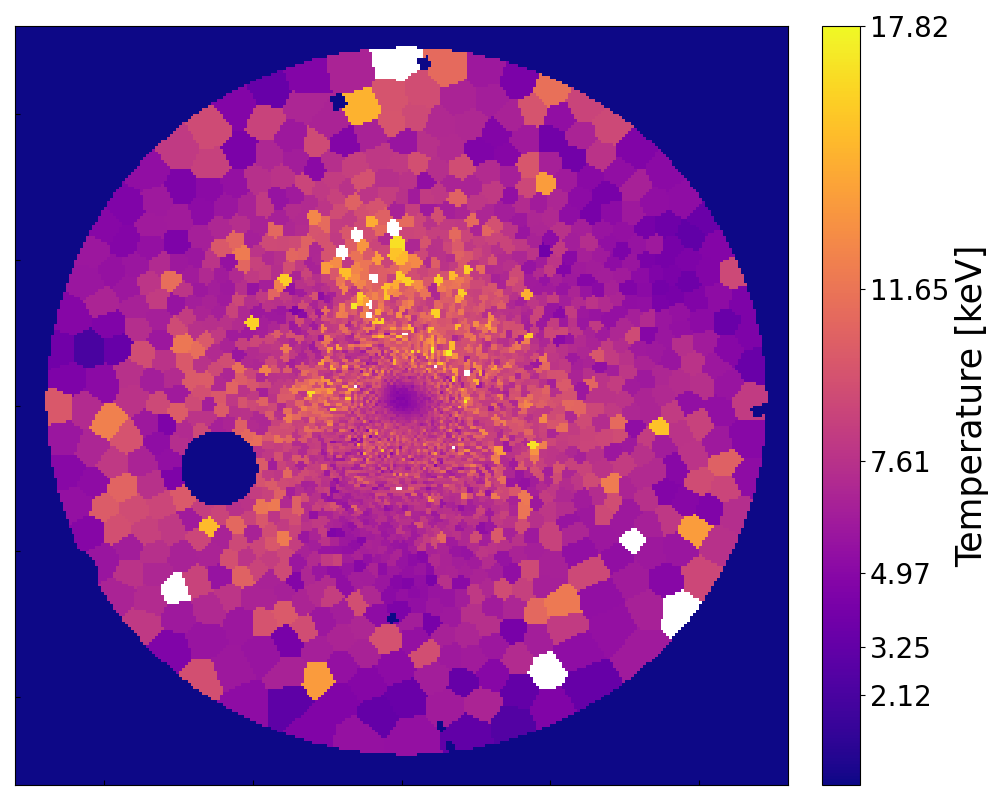}
    \caption{Voronoi-tessellated temperature map from the 2Ms XMM-Newton mock observation of the CL0030.1.115 cluster in \texttt{The300} simulation.}
    \label{fig:2Ms_Txmap}
\end{figure}

\begin{figure}
    \centering
    \includegraphics[width=\columnwidth]{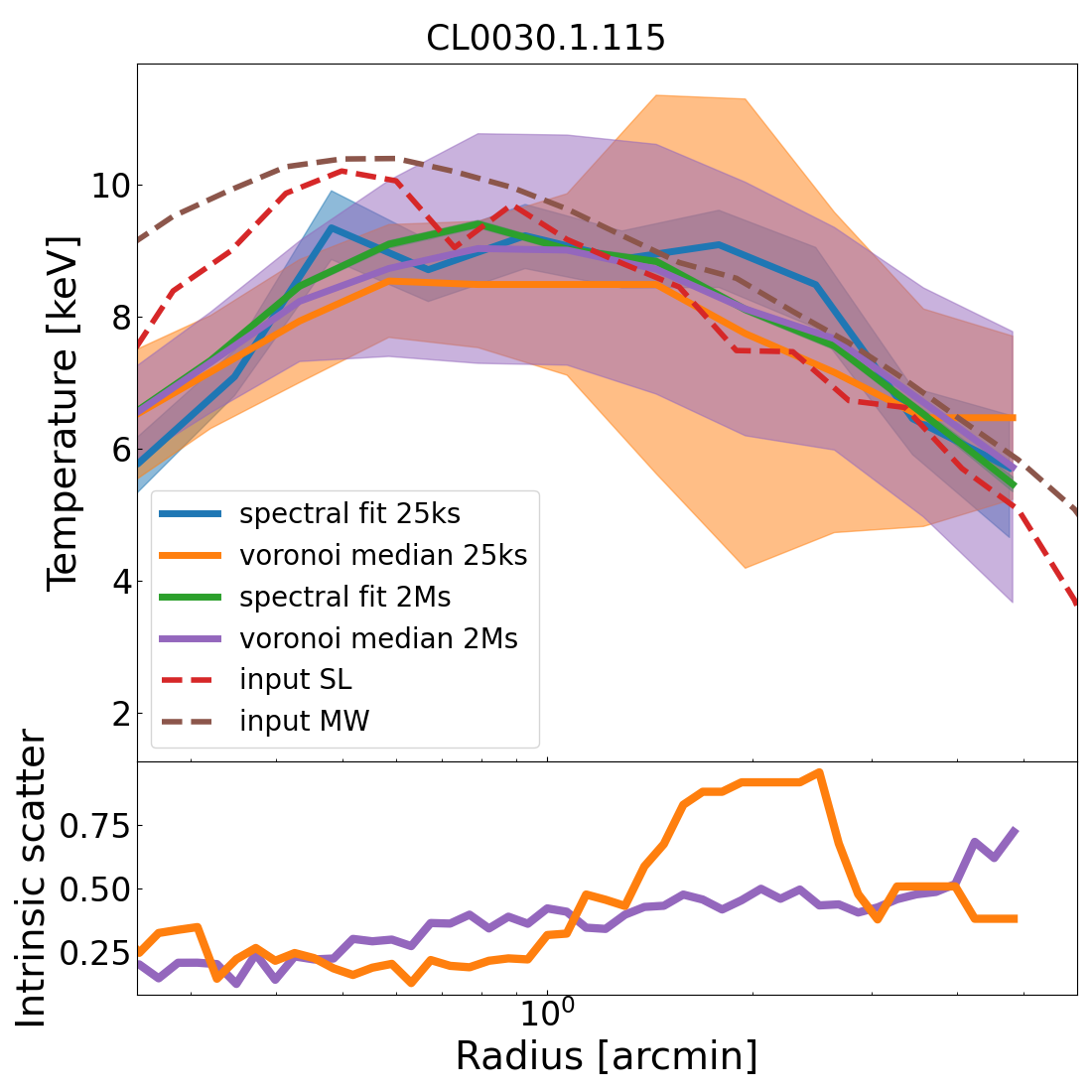}
    \caption{Temperature profiles extracted from the direct spectral fit and from the median of the voronoi maps for the 25ks and 2Ms simulations of CL0030.1.115.}
    \label{fig:kT_profs_comparison}
\end{figure}

The median profiles extracted from the maps and the ones from X-ray spectral fitting in annular bins are reported in the upper panel of Fig. \ref{fig:kT_profs_comparison}, together with the input SL and MW ones. The bottom panel show the relative intrinsic scatter in the voronoi median profiles extracted from the 25 ks and 2 Ms simulations. It is computed as the difference between the 84th and 16th percentile points of the distribution within each bin, divided by its median. This cluster has an R$_{\rm 500c}$=1491.5 kpc and is located at z=0.33, so that R$_{\rm 500c}$ covers an angular scale of about 5.05 arcmin. We find good agreement between the result of the X-ray spectral fitting in annular bins between the 25 ks and 2 Ms simulations, with the latter being smoother and with lower uncertainties than the former by about one order of magnitude, the typical error bar is about 0.05 keV compared to 0.5 keV. This specific systems shows a small inconsistency with the SL temperature in the inner core within 0.8 arcmin, corresponding to about 0.15$\times$R$_{\rm 500c}$. However, the same trend is visible in both the 25 ks and 2 Ms simulations. Similarly to the spectral fit in radial bins, also the median temperature profiles extracted from the voronoi maps are in agreement, with a discrepancy of about 6$\%$ at most, but still very well within the 1$\sigma$ interval. Finally, in both simulation we see an increasing trend of the intrinsic scatter as a function of radius, which is about 0.25 in the cluster core and get closer to 0.5 towards R$_{\rm 500}$. This is expected, as small regions in the core are more likely to include gas that is well described by a single temperature, whereas larger bins towards the outskirts are more likely to contain multi-phase gas, especially accounting for azimuthal variations in a single bin. The trend is smoother for the 2 Ms simulation, while the 25 ks one shows a big jump in the intrinsic scatter at about 2 arcmin, likely due to the low spatial resolution as a consequence of the large voronoi bins to achieve the desired signal to noise ratio. Nonetheless, the median properties of the profiles are overall in agreement between the 25 ks and the 2 Ms simulations.

%\rs{I tried to fit the same spectrum <r500c with apec and with gadem (Vittorio's comment). Here is what I get: from apec: T keV 7.6974 0.0212 0.0115 FFFFFFFFF Z solar 0.2512 0.0028 0.0033 FFFFFFFFF Norm 1.2651e-04 8.0000e-08 8.0000e-08 FFFFFFFFF and from gadem: T keV 8.7157 0.0271 0.0258 FFFFFFFFF Z solar 0.2740 0.0035 0.0035 FFFFFFFFF Norm 1.2874e-04 1.2000e-07 1.1000e-07 FFFFFFFFF sig 4.4972 0.0271 0.0258 FFFFFFFFF}

\subsection{Removing temperature fluctuations}

\begin{figure}
    \centering
    \includegraphics[width=\columnwidth]{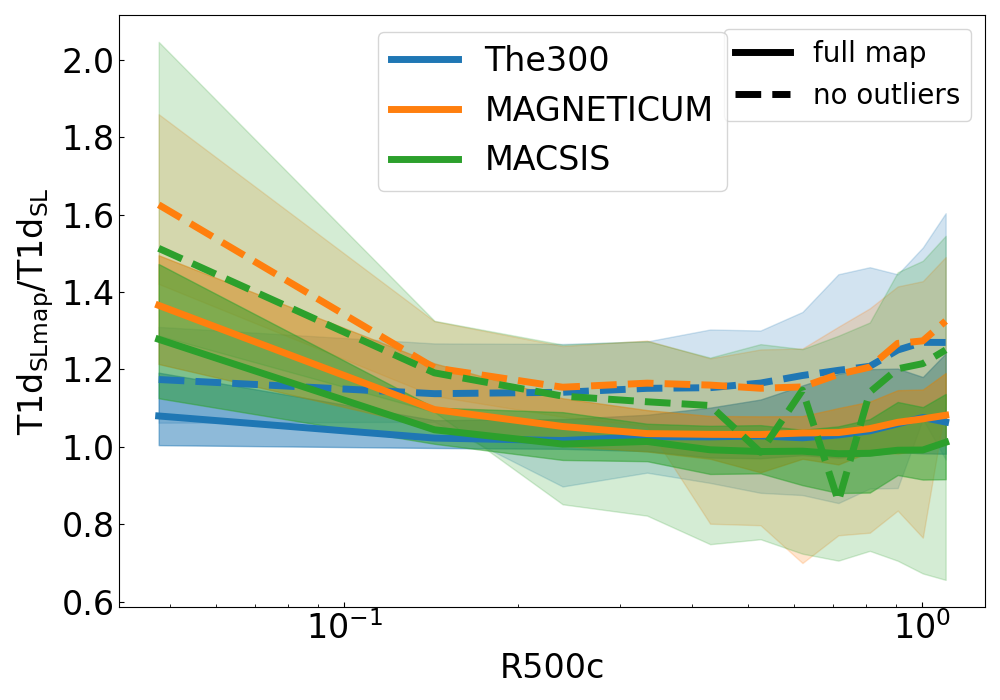}
    \caption{Comparison between the median of the 2D spectroscopic-like temperature maps and the 1D radial profile, before and after removing the most deviating regions.}
    \label{fig:Toutlier}
\end{figure}

We test the impact of removing the most deviating regions from the 2D temperature distribution on the projected radial temperature profile. This is an attempt of replicating a similar strategy applied to CHEX-MATE by \citet{Lovisari2024A&A_CXMATETX} from the perspective of the simulation. The authors identified regions deviating more than 1$\sigma$ from the ratio between the local 2D temperature map and the projected 1d profile. Such regions are likely associated to hot and cold gas clumps. We replicate the approach using 2D $T_{\rm SL}$ maps and 1D $T_{\rm SL}$ profiles, removing any systematic related to the temperature measurement by construction. In practice, for each pixel in the $T_{\rm SL}$ maps we compute the quantity S$_i$ = $\dfrac{T_{\rm 1D,SL} - T_{\rm 1D, SLmap}}{\sigma_{\rm T2D}}$, where $\sigma_{\rm T2D}$ is the standard deviation of the temperature map at a given radius. We then select and remove pixels with |S$_i$|>1 and recompute the 1D median profile without outliers.  The result is shown in Fig. \ref{fig:Toutlier}. We find that after removing outliers the temperature estimate increases by about 15-20$\%$, which means that we are preferentially removing clumps of cold gas that are easier to detect. This matches the result obtained by \citet{Lovisari2024A&A_CXMATETX}, who observed a similar trend with effects up to 10-20$\%$.

\section{Mis-centering}
\label{appendix:miscentering}

In this appendix we test whether any inconsistencies in the reconstruction of gas density profiles in Sect. \ref{sec:results} is due to mis-centering. In fact, from the observer's perspective, the profile is computed from the peak of the X-ray emission, while the input profiles use the position of the most bound particle, i.e. the deepest point of the potential well, as centre. 

From the full population in each simulation we subselect well-centred systems where the the offset between the peak of the X-ray emission and the dark matter is  $\Delta_{\rm X}$ < 0.007$\times$R$_{\rm 500c}$ and study the ratio between inferred and true density profiles. We obtain respectively 25, 15, and 15 well-centred systems in \texttt{The300}, \texttt{Magneticum}, and \texttt{MACSIS}. The result is shown in Fig. \ref{fig:ne_miscenter}, with the solid lines denoting the full samples and the dashed ones the well centred ones. The shaded areas denote the 16th-84th percentile points of the selected population. The ratio is basically unchanged in \texttt{The300}, which does not show significant inconsistencies even with the full sample. The issue is fully solved in \texttt{MACSIS}, where the ratio to true density is within 5$\%$ down to 0.06$\times$R$_{\rm 500c}$, whereas it crosses this threshold already at about 0.15$\times$R$_{\rm 500c}$ for the full sample. For \texttt{Magneticum} the discrepancy decreases drastically, reaching a radius of about 0.08$\times$R$_{\rm 500c}$ where the gas density is within 5$\%$ of the true value, compared to about 0.13$\times$R$_{\rm 500c}$ for the full sample. Additional differences may be due to deviations from sphericity. In any case, the gas mass enclosed in such regions is a small fraction of the total one and does not impact the total gas mass reconstruction, as shown in Fig. \ref{fig:Mgas}.

\begin{figure}
    \centering
    \includegraphics[width=\columnwidth]{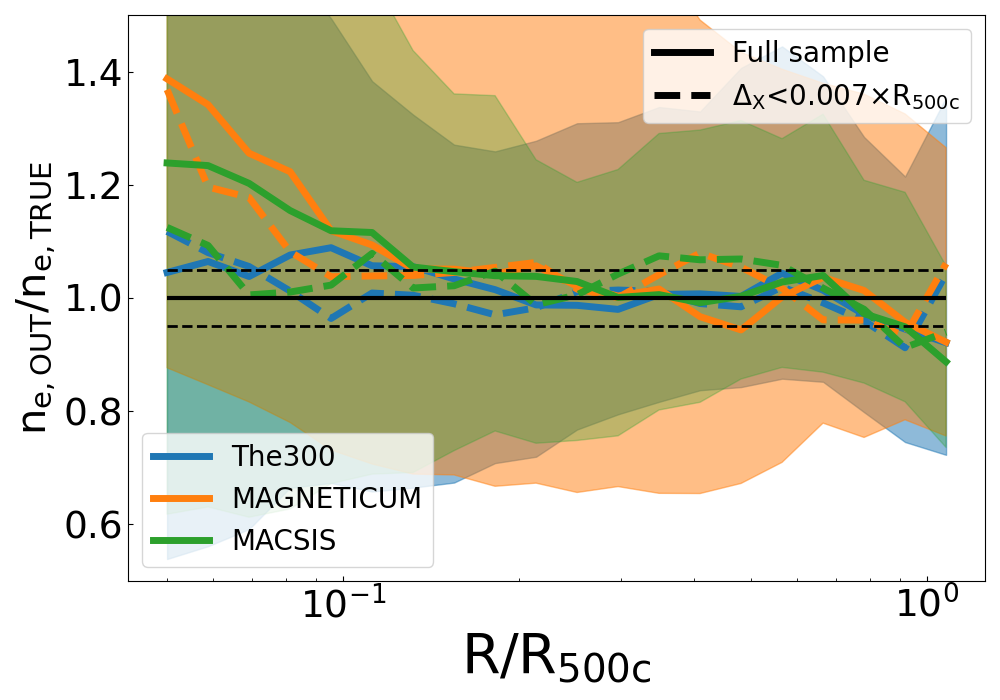}
    \caption{Ratio between reconstructed and true density profiles for all three simulations. The solid lines denote the full samples, the dashed ones refer only to the well centred objects, where $\Delta_{\rm X}$ is the offset between the identified X-ray centre and the dark matter one. The black dashed lines denote the $\pm5\%$ ratio.}
    \label{fig:ne_miscenter}
\end{figure}

\end{document}